\documentclass[letterpaper,showpacs,aps]{revtex4-2}

\usepackage[utf8]{inputenc}
\usepackage[english]{babel}

\usepackage{latexsym}
\usepackage{epsfig}
\usepackage{amsmath}
\usepackage{amssymb}
\usepackage{amsthm}
\usepackage{stackrel}
\usepackage{mathrsfs}
\usepackage{exercise}
\usepackage{eucal}
\usepackage{bm}
\usepackage{slashed}
\usepackage{tensor}
\usepackage[makeroom]{cancel}

\numberwithin{equation}{section}

\usepackage{epsfig}
\usepackage{graphicx}
\usepackage{float}

\usepackage{color}

\begin{document}


\title{The Dirac equation in General Relativity and the 3+1 formalism}

\author{Miguel Alcubierre}
\email{malcubi@nucleares.unam.mx}

\affiliation{Instituto de Ciencias Nucleares, Universidad Nacional
Aut\'onoma de M\'exico, A.P. 70-543, M\'exico D.F. 04510, M\'exico.}


\date{\today}


\begin{abstract}
I present a review of the Dirac equation in general relativity.
Although the generalization of the Dirac equation to a curved
spacetime is well known, it is not usually part of the standard
toolkit of techniques known to people working on classical general
relativity.  Recently, there has been some renewed interest in
studying solutions of the Einstein--Dirac system of equations,
particularly in the context of the so-called ``Dirac stars''.
Motivated by this, here I present a review of the Dirac equation in
general relativity, starting from Minkowski spacetime, and then
considering the Lorentz group and the tetrad formalism in order to
generalize this equation to the case of a curved spacetime.  I also
derive the form of the Dirac equation and its associated
stress--energy tensor for the case of the 3+1 formalism of general
relativity, which can be useful for the study of the evolution of the
Dirac field in a dynamical spacetime.
\end{abstract}




\maketitle


\section{Introduction}
\label{sec:intro}

The relativistic description of spin $1/2$ particles is given in terms
of the Dirac equation. This equation, proposed by Dirac in
1928~\cite{Dirac:1928}, is a first order equation in both time and
space that is fully Lorentz covariant and does not suffer the problem
of having negative probability densities as in the case of the
Klein--Gordon equation.  The price to pay is the need to introduce a
new type of geometric object different from vectors and tensors: a
four-component spinor that transforms according to its own special set
of rules with respect to a general Lorentz transformation.

The original form of the Dirac equation is perfectly consistent in
special relativity, but since Einstein's work in 1915 we know that in
the presence of gravity our Universe is not correctly described by
Minkowski spacetime, and one must use instead the curved spacetime
formalism of general relativity. A generalization of the Dirac
equation to the case of curved spacetimes was quickly found by Fock
and Ivanenko in 1929~\cite{Fock29a,Fock29b}, and later studied by
Bargmann~\cite{Bargmann:1932} and even
Schroedinger~\cite{Schroedinger:1932}.  However, due to the fact that
the gravitational field can be safely ignored when studying atomic
physics, this generalization was regarded for a long time as an
academic exercise with little practical applications.

Interest in the study of Dirac equation in curved spacetimes increased
in the 1970's with the work of Hawking on quantum field theory on
curved spacetimes, and since then one can find some discussion
(usually quite short) of the general relativistic version of the Dirac
equation in modern textbooks (see
e.g.~\cite{Birrel:1982,Freedman:2012}).  Still, the formulation of the
Dirac equation in a curved spacetime remains, even today, as something
that most researchers working in the field of general relativity never
study.

More recently, a revived interest in this subject has arisen related
to the work on exotic compact objects (ECO), in particular
self-gravitating stationary solutions of the Einstein equations
coupled to some matter field.  The case of ECO's formed by scalar
fields corresponds to the well-known boson stars initially introduced
by Kaup and Ruffini in the late 1960's~\cite{Kaup68,Ruffini69} (a
recent review can be found in~\cite{Liebling:2012fv}), while ECO's
formed from vector fields correspond to the more recently introduced
Proca stars of Brito et al.~\cite{Brito:2016}. The so-called ``Dirac
stars'', that is self-gravitating stationary solutions of the
Einstein--Dirac system, have also been considered, originally by
Finster in 1998~\cite{Finster:1999}, and more recently by Herdeiro et
al.~\cite{Herdeiro:2017} (see
also~\cite{Daka:2019,Sun2024a,Liang2024a}).

Motivated by these developments, here I present a review of the Dirac
equation in general relativity, starting from Minkowski spacetime, and
then considering the Lorentz group and the tetrad formalism in order
to generalize this equation to the case of a curved spacetime.  Other
reviews on this subject already exist in the literature (see
e.g.~\cite{Chapman:1976,Pollock:2010,Collas:2018,Shapiro:2022}), but
in my opinion none are fully comprehensive. Though most of the
material presented here is known, I try to present it in a pedagogical
way starting from first principles.  I also derive the form of the
Dirac equation and its associated stress--energy tensor for the
particular case of the 3+1 formalism of general relativity.  To my
knowledge, these last sections include new material which can be very
useful for the study of the evolution of the Dirac field in a
dynamical spacetime. Finally, I consider the particular example of
spherical symmetry: First the general case of the Dirac equation in a
spherically symmetric spacetime, and later the case of self-consistent
spherically symmetric solutions of the Einstein--Dirac system, and the
special configurations corresponding to Dirac stars.

A word about my conventions, throughout this paper I use the metric
signature $(-,+,+,+)$ and Planck units such that $c=\hbar=G=1$.


\section{The Dirac equation in special relativity}
\label{sec:DiracSR}


\subsection{Dirac equation}

The Dirac equation is a relativistic generalization of the
Schroedinger equation that describes the behaviour of spin 1/2
particles (the material presented in this section is well known, and
can be found in any standard text book on quantum field theory, see
e.g.~\cite{Itzykson1980,Kaku1993,Weinberg1995}). Before Dirac's work
in 1928~\cite{Dirac:1928}, there was already a relativistic
generalization of Schroedinger's equation, namely the Klein--Gordon
equation, which takes the form:
\begin{equation}
\Box \psi - m^2 \psi = 0 \; ,
\label{eq:KleinGordon}
\end{equation}
where $\Box$ is the standard d'Alambertian operator in special
relativity (though the equation takes exactly the same form in general
relativity (GR) with the curved d'Alambertian), and $m$ is a mass
parameter which corresponds to the mass of the associated particle
when the theory is quantized.  For a complex wave function $\psi$ one
can show that there is a conserved current given by:
\begin{equation}
j_\mu := - i \left( \psi^* \partial_\mu \psi - \psi \partial_\mu \psi^* \right) \; ,
\label{eq:KGcurrent}
\end{equation}
where $\psi^*$ denotes the complex conjugate of $\psi$, such that:
\begin{equation}
\nabla_\mu j^\mu = 0 \; .
\end{equation}

The main problem with interpreting the Klein--Gordon equation as a
quantum equation comes from the fact that the density associated with
the above conserved current is given by:
\begin{equation}
\rho \equiv j^0 = i \left( \psi^* \partial_t \psi - \psi \partial_t \psi^* \right) \; .
\label{eq:KGdensity}
\end{equation}
This expression is clearly not positive definite, so it can not be
associated with a probability density. This problem can be traced back
to the fact that, in contrast with the Schroedinger equation, the
Klein--Gordon equation is of second order in time. Because of this the
Klein--Gordon equation was initially rejected as a valid quantum
equation, which motivated Dirac to look for a relativistic equation
that was first order in time. The problem with the Klein--Gordon
equation was later solved in quantum field theory by associating the
density $\rho$ above not with a probability density, but rather with a
charge density, allowing it to describe particles and antiparticles of
opposite charge. However, this equation does not include the effects
of the spin of the particles, so today it is considered to describe
only scalar (spin 0) particles such as for example the Higgs
boson. But we will not go any deeper into quantum field theory here
and we will instead regard both the Klein--Gordon equation and the
Dirac equation below as purely ``classical'' field equations.

Since having an equation that is first order in time and second order
in space, such as the Schroedinger equation, clearly violates Lorentz
invariance, Dirac proposed a purely first order expression for the
Hamiltonian operator of the form:
\begin{equation}
\hat{H} \psi = \left( \bm{\alpha} \cdot \hat{\bm{p}} + \beta m \right) \psi \; .
\label{eq:Dirac1}
\end{equation}
where bold letters indicate three-dimensional objects, with
$\hat{p}_i$ the usual momentum operator, and where the quantities
$\alpha_i$ and $\beta$ are constant coefficients to be determined.  We
can immediately see that, if the $\alpha_i$ were the components of a
simple three-dimensional vector, the above equation would give
preference to a specific direction in space, in clear violation of
relativistic invariance, so they must be other type of objects.

If we now want our Hamiltonian operator to satisfy the relativistic
energy--momentum relation we must ask for:
\begin{equation}
\hat{H}^2 \psi = \left( \hat{\bm{p}}^2 + m^2 \right) \psi \; .
\label{eq:relation-EP}
\end{equation}
Taking the square of equation~\eqref{eq:Dirac1} we now find (notice
that here we are not assuming that the $\alpha_i$ and $\beta$
commute):
\begin{equation}
\hat{H}^2 = \sum_i \alpha_i^2 \hat{p}^2_i
+ \sum_{i > j} \left( \alpha_i \alpha_j + \alpha_j \alpha_i \right) \hat{p}_i \hat{p}_j
+ \sum_i m \left( \alpha_i \beta + \beta \alpha_i \right) \hat{p}_i + \beta^2 m^2 \; .
\end{equation}
Comparing this with~\eqref{eq:relation-EP} we find that we must have
$\alpha^2_i = \beta^2 = 1$, plus the objects
$(\alpha_1,\alpha_2,\alpha_3,\beta)$ must all anti-commute with each
other. Given these anti-commutation relations we must conclude that
these objects are not simple numbers, but must be matrices of some
dimension.  One can also show that in order to satisfy all these
relationships such matrices must be at least of dimension $4 \times
4$. One such set of matrices are known as the {\em Dirac--Pauli}\/
matrices, and are given by:
\begin{equation}
\alpha_i = \left(
\begin{array}{cc}
0 & \sigma_i \\
\sigma_i & 0 
\end{array}
\right) \; ,
\qquad
\beta = \left(
\begin{array}{cc}
I_2 & 0 \\
0 & -I_2 
\end{array}
\right) \; ,
\label{eq:Dirac-AB}
\end{equation}
where $I_2$ is the $2 \times 2$ identity matrix, and where the
$\sigma_i$ are the usual $2 \times 2$ Pauli matrices:
\begin{equation}
\sigma_1 = \left(
\begin{array}{cc}
0 & 1 \\
1 & 0 
\end{array}
\right) \; ,
\qquad
\sigma_2 = \left(
\begin{array}{cc}
0 & -i \\
i & 0 
\end{array}
\right) \; ,
\qquad
\sigma_3 = \left(
\begin{array}{cc}
1 & 0 \\
0 & -1 
\end{array}
\right) \; .
\label{eq:PauliMatrices}
\end{equation}

Given the fact that the Pauli matrices are hermitian (i.e. equal to
their conjugate transpose), the $\alpha_i$ and $\beta$ are also
hermitian.  It is important to remember at this point that the Pauli
matrices anti-commute with each other and are such that
$\sigma_i^2=1$.  Both these properties can be combined into the
expression:
\begin{equation}
\left\{ \sigma_i , \sigma_j \right\} = 2 \delta_{ij} I_2 \; ,
\label{eq:Pauli-anticommute}
\end{equation}
where $\{,\}$ denotes the anticommutator defined as
$\{\sigma_i,\sigma_j\}:= \sigma_i \sigma_j + \sigma_j \sigma_i$.

One should mention the fact that the above choice for the matrices
$\alpha_i$ and $\beta$ is clearly not unique.  In fact, any set of matrices
related to the above choice by a transformation of the form:
\begin{equation}
\alpha'_k = U \alpha_k U^{-1} \; , \qquad
\beta' = U \beta U^{-1} \; ,
\end{equation}
with $U$ a unitary matrix, would satisfy the same relations. Another
common representation is the {\em Weyl}\/ or {\em quiral}\/ representation,
and is given by:
\begin{equation}
\alpha_i = \left(
\begin{array}{cc}
- \sigma_i & 0 \\
0 & \sigma_i 
\end{array}
\right) \; ,
\qquad
\beta = \left(
\begin{array}{cc}
0 & I_2 \\
I_2 & 0 
\end{array}
\right) \; .
\label{eq:Weyl-AB}
\end{equation}

Now, given the fact that the Dirac equation~\eqref{eq:Dirac1} involves
matrix operators of dimension $4 \times 4$, we conclude that the
``wave function'' $\psi$ must in fact be a complex column vector with
four components known as a {\em Dirac spinor}.  The Dirac equation
then represents a set of four coupled equations for the four complex
components of $\psi$. Notice that even if the Dirac spinor has four
components, it is not a 4-vector in the usual spacetime sense, but
rather a collection of four complex numbers that transform in a
special way under rotations, as we will see below.


\subsection{Covariant form}

The Dirac equation~\eqref{eq:Dirac1} can easily be rewritten in a
manifestly covariant form. In order to do this we first write the
standard energy and momentum operators in the usual form (remember that
we are taking $\hbar=1$):
\begin{equation}
\hat{H} = i \partial_t \; , \qquad
\hat{P}_i = - i \partial_i \; .
\end{equation}
The equation now takes the form:
\begin{equation}
i \partial_t \psi = - i \alpha^i \partial_i \psi + \beta m \psi \; ,
\label{eq:DiracCov0}
\end{equation}
where we have defined $\alpha^i \equiv \alpha_i$.  From here on we
will adopt the convention that Greek indices take values from 0 to 3,
with 0 corresponding to the time coordinate, while Latin indices only
take values from 1 to 3. We also adopt the Einstein summation
convention: repeated indices in the same term, once covariant and once
contravariant, are to be summed over all their allowed
values. Multiplying the above equation with $-i$ we obtain:
\begin{equation}
\partial_t \psi = - \alpha^i \partial_i \psi - i \beta m \psi \; .
\label{eq:DiracEvol}
\end{equation}
Written in this form the Dirac equation can be interpreted as an
evolution equation in time for $\psi$. This is important, for example,
if one is interested in dynamical simulations of solutions of the
Dirac equation.  Multiplying now~\eqref{eq:DiracCov0} with $\beta$
from the left, and rearranging terms we find:
\begin{equation}
i \beta \partial_t \psi + i \beta \alpha^i \partial_i \psi
- m \psi = 0 \; ,
\end{equation}
where we used the fact that $\beta^2 = 1$. The above equation can now
be written in covariant form as:
\begin{equation}
i \gamma^\mu \partial_\mu \psi - m \psi = 0 \; .
\label{eq:DiracCov1}
\end{equation}
where we defined the $\gamma^\mu$ matrices as:
\begin{equation}
\gamma^0 := \beta \; , \qquad \gamma^k := \beta \alpha^k \; .
\end{equation}
The above matrices are the so-called {\em Dirac matrices}. In
the standard representation they take the form:
\begin{equation}
\gamma^0 = \left(
\begin{array}{cc}
I_2 & 0 \\
0 & -I_2 
\end{array}
\right) \; ,
\qquad
\gamma^k = \left(
\begin{array}{cc}
0 & \sigma_k \\
-\sigma_k & 0 
\end{array}
\right) \; ,
\label{eq:DiracMatrices}
\end{equation}
while in the Weyl representation they are instead:
\begin{equation}
\gamma^0 = \left(
\begin{array}{cc}
0 & I_2 \\
I_2 & 0 
\end{array}
\right) \; ,
\qquad
\gamma^k = \left(
\begin{array}{cc}
0 & \sigma_k \\
-\sigma_k & 0 
\end{array}
\right) \; .
\label{eq:WeylMatrices}
\end{equation}
It is usual to define the operator $\slashed{\partial} := \gamma^\mu
\partial_\mu$, so that the Dirac equation takes the more compact form:
\begin{equation}
\left( i \slashed{\partial} - m \right) \psi = 0 \; .
\label{eq:DiracCov2}
\end{equation}

From the definition of the $\gamma^\mu$ matrices one can show that:
\begin{equation}
\gamma^\mu \gamma^\nu + \gamma^\nu \gamma^\mu = - 2 \eta^{\mu \nu} I_4 \; ,
\label{eq:DiracMatrices-metric}
\end{equation}
with $\eta^{\mu \nu}$ the Minkowski tensor, and where here $I_4$
denotes the $4 \times 4$ identity matrix.  Here it is
important to mention the fact that the above relation is obtained when
using a signature $(-,+,+,+)$ for the metric. In quantum field theory
text books it is common to use the opposite signature, so that the
term on the right hand side above changes sign. The previous relation
defines what is known as a {\em Clifford algebra}.  In particular we
have:
\begin{equation}
\left( \gamma^0 \right)^2 = I_4 \; , \qquad \left( \gamma^k \right)^2 = - I_4 \; ,
\label{eq:gamma2}
\end{equation}
and:
\begin{equation}
\gamma^0 \gamma^k + \gamma^k \gamma^0 = 0 \; ,
\end{equation}
that is, $\gamma^0$ and $\gamma^k$ anti-commute with each other.
One can also show that:
\begin{equation}
{\gamma^0}^\dag = \gamma^0 \; , \qquad {\gamma^k}^\dag = - \gamma^k \; ,
\end{equation}
where the symbol ${}^\dag$ denotes the transpose conjugate. That is,
$\gamma^0$ is hermitian while the $\gamma^k$ are anti-hermitian.  These
last relations can be summarized as:
\begin{equation}
\gamma^0 \gamma^\mu \gamma^0 = {\gamma^\mu}^\dag \; .
\label{eq:gamma-0mu0}
\end{equation}

It is common to also define the matrix $\gamma^5$ as follows:
\begin{equation}
\gamma^5 := i \gamma^0 \gamma^1 \gamma^2 \gamma^3 \; .
\end{equation}
The use of the number 5 comes from the fact that many older texts take
the spacetime coordinates to run from 1 to 4 instead of from 0 to 3 as
we do here. This matrix is useful for many calculations in quantum
field theory, but we will not consider it further here.

\vspace{5mm}

There is another useful relation that can be obtained from the
Clifford algebra that allows us to commute two pairs of Dirac
matrices.  From~\eqref{eq:DiracMatrices-metric} one can show,
after some algebra:
\begin{equation}
\gamma^\alpha \gamma^\beta \gamma^\mu \gamma^\nu = \gamma^\mu \gamma^\nu \gamma^\alpha \gamma^\beta
+ 2 \left( \eta^{\alpha \mu} \gamma^\beta \gamma^\nu - \eta^{\beta \mu} \gamma^\alpha \gamma^\nu
+ \eta^{\alpha \nu} \gamma^\mu \gamma^\beta - \eta^{\beta \nu} \gamma^\mu \gamma^\alpha \right) \; .
\label{eq:fourgammas}
\end{equation}


\subsection{Adjoint equation, conserved current and the Klein--Gordon equation}

In order to find a conserved current associated with the Dirac
equation we start from considering its hermitian conjugate.  We first
write the Dirac equation in extended form as:
\begin{equation}
i \gamma^0 \partial_t \psi + i \gamma^k \partial_k \psi - m \psi = 0 \; ,
\end{equation}
so that its hermitian conjugate takes the form:
\begin{equation}
- i \left( \partial_t \psi^\dag \right) \gamma^0
- i \left( \partial_k \psi^\dag \right) (-\gamma^k) - m \psi^\dag = 0 \; .
\end{equation}
Notice that in the above equation $\psi^\dag$ is now a row vector,
while $\psi$ is a column vector. Multiplying the last equation
with $\gamma^0$ from the right, and defining the {\em adjoint spinor}\/
as $\bar{\psi} := \psi^\dag \gamma^0$, we find:
\begin{equation}
i \left( \partial_\mu \bar{\psi} \right) \gamma^\mu + m \bar{\psi} = 0 \; ,
\label{eq:DiracAdjoint}
\end{equation}
where we used the fact that $\gamma^0$ and $\gamma^k$ anti-commute.
The last equation is known as the adjoint Dirac equation.

We can now multiply the Dirac equation~\eqref{eq:DiracCov1} with
$\bar{\psi}$ on the left, and the adjoint
equation~\eqref{eq:DiracAdjoint} with $\psi$ on the right, and add the
resulting equations together (remembering that the $\gamma^\mu$
matrices are constant).  We then find:
\begin{equation}
\bar{\psi} \left( \gamma^\mu \partial_\mu \psi \right)
+ \left( \partial_\mu \bar{\psi} \right) \gamma^\mu \psi
= \partial_\mu \left( \bar{\psi} \gamma^\mu \psi \right) = 0 \; .
\end{equation}
This result implies that we have a conserved current of the form:
\begin{equation}
j^\mu = \bar{\psi} \gamma^\mu \psi \; .
\label{eq:DiracCurrent}
\end{equation}
In particular, the associated density is now given by:
\begin{equation}
\rho = \bar{\psi} \gamma^0 \psi = \psi^\dag \psi
= \sum_{i=1}^4 \left| \psi_i \right|^2 \; ,
\label{eq:DiracDensity}
\end{equation}
where here (and in similar expressions below) the index $i$ labels
spinor components.  Clearly $\rho$ is now positive definite and can be
interpreted as a probability density, which was Dirac's main
motivation.

\vspace{5mm}

Let us now go back to equation~\eqref{eq:DiracCov1}.  Applying the
operator $-i \gamma^\nu \partial_\nu$ from the left we obtain:
\begin{equation}
\gamma^\nu \gamma^\mu \partial_\nu \partial_\mu \psi 
+ i m \gamma^\nu \partial_\nu \psi = 0 \; ,
\end{equation}
where we again used the fact that the $\gamma^\mu$ matrices are
constant. Using again~\eqref{eq:DiracCov1} in the second term this
reduces to:
\begin{equation}
\gamma^\nu \gamma^\mu \partial_\nu \partial_\mu \psi + m^2 \psi = 0 \; .
\label{eq:KG0}
\end{equation}
On the other hand, since partial derivatives commute we can write:
\begin{equation}
\left( \gamma^\mu \gamma^\nu + \gamma^\nu \gamma^\mu \right) \partial_\mu \partial_\nu \psi
= 2 \gamma^\mu \gamma^\nu \partial_\mu \partial_\nu \psi \; .
\end{equation}
Using now the Clifford algebra,
relation~\eqref{eq:DiracMatrices-metric}, this implies that:
\begin{equation}
\gamma^\mu \gamma^\nu \partial_\mu \partial_\nu \psi = - \eta^{\mu \nu} \partial_\nu \psi 
= - \Box \psi \; ,
\end{equation}
so that equation~\eqref{eq:KG0} finally reduces to:
\begin{equation}
\Box \psi - m^2 \psi = 0 \; ,
\end{equation}
which is nothing more than the Klein--Gordon equation.  That is, each
of the individual components of the spinor $\psi$ obey the
Klein--Gordon equation separately.  However, will see in the following
sections that this is strictly true only in a flat spacetime and in
Cartesian coordinates.


\section{The Lorentz group}
\label{sec:LorentzGroup}

\subsection{Tensor representation}
\label{sec:Lorentz-tensor}

In order to study the behaviour of the Dirac equation under Lorentz
transformations we must first understand in some detail the {\em
  Lorentz group}, which includes the proper Lorentz transformations as
well as the three-dimensional rotations in space.  A general Lorentz
transformation is defined as a linear (and real) coordinate
transformation that leaves the Minkowski interval invariant.  Such a
transformation can be represented in the general form:
\begin{equation}
{x'}^\alpha = {\Lambda^\alpha}_\beta \: x^\beta \; ,
\label{eq:Lorentz1}
\end{equation}
where $\{ x^\alpha \}$ are the original coordinates, $\{ {x'}^\alpha
\}$ are the new coordinates, and where ${\Lambda^\alpha}_\beta :=
\partial {x'}^\alpha/ \partial x^\beta$ is the jacobian matrix that
must be constant for a linear transformation.  Notice that in the
above expression the order of the indices in the matrix
${\Lambda^\alpha}_\beta$ matters, as we will see in a moment.

It is important to mention the fact that in general we are not
assuming any symmetry properties for the matrix
${\Lambda^\alpha}_\beta$.  For example, for as Lorentz ``boost'' along
the $x$ direction the matrix turns out to be symmetric, while for a
three-dimensional rotation in space around the $x$ axis it is
antisymmetric. We will return to this point below.

We can now raise and lower indices of the jacobian matrix using the
Minkowski tensor $\eta_{\alpha \beta}$ to construct, for example:
\begin{equation}
\Lambda_{\alpha \beta} = \eta_{\alpha \mu} {\Lambda^\mu}_\beta \; , \qquad
\Lambda^{\alpha \beta} = \eta^{\beta \mu} {\Lambda^\alpha}_\mu \; , \qquad
{\Lambda_\alpha}^\beta = \eta_{\alpha \mu} \eta^{\beta \nu} {\Lambda^\mu}_\nu \; .
\end{equation}
In particular, since the Minkowski tensor is invariant under Lorentz transformations
by definition, we must have:
\begin{equation}
\eta^{\alpha \beta} = {\Lambda^\alpha}_\mu {\Lambda^\beta}_\nu \: \eta^{\mu \nu}
= \Lambda^{\alpha \nu} {\Lambda^\beta}_\nu \; ,
\end{equation}
which implies:
\begin{equation}
\Lambda^{\alpha \nu} \Lambda_{\beta \nu}
= {\Lambda^\alpha}_\nu {\Lambda_\beta}^\nu = \delta^\alpha_\beta \; .
\label{eq:deltas1}
\end{equation}

On the other hand, for the inverse transformation we have:
\begin{equation}
x^\alpha = {(\Lambda^{-1})^\alpha}_\beta \: {x'}^\beta \; ,
\label{eq:Lorentz2}
\end{equation}
so that:
\begin{equation}
{(\Lambda^{-1})^\alpha}_\mu {\Lambda^\mu}_\beta = \delta^\alpha_\beta \; , \qquad
{\Lambda^\alpha}_\mu {(\Lambda^{-1})^\mu}_\beta = \delta^\alpha_\beta \; .
\end{equation}
Comparing this with~\eqref{eq:deltas1} we find:
\begin{equation}
{(\Lambda^{-1})^\alpha}_\beta = {\Lambda_\beta}^\alpha \; ,
\label{eq:inversetransf}
\end{equation}
so the inverse transformation takes the form:
\begin{equation}
x^\alpha = {\Lambda_\beta}^\alpha \: {x'}^\beta \; .
\end{equation}
That is, in order to obtain the inverse of the jacobian matrix we must
lower the first index, raise the second index, and transpose the
matrix. In matrix notation this result can be written as $\Lambda^{-1}
= \eta \: \Lambda^T \eta$.  In particular, for a rotation in space
raising and lowering indices has no effect, so the inverse is simply
the transpose, and since the jacobian matrix is real we see that
rotations in space correspond to orthogonal matrices (with inverse
equal to the transpose).  In contrast, for a Lorentz boost raising
one index and lowering the other changes the sign of the first column
and first row (keeping the $00$ component unchanged), and leaves us
again with a symmetric matrix, so now taking the transpose has no
effect.  A Lorentz boost therefore does not correspond to an orthogonal
matrix.

Equation~\eqref{eq:inversetransf} also implies the following relations
(compare this with~\eqref{eq:deltas1}):
\begin{equation}
\Lambda^{\mu \alpha} \Lambda_{\mu \beta}
= {\Lambda_\mu}^\alpha {\Lambda^\mu}_\beta
= \delta^\alpha_\beta \; .
\label{eq:deltas2}
\end{equation}

Given the previous results, the Lorentz transformations of vectors and
1-forms take the form:
\begin{equation}
{v'}^\alpha = {\Lambda^\alpha}_\beta v^\beta \; , \qquad
{q'}_\alpha = {\Lambda_\alpha}^\beta q_\beta \; ,
\end{equation}
which can be generalized directly to tensors of arbitrary range.  In
particular, the coordinate basis vectors $(\vec{e}_\alpha)$ 
transform in the same way as the components of a 1-form:
\begin{equation}
{\vec{e}}\:'_\alpha = {\Lambda_\alpha}^\beta \vec{e}_\beta \; .
\end{equation}

On the other hand, since the determinant of the Minkowski tensor
is $-1$, we also find that:
\begin{equation}
\left[ \det \left( {\Lambda^\alpha}_\beta \right) \right]^2 = 1 \qquad
\implies \quad \left[ \det \left( {\Lambda^\alpha}_\beta \right) \right] =  \pm 1 \; .
\end{equation}

The group of Lorentz transformations in known as $O(3,1)$, which is
the general group that leaves the Minkowski interval invariant.  If we
restrict ourselves to those transformations that have determinant
equal to $+1$ we obtain the ``special'' or ``proper'' Lorentz group
$SO(3,1)$.  If, moreover, we ask for the direction of time to remain
the same, that is we ask for ${\Lambda^0}_0 \geq 1$, we obtain the
orthochronous Lorentz group.

\vspace{5mm}

Let us now consider an infinitesimal transformation of the form:
\begin{equation}
{\Lambda^\alpha}_\beta = {\delta^\alpha}_\beta + {\lambda^\alpha}_\beta \; ,
\end{equation}
with $|{\lambda^\alpha}_\beta| \ll 1$. Raising and lowering indices
we find:
\begin{equation}
\Lambda_{\alpha \beta} = \eta_{\alpha \beta} + \lambda_{\alpha \beta} \; , \qquad
\Lambda^{\alpha \beta} = \eta^{\alpha \beta} + \lambda^{\alpha \beta} \; .
\end{equation}
These results imply that, to first order in small quantities, we must have:
\begin{equation}
\lambda^{\alpha \beta} + \lambda^{\beta \alpha} = 0 \; ,
\end{equation}
that is, $\lambda^{\alpha \beta}$ must be antisymmetric.  In four dimensions
such a matrix has only 6 independent components which correspond to
the three spatial rotations and the three possible boosts.

The next step is to introduce a basis for the space of the $4 \times
4$ antisymmetric matrices.  This basis must be clearly formed from 6
antisymmetric matrices $M^A$ that are linearly independent from each
other, with $A=1,...,6$.  In fact, it turns out to be very convenient
to replace the index $A$ with a new pair of antisymmetric indices, so
that our 6 basis matrices will now be $M^{\rho \sigma}$, with elements
given by $(M^{\rho \sigma})^{\alpha \beta}$.  This notation can seem
somewhat cumbersome at first, but notice that we can now find an
explicit expression for our basis matrices as:
\begin{equation}
(M^{\rho \sigma})^{\alpha \beta} = - \eta^{\rho \alpha} \eta^{\sigma \beta}
+ \eta^{\sigma \alpha} \eta^{\rho \beta} \; .
\label{eq:antisym-basis1}
\end{equation}
In the above expression $(\alpha,\beta)$ denote the different elements
of a given matrix, while $(\rho,\sigma)$ denote which particular
matrix we are considering.  By direct computation it is not difficult
to show that the basis introduced above corresponds to taking, for the
matrices $M^{0i}$ with $i=1,2,3$:
\begin{equation}
(M^{0i})^{0i} = - (M^{0i})^{i0} = + 1 \; ,
\end{equation}
with all other components equal to zero, and for the matrices $M^{ij}$
with $i,j=1,2,3$:
\begin{equation}
(M^{ij})^{ij} = - (M^{ij})^{ji} = -1 \; ,
\end{equation}
again with all other components equal to zero.  That is, each basis
matrix takes one of the six independent components of a general $4
\times 4$ antisymmetric matrix equal to $\pm 1$, with all other other
independent components equal to zero.

The matrix $\lambda^{\alpha \beta}$  associated with an infinitesimal
Lorentz transformation can now be written as a linear combination
of our basis matrices in the form:
\begin{equation}
\lambda^{\alpha \beta} = \frac{1}{2} \: C_{\rho \sigma}
(M^{\rho \sigma})^{\alpha \beta} \; ,
\label{eq:Lorentz-infinitesimal1}
\end{equation}
where the coefficients $C_{\rho \sigma}$ are six small parameters
(antisymmetric in $\rho$ and $\sigma$) that identify the type of
transformation we are doing, that is, which precise combination of
rotations and boosts.  The factor $1/2$ is there to compensate for the
fact that the sum over $(\rho,\sigma)$ counts each independent term
twice.  In practice, in order to apply a Lorentz transformation we
need to lower one index and use ${\lambda^\alpha}_\beta$, where one
should remember that the matrices ${\lambda^\alpha}_\beta$ are no
longer necessarily antisymmetric (what we actually have is
${\lambda_\alpha}^\beta + {\lambda^\beta}_\alpha = 0$). It is
interesting to note that, given the form of the matrices $M$
in~\eqref{eq:antisym-basis1}, the infinitesimal Lorentz transformation
simply reduces to:
\begin{equation}
\lambda^{\alpha \beta} = - C^{\alpha \beta} \; .
\label{eq:Lorentz-infinitesimal2}
\end{equation}
The matrices $M^{\rho \sigma}$ are known as the {\em generators}\/ of
the Lorentz group, and satisfy the Lie algebra:
\begin{equation}
\left[ M^{\rho \sigma}, M^{\mu \nu} \right] = \eta^{\rho \mu} M^{\sigma \nu}
- \eta^{\rho \nu} M^{\sigma \mu} + \eta^{\sigma \nu} M^{\rho \mu}
- \eta^{\sigma \mu} M^{\rho \nu} \; .
\label{eq:LorentzAlgebra0}
\end{equation}

We can now express a finite Lorentz transformation by using the
standard exponential mapping given by:
\begin{equation}
{\Lambda^\alpha}_\beta = \exp \left( \frac{1}{2} \: C_{\rho \sigma}
{(M^{\rho \sigma})^\alpha}_\beta \right) \; .
\label{eq:L-exp}
\end{equation}

Is it common to give alternative names to the matrices $M$ as follows:
\begin{equation}
{\left( B_i \right)^\alpha}_\beta := {\left( M^{0i} \right)^\alpha}_\beta \; , \qquad
{\left( R_i \right)^\alpha}_\beta = \frac{1}{2} \: \epsilon_{ijk}
{\left( M^{jk} \right)^\alpha}_\beta \; ,
\end{equation}
where the Latin indices $(i,j,k)$ only take values from 1 to 3, and
with $\epsilon^{ijk}$ the Levi--Civita symbol in three dimensions. The
names of these new matrices are clearly associated with spatial
rotations, $R$, and Lorentz boosts, $B$. We find explicitly:
\begin{equation}
\rule{-3mm}{0mm} B_1 = \left(
\begin{array}{rrrr}
0 & +1 & 0 & 0 \\
+1 & 0 & 0 & 0 \\
0 & 0 & 0 & 0 \\
0 & 0 & 0 & 0 
\end{array}
\right) , \quad
B_2 = \left(
\begin{array}{rrrr}
0 & 0 & +1 & 0 \\
0 & 0 & 0 & 0 \\
+1 & 0 & 0 & 0 \\
0 & 0 & 0 & 0 
\end{array}
\right) , \quad
B_3 = \left(
\begin{array}{rrrr}
0 & 0 & 0 & +1 \\
0 & 0 & 0 & 0 \\
0 & 0 & 0 & 0 \\
+1 & 0 & 0 & 0 
\end{array}
\right) ,
\end{equation}
and:
\begin{equation}
\rule{-3mm}{0mm} R_1 = \left(
\begin{array}{rrrr}
0 & 0 & 0 & 0 \\
0 & 0 & 0 & 0 \\
0 & 0 & 0 & -1 \\
0 & 0 & +1 & 0 
\end{array}
\right) , \quad
R_2 = \left(
\begin{array}{rrrr}
0 & 0 & 0 & 0 \\
0 & 0 & 0 & +1 \\
0 & 0 & 0 & 0 \\
0 & -1 & 0 & 0 
\end{array}
\right) , \quad
R_3 = \left(
\begin{array}{rrrr}
0 & 0 & 0 & 0 \\
0 & 0 & -1 & 0 \\
0 & +1 & 0 & 0 \\
0 & 0 & 0 & 0 
\end{array}
\right) .
\end{equation}

We can now define the complex matrices $J_i := i R_i$ and $K_i := i B_i$,
which obey the following algebra:
\begin{equation}
[J_i,J_j] = i {\epsilon_{ij}}^k J_k \; , \qquad
[J_i,K_j] = i {\epsilon_{ij}}^k K_k \; , \qquad
[K_i,K_j] = - i {\epsilon_{ij}}^k J_k \; .
\label{eq:LorentzAlgebra1}
\end{equation}
These relations are equivalent to~\eqref{eq:LorentzAlgebra0} and define
the algebra of the Lorentz group.  The matrices $J$ generate spatial
rotations, while the matrices $K$ generate Lorentz boosts. Notice here
that the $J$ matrices are purely imaginary and antisymmetric, so they
are hermitian, while the matrices $K$ are purely imaginary and
symmetric, and as such are anti-hermitian.

With the previous definitions, a general Lorentz transformation can be
expressed as:
\begin{equation}
\Lambda = \exp \left( \vec{\theta} \cdot \vec{R}_l - \vec{\varphi} \cdot \vec{B}_l \right)
= \exp \left( - i \vec{\theta} \cdot \vec{J} + i \vec{\varphi} \cdot \vec{K} \right) \; ,
\label{eq:L-exp2}
\end{equation}
with $\theta_i$ and $\varphi_i$ parameters associated with spatial
rotations and boosts, respectively, and where the dot product is the
usual one in three dimensions (the negative sign on the term with the
$\varphi_i$ is necessary in order to recover the Lorentz boosts in
their usual form).  The $\theta_i$ represent directly rotation angles,
while the $\varphi_i$ are in fact velocity parameters as will be clear
below.

To continue we will now define 6 ``partial identity matrices''
$I_{\rho \sigma}$ as diagonal matrices such that their only non-zero
components are the $(\rho,\rho)$ and $(\sigma,\sigma)$ components which
are equal to 1. A little algebra now allows us to show that the square
of the $B$ and $R$ matrices is given by:
\begin{align}
\left( B_1 \right)^2 &= + I_{01} \; , \quad
\left( B_2 \right)^2 = + I_{02} \; , \quad
\left( B_3 \right)^2 = + I_{03} \; , \\
\left( R_1 \right)^2 &= - I_{23} \; , \quad
\left( R_2 \right)^2 = - I_{13} \; , \quad
\left( R_3 \right)^2 = - I_{12} \; . 
\end{align}
These relationships allow us to find recurrent formulas for any power
of the $B$ and $R$ matrices.  In particular, even powers of any $B_i$
are just the respective partial identity matrix, while odd powers are
the same $B_i$ again.  For the $R_i$ matrices the situation is similar
but with alternating signs.

We can then expand the exponential mapping in a Taylor series, so that
for the $B_i$ matrices we get only positive signs that can be grouped
into hyperbolic sines and cosines.  For example, for a boost along the
$x$ direction ($\varphi_1 = \varphi$, $\varphi_2=\varphi_3=0$,
$\theta_i=0$) we find:
\begin{equation}
\Lambda = \exp \left( - \varphi B_1 \right) 
= \left(
\begin{array}{cccc}
 \cosh(\varphi) & - \sinh(\varphi) & 0 & 0 \\
- \sinh(\varphi) & \cosh(\varphi) & 0 & 0 \\
0 & 0 & 0 & 0 \\
0 & 0 & 0 & 0 
\end{array}
\right) \; ,
\label{eq:L-BoostX}
\end{equation}
and similarly for the $y$ and $z$ directions. This can be immediately
recognized as a usual Lorentz boost written in terms of the velocity
parameter $\varphi$ defined as $v = \tanh(\varphi)$, with $v$ the
speed associated with the boost.

On the other hand, for the $R_i$ matrices the alternating signs result
in the Taylor expansion for the standard trigonometric sines and
cosines.  For example, for a rotation around the $x$ axis ($\varphi_i
= 0$, $\theta_1=\theta, \theta_2 = \theta_3 = 0$) we now find:
\begin{equation}
\Lambda = \exp \left( \theta R_1 \right)
= \left(
\begin{array}{cccc}
0 & 0 & 0 & 0 \\
0 & 0 & 0 & 0 \\
0 & 0 & \cos(\theta) & - \sin(\theta) \\
0 & 0 &  \sin(\theta) & \cos(\theta)
\end{array}
\right) \; ,
\label{eq:L-RotX}
\end{equation}
which is the usual rotation matrix around the $x$ axis.


\subsection{Spinor representation}
\label{sec:Lorentz-spinor}

The algebra of the Lorentz group is defined by the commutation
relations~\eqref{eq:LorentzAlgebra0}, but the specific form of the
matrices, and even their rank, can change when we pass from one
representation of the algebra to another. In order to see how this
is related to the Dirac equation, let us define a new set
of matrices  $\sigma^{\mu \nu}$ of the form:
\begin{equation}
\sigma^{\mu \nu} := \frac{1}{4} \left[ \gamma^\mu , \gamma^\nu \right] 
= \frac{1}{2} \left( \gamma^\mu \gamma^\nu + \eta^{\mu \nu} I_4 \right) \; ,
\label{eq:Sigma-munu}
\end{equation} 
with $\gamma^\mu$ the Dirac matrices we defined before
in~\eqref{eq:DiracMatrices}, and where
$[\gamma^\mu,\gamma^\nu]:=\gamma^\mu \gamma^\nu -\gamma^\nu
\gamma^\mu$ now denotes the commutator, and in the second equality we
used the Clifford algebra~\eqref{eq:DiracMatrices-metric}. Notice in
particular that we have $\sigma^{\mu \mu}=0$. From this definition it
is not difficult to show that:
\begin{equation}
\left[ \sigma^{\mu \nu} , \gamma^\rho \right]
= \eta^{\mu \rho} \gamma^\nu - \eta^{\nu \rho} \gamma^\mu \; .
\label{eq:N-gamma}
\end{equation}
And using this last result we find:\begin{equation}
\left[ \sigma^{\rho \sigma}, \sigma^{\mu \nu} \right] = \eta^{\rho \mu} \sigma^{\sigma \nu}
- \eta^{\rho \nu} \sigma^{\sigma \mu} + \eta^{\sigma \nu} \sigma^{\rho \mu}
- \eta^{\sigma \mu} \sigma^{\rho \nu} \; .
\label{eq:sigma-sigma}
\end{equation}
But these are precisely the same commutation relations that we had
before for the $M$ matrices in~\eqref{eq:LorentzAlgebra0}.  We then
conclude that the $\sigma$ matrices are a different representation of
the Lorentz group (notice in particular that the $M$ matrices are
real, while the $\sigma$ matrices are complex).

Just as we did before, we can now introduce an infinitesimal Lorentz
transformation as:
\begin{equation}
s = \frac{1}{2} \: C_{\rho \sigma} \sigma^{\rho \sigma} \; ,
\label{eq:S-inf}
\end{equation}
where the coefficients $C_{\rho \sigma}$ are the same as before, 
and the corresponding exponential map as:
\begin{equation}
S = \exp \left( \frac{1}{2} \: C_{\rho \sigma}
\sigma^{\rho \sigma} \right) \; ,
\label{eq:S-exp}
\end{equation}
where now $S$ represents a finite Lorentz transformation. Here one must
remember that both $s$ and $S$ are $4 \times 4$ matrices. Just as the
original Lorentz transformation in the representation $\Lambda$ acts
on vectors $v^\alpha$ (and tensors) as:
\begin{equation}
{v'}^\alpha = {\Lambda^\alpha}_\beta \: v^\beta \; ,
\end{equation}
the Lorentz transformation in the representation $S$ acts on
4-component Dirac spinors in the following way (as will be shown in
Sec.~\ref{sec:LorentzTranf} below):
\begin{equation}
{\psi'}^i = {S^i}_j \: \psi^j \; ,
\end{equation}
where in the last equation the indices $(i,j)$ are not spacetime
indices, but rather spinor indices.

In order to find the explicit form of the $\sigma^{\alpha \beta}$
matrices, let us first consider the purely spatial components
associated with spatial rotations. From the definition we find, for $i
\neq j$:
\begin{equation}
\sigma^{ij} = \frac{1}{4} \left[ \gamma^i , \gamma^j \right] 
= \frac{1}{2} \: \gamma^i \gamma^j = \frac{1}{2}
\left(
\begin{array}{cc}
0 & \sigma_i \\
- \sigma_i & 0 
\end{array}
\right)
\left(
\begin{array}{cc}
0 & \sigma_j \\
- \sigma_j & 0 
\end{array}
\right)
= - \frac{i}{2} \: \epsilon^{ijk}
\left(
\begin{array}{cc}
\sigma_k & 0 \\
0 & \sigma_k 
\end{array}
\right) \; ,
\label{eq:sigma_ij}
\end{equation}
where we used the fact that the $\gamma^i$ and $\gamma^j$ anticommute,
and in the last steps we used the form of the $\gamma^i$ matrices in
Dirac's representation (though in this case one in fact finds the same
result in the Weyl representation). If we now define the angles
$\theta^k$ such that $C_{ij} = - \epsilon_{ijk} \theta^k$, the
rotation matrix can be written as:
\begin{equation}
S = \left(
\begin{array}{cc}
e^{i \vec{\theta} \cdot \vec{\sigma}/2} & 0 \\
0 & e^{i \vec{\theta} \cdot \vec{\sigma}/2} 
\end{array}
\right) \; ,
\label{eq:S-Rot}
\end{equation}
where the dot product is again the standard in three dimensions.
Let us now consider, for example, a rotation around the $x$ axis
by an angle $\theta$, in that case we will have:
\begin{equation}
S = \left(
\begin{array}{cc}
e^{+i (\theta/2) \sigma_1} & 0 \\
0 & e^{+i (\theta/2) \sigma_1} 
\end{array}
\right)
= \cos(\theta/2)
\left(
\begin{array}{cc}
I_2 & 0 \\
0 & I_2 
\end{array}
\right)
+ i \sin(\theta/2)
\left(
\begin{array}{cc}
\sigma_1 & 0 \\
0 & \sigma_1
\end{array}
\right) \; .
\label{eq:S-RotX}
\end{equation}
where in the last step we use the fact that $(\sigma_1)^2=I_2$ in
order to expand the exponential in a Taylor series and regroup terms
into sines and cosines.  Notice now that the form of this matrix is
quite different from the matrix $\lambda$ associated to the same
rotation given by~\eqref{eq:L-RotX}. In particular, if we take $\theta
= 2 \pi$ we find $S=-I_4$, so that $\psi$ now changes sign after a full
rotation, in contrast with what happens with vectors and tensors that
are invariant under a full rotation.  This is a well known
characteristic of spinors.

On the other hand, for a Lorentz boost we find, in the Dirac
representation:
\begin{equation}
\sigma^{0i} = \frac{1}{2} \: \gamma^0 \gamma^i = \frac{1}{2}
\left(
\begin{array}{cc}
I_2 & 0 \\
0 & -I_2 
\end{array}
\right)
\left(
\begin{array}{cc}
0 & \sigma_i \\
- \sigma_i & 0 
\end{array}
\right)
= \frac{1}{2}
\left(
\begin{array}{cc}
0 & \sigma_i \\
\sigma_i & 0 
\end{array}
\right) \; .
\end{equation}
Defining now the velocity parameter as $\varphi := - C_{0i}$
we find, for an arbitrary boost:
\begin{equation}
S = \left(
\begin{array}{cc}
0 & e^{-\vec{\varphi} \cdot \vec{\sigma}/2} \\
e^{-\vec{\varphi} \cdot \vec{\sigma}/2} & 0
\end{array}
\right) \; .
\label{eq:S-Boost}
\end{equation}
Using now the fact that:
\begin{equation}
\left(
\begin{array}{cc}
0 & \sigma_i \\
\sigma_i & 0
\end{array}
\right)^2 = I_4 \; ,
\label{eq:sigmas-squared}
\end{equation}
one can show that a boost along the direction $x$ takes the form:
\begin{equation}
S = \left(
\begin{array}{cc}
0 & e^{-(\varphi/2) \sigma_1} \\
e^{-(\varphi/2) \sigma_1} & 0  
\end{array}
\right)
= \cosh(\varphi/2)
\left(
\begin{array}{cc}
I_2 & 0 \\
0 & I_2 
\end{array}
\right)
- \sinh(\varphi/2)
\left(
\begin{array}{cc}
0 & \sigma_1 \\
\sigma_1 & 0 
\end{array}
\right) \; .
\label{eq:S-BoostX}
\end{equation}
One should be very careful when applying a Lorentz boost (also a rotation)
using the exponential map.  The exponential of a matrix is really
defined in terms of the Taylor expansion.  This implies, for example,
that even if the matrix $S$ given in~\eqref{eq:S-Boost} apparently has
no elements in the diagonal, once we do the Taylor expansion such
diagonal elements do appear (because of
equation~\eqref{eq:sigmas-squared}).  That is the reason why it is
easier to use the Lorentz transformation along a specific
direction as shown in~~\eqref{eq:S-BoostX}.

It turns out that for the case of a Lorentz boost it is more convenient
to work in the Weyl representation, in which case we have:
\begin{equation}
\sigma^{0i} = \frac{1}{2} \: \gamma^0 \gamma^i = \frac{1}{2}
\left(
\begin{array}{cc}
0 & I_2 \\
I_2 & 0 
\end{array}
\right)
\left(
\begin{array}{cc}
0 & \sigma_i \\
- \sigma_i & 0 
\end{array}
\right)
= \frac{1}{2}
\left(
\begin{array}{cc}
- \sigma_i & 0 \\
0 & \sigma_i 
\end{array}
\right) \; ,
\end{equation}
and taking again $\varphi := - C_{0i}$ we find:
\begin{equation}
S = \left(
\begin{array}{cc}
e^{+\vec{\varphi} \cdot \vec{\sigma}/2} & 0 \\
0 & e^{-\vec{\varphi} \cdot \vec{\sigma}/2}
\end{array}
\right) \; .
\end{equation}
For a boost along the $x$ direction we now have:
\begin{equation}
S = \left(
\begin{array}{cc}
e^{(\varphi/2) \sigma_1} & 0 \\
0 & e^{-(\varphi/2) \sigma_1} 
\end{array}
\right)
= \cosh(\varphi/2)
\left(
\begin{array}{cc}
I_2 & 0 \\
0 & I_2 
\end{array}
\right)
+ \sinh(\varphi/2)
\left(
\begin{array}{cc}
\sigma_1 & 0 \\
0 & - \sigma_1
\end{array}
\right) \; .
\end{equation}
The reason for which the Weyl representation is better in this case is
that in the Dirac representation the components $(1,2)$ of the spinor
are mixed with the components $(3,4)$ for a Lorentz boost, but this
does not happen in the Weyl representation since the matrix in now
block diagonal. Here it is important to notice that the $S$
transformations in general are {\em not}\/ unitary (i.e. with inverse
equal to their transpose conjugate).  A three-dimensional rotation is
unitary as can be easily seen from~\eqref{eq:S-Rot} and the fact that
the Pauli matrices $\sigma_i$ are hermitian, but a Lorentz boost is
not unitary.

\vspace{5mm}

To finish this section we will show a very important relation between
the matrices $S$ and $\Lambda$ associated with the same Lorentz
transformation.  It turns out that in general one has:
\begin{equation}
S^{-1} \gamma^\mu S = {\Lambda^\mu}_\nu \gamma^\nu \; .
\label{eq:gamma-transf1}
\end{equation}
The previous expression has to be understood with some care.  On the
left hand side we have the product of three matrices, $S^{-1}
\gamma^\mu S$, while on the right hand side we have a linear
combination of matrices $\gamma^\nu$ with coefficients given by the
${\Lambda^\mu}_\nu$. In order to prove this relation we shall work in
the limit of infinitesimal transformations.  In that case we have,
from~\eqref{eq:L-exp} and~\eqref{eq:S-exp} (here and in what follows
$\simeq$ denotes approximately equal to):
\begin{equation}
\Lambda \simeq I_4 + \frac{1}{2} \: C_{\rho \sigma} M^{\rho \sigma} \; , \qquad
S \simeq I_4 + \frac{1}{2} \: C_{\rho \sigma} \sigma^{\rho \sigma} \; .
\end{equation}
This implies:
\begin{equation}
{\Lambda^\mu}_\nu \gamma^\nu \simeq \gamma^\mu + \frac{1}{2} \:
C_{\rho \sigma} {(M^{\rho \sigma})^\mu}_\nu \gamma^\nu \; ,
\end{equation}
and:
\begin{equation}
S^{-1} \gamma^\mu S \simeq \gamma^\mu - \frac{1}{2} \: C_{\rho \sigma}
\left( \sigma^{\rho \sigma} \gamma^\mu - \gamma^\mu \sigma^{\rho \sigma} \right) \; ,
\end{equation}
where in the last expression we kept only linear terms in the
coefficients $C_{\rho \sigma}$, which we assume are small.
In order to prove~\eqref{eq:gamma-transf1} we must then ask for:
\begin{equation}
\left[ \sigma^{\rho \sigma} , \gamma^\mu \right] = - {(M^{\rho \sigma})^\mu}_\nu \gamma^\nu \; .
\label{eq:N-gamma-2}
\end{equation}
But, from~\eqref{eq:antisym-basis1} we have:
\begin{equation}
{(M^{\rho \sigma})^\mu}_\nu \gamma^\nu
= - \eta^{\rho \mu} \gamma^\sigma + \eta^{\sigma \mu} \gamma^\rho \; ,
\end{equation}
and using now~\eqref{eq:N-gamma} we obtain
precisely~\eqref{eq:N-gamma-2}.  Equation~\eqref{eq:gamma-transf1} can
be easily inverted to find the equivalent expression:
\begin{equation}
S \gamma^\mu S^{-1} = {\Lambda_\nu}^\mu \gamma^\nu \; .
\label{eq:gamma-transf2}
\end{equation}


\section{Lorentz invariance of the Dirac equation}
\label{sec:LorentzTranf}

We are now in a position to show the invariance of Dirac's equation
under a Lorentz transformation. We start from Dirac's equation written
in covariant form:
\begin{equation}
i \gamma^\mu \partial_\mu \psi - m \psi = 0 \; .
\end{equation}
Under a Lorentz transformation $x'^\mu = {\Lambda^\mu}_\nu x^\nu$
this equation must keep the exact same form, with
the same $\gamma^\mu$ matrices, so we must have:
\begin{equation}
i \gamma^\mu \partial'_\mu \psi' - m \psi' = 0 \; ,
\label{eq:Dirac-prime}
\end{equation}
with $\partial'_\mu = {\Lambda_\mu}^\nu \partial_\nu$.  We now propose
that the Dirac spinor transforms with a transformation matrix $S$ as:
\begin{equation}
\psi' = S \psi \; ,
\end{equation}
where at the moment we are not assuming anything about the form of
$S$, except for the fact that it must be a constant matrix (because of
the homogeneity and isotropy of spacetime).  Substituting this into
equation~\eqref{eq:Dirac-prime} we find:
\begin{equation}
i {\Lambda_\mu}^\nu \gamma^\mu S \: \partial_\nu \psi - m S \psi = 0 \; ,
\end{equation}
where we used the fact that $S$ is constant and the
${\Lambda_\mu}^\nu$ are real numbers so they commute with the
matrices. Multiplying now from the left with $S^{-1}$
we find:
\begin{equation}
i {\Lambda_\mu}^\nu \left( S^{-1} \gamma^\mu S \right) \partial_\nu \psi - m \psi = 0 \; ,
\end{equation}
and comparing with the original Dirac equation we conclude that we
must have:
\begin{equation}
{\Lambda_\mu}^\nu \left( S^{-1} \gamma^\mu S \right) = \gamma^\nu \; .
\end{equation}
Using now the fact that  ${\Lambda_\mu}^\nu = {(\Lambda^{-1})^\nu}_\mu$,
we can see that the previous expression is equivalent to:
\begin{equation}
S^{-1} \gamma^\mu S = {\Lambda^\mu}_\nu \gamma^\nu \; .
\end{equation}
But this is nothing more than equation~\eqref{eq:gamma-transf1} from
the previous section, so we conclude that the Dirac spinor $\psi$
transforms precisely with the $S$ matrices we discussed before.

From the transformation law for $\psi$ one can easily show
that the adjunct spinor $\bar{\psi}$  transforms as:
\begin{equation}
\bar{\psi}' = {\psi'}^\dag \gamma^0 = \psi^\dag S^\dag \gamma^0 \; .
\end{equation}
Using now the relations~\eqref{eq:gamma-0mu0} it is not difficult
to show that the matrices $\sigma^{\mu \nu}$ defined in~\eqref{eq:Sigma-munu} 
satisfy:
\begin{equation}
(\sigma^{\mu \nu})^\dag = - \gamma^0 \sigma^{\mu \nu} \gamma^0 \; ,
\end{equation}
which in turn implies:
\begin{equation}
S^\dag = \gamma^0 S^{-1} \gamma^0 \; .
\end{equation}
The transformation of $\bar{\psi}$ then takes the final form:
\begin{equation}
\bar{\psi}' = \psi^\dag \gamma^0 S^{-1} = \bar{\psi} S^{-1} \; ,
\end{equation}
where we used the fact that $(\gamma^0)^2 = 1$. The result implies, in
particular, that the object $(\bar{\psi} \psi)$ is a Lorentz scalar,
while the conserved current $(\bar{\psi} \gamma^\mu \psi)$ transforms
as a vector.  Moreover, $(\bar{\psi} \gamma^\mu \gamma^\nu \psi)$
transforms as a rank 2 tensor, or strictly speaking its antisymmetric
part transforms as a tensor (actually a 2-form), while its symmetric
part is Lorentz invariant since it is proportional to the Minkowski
tensor (see equation~\eqref{eq:DiracMatrices-metric}).


\section{Dirac equation in general relativity}
\label{sec:DiracGR}

\subsection{The tetrad formalism}

In order to write the Dirac equation in the case of a general curved
spacetime we first need to introduce the formalism of tetrads.  We
will not describe here that formalism in great detail as it is quite
standard in advanced general relativity text books, and we will limit
ourselves to discuss some of its more important properties.

The basic idea behind the tetrad formalism is to choose, at each point
of spacetime, a set of four orthonormal vectors $\{ \vec{e}_A \}$ that
define a local inertial frame, where the index $A$ identifies each of
the four vectors ($A=0,1,2,3$). It is important to emphasize that this
orthonormal basis is in principle completely independent from our
coordinate system.  The basis $\{ \vec{e}_A \}$ is known as a {\em
  tetrad} (also frequently called a {\em vierbein} from the German
word for ``four legs'', or even a {\em vielbein} meaning ``many
legs'').

Assume now that we have a general coordinate system $\{ x^\mu \}$, the
corresponding coordinate basis components of the tetrad will then be
$e_A^\mu$, and since the tetrad basis is orthonormal by definition we
will have:
\begin{equation}
\vec{e}_A \cdot \vec{e}_B
= g_{\mu \nu} e_A^\mu \: e_B^\nu
= \eta_{AB} \; ,
\end{equation}
with $g_{\mu \nu}$ the components of the metric tensor and $\eta_{AB}$
the usual Minkowski tensor. Notice that now we have two different
types of indices: indices associated with the spacetime coordinates,
denoted by Greek letters, and indices associated to the tetrad basis
vectors denoted by uppercase Latin letters and usually called
``Lorentz indices''.  From here on we also take the convention that
Greek indices are raised and lowered with $g_{\mu \nu}$, while
uppercase Latin indices are raised and lowered with $\eta_{AB}$.  For
example we will have:
\begin{equation}
e_{\mu A} = g_{\mu \nu} e_A^\nu \; , \qquad
e^{\mu A} = \eta^{AB} e_B^\mu \; .
\end{equation}
In particular, this implies:
\begin{equation}
e_{\mu A} e_B^\mu = \eta_{AB} \; .
\label{eq:tetrad-contraction1}
\end{equation}
Somewhat more formally, we define a set of four 1-forms $\{ \tilde{e}^A \}$
associated with our tetrad such that:
\[
\tilde{e}^A \left( \vec{e}_B \right) = e^A_\mu e_B^\mu = \delta^A_B \; .
\]
The 1-forms $\{ \tilde{e}^A \}$ are known as the ``co-tetrad''.
However, it is easier just to think of raising and lowering indices
with $\eta_{AB}$ and $g_{\mu \nu}$.

We can now project arbitrary vectors onto the tetrad.  For example,
for a vector $\vec{v}$ we will have:
\begin{equation}
v^A = v^\mu e^A_\mu \; , \qquad
v_A = v^\mu e_{\mu A} \; ,
\label{eq:estrangular}
\end{equation}
and the original vector can be written as:
\begin{equation}
\vec{v} = v^A \vec{e}_A \; .
\end{equation}
This relation implies then that:
\begin{equation}
v^\mu = v^A e_A^\mu \; .
\label{eq:resucitar}
\end{equation}
The dot product of two vectors can then be written as:
\begin{equation}
\vec{v} \cdot \vec{u}
= \left( v^A \vec{e}_A \right) \cdot \left( u^B \vec{e}_B \right)
= v^A u^B \eta_{AB} = v^A u_A \; .
\end{equation}
That is, we have $\vec{v} \cdot \vec{u} = v^\mu u_\mu = v^A u_A$.

Let now $\{ \vec{z}_\mu \}$ denote the coordinate basis vectors.
We can then express this basis in terms of the tetrad as:
\begin{equation}
\vec{z}_\mu = z_\mu^A \vec{e}_A \; .
\end{equation}
The dot product of two coordinate basis vectors will then be:
\begin{equation}
\vec{z}_\mu \cdot \vec{z}_\nu = z_{\mu A}  z_\nu^A
= \left( e_{\lambda A} z_\mu^\lambda \right)
\left( e^A_\sigma z_\nu^\sigma \right)
= \left( e_{\lambda A} \delta_\mu^\lambda \right)
\left( e^A_\sigma \delta_\nu^\sigma \right)
= e_{\mu A} e^A_\nu \; ,
\end{equation}
where we used the fact that the spacetime components of the coordinate
basis are just the Kronecker delta. But the dot product of coordinate
basis vectors is precisely the definition of the components of the
metric tensor, so we finally have (compare this
with~\eqref{eq:tetrad-contraction1}):
\begin{equation}
e_{\mu A} e^A_\nu = g_{\mu \nu} \; .
\label{eq:tetrad-contraction2}
\end{equation}
This is a very important result, and shows that in some sense the
tetrad can be understood as a ``square root'' of the metric tensor.
In particular, the last expression implies that the determinant of the
metric can be written in terms of the tetrad as:
\begin{equation}
g = \det \left( e^A_\mu e^B_\nu \eta_{AB} \right)
= \left[ \det (e^A_\mu) \right]^2 \det (\eta_{AB})
= - \left[ \det (e^A_\mu) \right]^2 \; ,
\end{equation}
so the volume element takes the form:
\begin{equation}
|g|^{1/2} = \det (e^A_\mu) \; .
\end{equation}

It is interesting to notice that the components of the tetrad
$e^\mu_A$ can also be interpreted as the jacobian of the
transformation from the coordinates $\{ x^\mu \}$ to a new set of
locally flat coordinates $\{ X^A \}$.  If we define $e^\mu_A :=
\partial x^\mu / \partial X^A$ we will have, for an arbitrary vector:
\begin{equation}
v^\mu = \left( \frac{\partial x^\mu}{\partial X^A} \right) v^A
= e^\mu_A v^A \; ,
\end{equation}
which corresponds to the rule~\eqref{eq:resucitar} that we saw above.

We can also project tensors of arbitrary rank onto the tetrad.  For
example, for a rank 2 tensor we will have:
\begin{equation}
T_{AB} := e_A^\mu e_B^\nu T_{\mu \nu} \; .
\end{equation}
In particular, if we take $T$ as the metric tensor we find immediately
$g_{AB} = \eta_{AB}$, as should be expected.  The operation that takes
a spacetime index into a Lorentz (tetrad) index is usually known as
``strangulation'' ($v^A = e^A_\mu v^\mu$), while the opposite
operation that takes a Lorentz index into a spacetime index is known
as ``resurrection'' ($v^\mu = e_A^\mu v^A$).

The importance of the tetrad formalism comes from the fact that, once
projected onto the tetrad, the components of vectors and tensors
behave as scalar functions under changes of coordinates.  In the
tetrad formalism we then have two different types of transformations:

\begin{enumerate}

\item General coordinate transformations for which the tetrad basis
  vectors behave in the usual manner, but the components of different
  geometric objects (vectors, tensors and spinors) projected onto the
  tetrad behave as scalars.

\item Local Lorentz transformations that take an initial tetrad onto a
  new tetrad, for which the components of geometric objects transform
  as tensors in special relativity (or spinors, see next sections).

\end{enumerate}

\vspace{5mm}

The next step is to consider the derivatives of vectors and tensors in
the tetrad formalism.  For this we need to define the so-called {\em
  spin connection coefficients}, also known as the {\em Ricci rotation
  coefficients}, as:
\begin{equation}
\omega_{\mu B \nu} :=  \nabla_\nu e_{\mu B} \equiv e_{\mu B ; \nu} \; .
\label{eq:spinconnection0}
\end{equation}
The order of the indices for the $\omega$'s is chosen here such that
it coincides with the natural order when using the ``$;$'' notation
for covariant derivatives (this convention is common, but not
universal, so care must be taken when reading different references).
It is usual to work with these coefficients with one or both of the
spacetime indices projected onto the tetrad (strangled):
\begin{equation}
\omega_{AB \nu} = e_A^\lambda \: \omega_{\lambda B \nu} \;, \qquad
\omega_{ABC} = e_A^\lambda e_C^\sigma \: \omega_{\lambda B \sigma} \; .
\label{eq:spinconnection1}
\end{equation}
One often finds that the name ``spin connection coefficients'' is
reserved for the fully strangled version $\omega_{ABC}$.
Similarly, one can work with all three indices of spacetime type
(resurrected):
\begin{equation}
\omega_{\mu \lambda \nu} = e_\lambda^B \: \omega_{\mu B \nu} \; .
\end{equation}
Notice that with our notation the third index always corresponds to
the derivative (but many references put the derivative index as the
first one), either directly as in $\omega_{\mu A \nu}$ and $\omega_{AB
  \nu}$, or strangled as in $\omega_{ABC}$.

From the above definition it is not difficult to show that in general
we have:
\begin{equation}
\partial_\mu \vec{e}_A \equiv \left( e\indices{^\nu_A_{;\mu}} \right) \vec{z}_\nu
= \omega\indices{^\nu_A_\mu} \vec{z}_\nu 
= \omega\indices{^B_A_\mu} \vec{e}_B \; .
\label{eq:tetrad-covder}
\end{equation}

The spin connection that we have just defined turns out to be antisymmetric
in the first two indices when they are of the same type.  In order to see this
one must remember that  \mbox{$\eta_{AB} = g_{\mu \nu} {e_A}^\mu
  {e_B}^\nu$}, which implies:
\begin{align}
0 &= \eta_{AB;\lambda} = g_{\mu \nu}
\left( e_A^\mu e^\nu_{B;\lambda} + e_B^\nu e^\mu_{A;\lambda} \right)
\nonumber \\
&= e_{\nu A} e^\nu_{B;\lambda} +  e_{\mu B} e^\mu_{A;\lambda}
= \omega_{AB \lambda} + \omega _{BA \lambda} \; ,
\end{align}
and finally:
\begin{equation}
\omega_{AB \alpha} = - \omega_{BA \alpha} \; .
\end{equation}
From this we immediately also find $\omega_{\alpha \beta \lambda} = -
\omega_{\beta \alpha \lambda}$. This property ensures that when we
express the covariant derivative of the metric tensor in terms of the
tetrad we will have $\nabla_\mu g_{\alpha \beta}=0$.  To see this
notice that from~\eqref{eq:tetrad-contraction2} we have:
\begin{align}
\nabla_\mu g_{\alpha \beta} &= \left( e_{\alpha A} e^A_\beta \right)_{; \mu}
= e_{\alpha A} e^A_{\beta ; \mu} + e^A_\beta e_{\alpha A ; \mu} \nonumber \\
&= e_{\alpha A} \omega\indices{_\beta^A_\mu} + e^A_\beta \omega\indices{_\alpha_A_\mu}
= \omega_{\beta \alpha \mu} + \omega_{\alpha \beta \mu} = 0 \; . 
\end{align}
Notice that, from equation~\eqref{eq:tetrad-covder} above, the Ricci
rotation coefficients play a similar role to the Christoffel symbols
when working with the tetrad instead of a coordinate basis.  But
crucially, while the Christoffel symbols are symmetric in two indices,
$\Gamma^\alpha_{\mu \nu} = \Gamma^\alpha_{\nu \mu}$, the Ricci
rotation coefficients are antisymmetric in two indices, $\omega_{AB
  \alpha} = - \omega_{BA \alpha}$. This implies that there are fewer
independent $\omega_{AB \alpha}$ than there are $\Gamma^\alpha_{\mu
  \nu}$.  For example, in 4 dimensions there are 40 independent
$\Gamma^\alpha_{\mu \nu}$, while there are only 24 independent
$\omega_{AB \alpha}$.  This is one of the advantages of the tetrad
formalism: one has to compute fewer quantities.

The name ``rotation coefficients'' comes from considering the change
of the tetrad for an infinitesimal displacement $\delta x^\mu$.  In
that case we have:
\begin{equation}
\delta e^\mu_A = e^\mu_{A ; \nu} \delta x^\nu
= {\omega^\mu}_{A \nu} \delta x^\nu
= \left( {\omega^B}_{A \nu} \delta x^\nu \right) e^\mu_B \; .
\end{equation}
If we now define ${\Lambda^B}_A := {\omega^B}_{A \nu} \delta x^\nu$
we find:
\begin{equation}
\delta e^\mu_A = {\Lambda^B}_A e^\mu_B \; .
\end{equation}
But this is just a rotation of the tetrad in 4 dimensions, that is a
general Lorentz transformation such as those we discussed above.

\vspace{5mm}

One can in fact define different concepts of derivatives of geometric
quantities in the tetrad formalism.  The first one is the standard
covariant derivative for which an object completely projected onto the
tetrad, i.e. completely strangled, behaves as a scalar.  To denote
this derivative we will continue to use the $\nabla$ (or $;$) symbol.

We can also define two different types of derivatives for geometric
objects that have indices of mixed type. The {\em intrinsic
  derivative}\/ is defined by first strangling all the spacetime
indices, then taking the covariant derivative of the resulting scalar,
and finally resurrecting back the indices that had been strangled.  To
denote this derivative we use a vertical bar instead of as semicolon,
for example:
\begin{equation}
T^\mu_{A | \nu} = \left( T_A^\lambda e^B_\lambda \right)_{; \nu} e^\mu_B
= T^B_{A ; \nu} e^\mu_B = T^B_{A , \nu} e^\mu_B \; ,
\label{eq:deriv-intr1}
\end{equation}
where in the last step we used the fact that fully strangled
components behave as scalars for the covariant derivative. The
intrinsic derivative then corresponds to the change of the tensor
components with respect to the tetrad basis.  That is, if the tetrad
changes from one point to another and the tensor also changes but in
such a way that its tetrad components are the same, then its intrinsic
derivative vanishes.  In particular, from the previous definition it
is easy to see that:
\begin{equation}
e^\mu_{A | \nu} = \left( e_A^\lambda e^B_\lambda \right)_{; \nu} e^\mu_B
= \left( \delta_A^B \right)_{; \nu} e^\mu_B = 0 \; ,
\end{equation}
that is, the intrinsic derivative of the tetrad is always zero, so
that such derivative commutes with the strangulation and resurrection
operations, for example:
\begin{equation}
e^B_\mu T^\mu_{A | \nu} = T^B_{A | \nu} \; , \qquad
e^A_\lambda T^\mu_{A | \nu} = T^\mu_{\lambda | \nu} \; .
\end{equation}

Equation~\eqref{eq:deriv-intr1} can also be written explicitly
in terms of the spin connection as follows:
\begin{equation}
T^\mu_{A | \nu} = T^\mu_{A ; \nu} + {{\omega_\sigma}^\mu}_\nu T_A^\sigma \; .
\end{equation}
The first term in the previous expression is the usual covariant
derivative, for which strangled indices behave as scalars (so that
$T^\mu_A$ behaves as a vector).  In the second term the spin
connection plays a similar role as the usual Christoffel symbols, but
remember that they have different symmetries.  It is also possible to
strangle the index in the derivative to define a directional
derivative along the tetrad, for example:
\begin{equation}
T^\mu_{A | C} = e^\nu_C  T^\mu_{A | \nu}
= T^\mu_{A ; C} + {{\omega_\sigma}^\mu}_C T_A^\sigma \; ,
\end{equation}
where we have defined  $T^\mu_{A ; C} := e^\nu_C  T^\mu_{A ; \nu}$.
Similarly, we also can strangle the $\mu$ index to find:
\begin{equation}
T^B_{A | C} = e^B_\mu T^\mu_{A ; C} + {{\omega_\sigma}^B}_C T_A^\sigma
= e^B_\mu T^\mu_{A ; C} + {{\omega_D}^B}_C T_A^D \; .
\end{equation}
Notice that for the first term we can not simply write $T^B_{A ; C}$,
since the covariant derivative of the tetrad in general does not
vanish.

The intrinsic derivative can be generalized to tensors with multiple
indices in the same way as the covariant derivative.  For example, for
an object with fully strangled indices we have:
\begin{equation}
T^A_{B | \nu} = T^A_{B , \nu} \: ,
\end{equation}
while for an object with one covariant spacetime index we will have:
\begin{equation}
T_{\mu A | \nu} = T_{\mu A ; \nu}
- {{\omega_\mu}^\sigma}_\nu T_{\sigma A} \; ,
\end{equation}
and for objects with mixed spacetime indices we find:
\begin{equation}
T^\mu_{\lambda A | \nu} = T^\mu_{\lambda A ; \nu}
+ {{\omega_\sigma}^\mu}_\nu T^\sigma_{\lambda A}
- {{\omega_\lambda}^\sigma}_\nu T^\mu_{\sigma A} \; .
\end{equation}

\vspace{5mm}

A different type of derivative is known as the {\em invariant
  derivative}.  In this case the definition is the opposite: we first
resurrect all tetrad indices, we then take the covariant derivative,
and finally we strangle again.  We denote this derivative by a dot,
for example:
\begin{equation}
T^\mu_{A \cdot \nu} = \left( T^\mu_B e^B_\lambda \right)_{; \nu} e^\lambda_A
= T^\mu_{\lambda ; \nu} e^\lambda_A \; .
\label{eq:invariantderiv}
\end{equation}
In the same way as with the intrinsic derivative, it turns out that the
invariant derivative of the tetrad also vanishes:
\begin{equation}
e^\mu_{A \cdot \nu} = \left( e_B^\mu e^B_\lambda \right)_{; \nu} e^\lambda_A
= \left( \delta_\mu^\lambda \right)_{; \nu} e^\lambda_A = 0 \; .
\label{eq:inaviantderiv-tetrad1}
\end{equation}
The invariant derivative then also commutes with strangulation and
resurrection, for example:
\begin{equation}
e^A_\lambda T^\mu_{A \cdot \nu} = T^\mu_{\lambda \cdot \nu} \; , \qquad
e^B_\mu T^\mu_{A \cdot \nu} = T^B _{A \cdot \nu} \; .
\end{equation}
In particular, this implies that when a tensor is expressed with
purely spacetime indices, its invariant derivative is simply equal to
its covariant derivative:
\begin{equation}
T^\mu_{\lambda \cdot \nu} = T^\mu_{\lambda ; \nu} \; .
\end{equation}
In other words, the invariant derivative is just the projection of the usual
covariant derivative of a tensor with purely spacetime indices onto the tetrad,
in one or several of its indices.  In particular we have, for example:
\begin{equation}
T^A_{B \cdot C} = e^A_\alpha e_B^\beta e_C^\nu T^\alpha_{\beta ; \nu} \; .
\end{equation}
The previous expression can in fact be inverted to find:
\begin{equation}
T^\alpha_{\beta ; \nu} = e^\alpha_A e_\beta^B e_\nu^C T^A_{B \cdot C} \; .
\end{equation}

For a tensor with mixed indices, the invariant derivative gives us the
change of the tensor as an abstract geometric object, that is already
reconstructed in terms of the corresponding basis.  For example, if
we have a tensor $\mathbf{T} \equiv T^\mu_A \vec{z}_\mu \vec{e}^A$,
then its derivative will be:
\begin{equation}
\partial_\mu \left( \mathbf{T} \right)
= \partial_\mu \left( T^\nu_A \vec{z}_\nu \vec{e}^A \right) 
= T^\nu_{A \cdot \mu} \vec{z}_\nu \vec{e}^A \; .
\end{equation}
It is because of this property that the invariant derivative is the
most natural generalization of the covariant derivative in the tetrad
formalism.

The invariant derivative can also be written in terms of the spin connection.
For example we have:
\begin{align}
T\indices{^\mu^A_\cdot_\nu} &= T\indices{^\mu^A_;_\nu} + {\omega^A}_{B \nu} T^{\mu B} \; , 
\label{eq:invariantderiv-indexup} \\
T\indices{^\mu_A_\cdot_\nu} &= T\indices{^\mu_A_;_\nu} - {\omega^B}_{A \nu} {T^\mu}_B \; .
\label{eq:invariantderiv-indexdown}
\end{align}
In a similar way, for an object with mixed covariant and contravariant
Lorentz indices we will have:
\begin{equation}
T^{\mu A}_{ \; B \cdot \nu} = T^{\mu A}_{\; B ; \nu}
+ {\omega^A}_{C \nu} T^{\mu C}_{\; B}
- {\omega^C}_{B \nu} {T^{\mu A}}_C \; .
\label{eq:invariantderiv-mixed}
\end{equation}
It is interesting to notice that, while in the case of the intrinsic
derivative the spin connection ``takes'' the spacetime indices of the
original tensor, in the case of the invariant derivative it takes the
Lorentz indices.  In particular, if the original tensor had no
spacetime indices the intrinsic derivative has no extra terms and just
reduces to the covariant derivative (in fact the partial derivative),
while the opposite happens for the case of the invariant derivative
where for a tensor with no Lorentz indices it reduces to the usual
covariant derivative.  Notice that for a tensor with no spacetime
indices the invariant derivative simplifies and the first term reduces
to a partial derivative, so that we have for example:
\begin{equation}
v^A_{\; \cdot B} = e_B^\mu v^A_{\; \cdot \mu}
= e_B^\mu \left( \partial_\mu v^A + {\omega^A}_{C \mu} v^C \right)
= \partial_B v^A + {\omega^A}_{CB} v^C \; ,
\end{equation}
where we defined $\partial_B := e_B^\mu \partial_\mu$. Similarly:
\begin{align}
v_{A \cdot B} &= \partial_B v_A - {\omega^C}_{A B} v_C \; , \\
T^A_{B \cdot C} &= \partial_C T^A_B
+ {\omega^A}_{DC} T^D_{\; B}
- {\omega^D}_{BC} {T^A}_D \; .
\end{align}

In particular if we apply the rule~\eqref{eq:invariantderiv-indexdown}
to the tetrad vectors we find:
\begin{equation}
e_{\mu A \cdot \alpha} = e_{\mu A ; \alpha} - {\omega^C}_{A \alpha} e_{\mu C} 
= \omega\indices{_\mu_A_\alpha} - \omega\indices{_\mu_A_\alpha} = 0
\; ,
\end{equation}
so that we recover~\eqref{eq:inaviantderiv-tetrad1}. This result is
usually called the ``tetrad postulate'', and can sometimes be a source
of some confusion. But notice that in this case the invariant
derivative does not correspond directly with the covariant derivative
of the 1-form $\tilde{e}_ A$, which is not zero in general, but rather
with the covariant derivative of the ``tensor'' $e_{\mu \nu}$ projected
onto the tetrad, and we have $e_{\mu \nu} = e_{\mu A} e^A_\nu = g_{\mu
  \nu}$.  That is, having the invariant derivative of the tetrad
vanish simply represents the fact that the covariant derivative of
the metric tensor is always zero.

\vspace{5mm}

We can now use the expression for the invariant derivative of a vector
to find a relation between the spin connection and the Christoffel symbols.
If we calculate $v^\mu_{; \alpha}$  directly we have:
\begin{equation}
v^\lambda_{; \mu} = \partial_\mu v^\lambda + \Gamma^\lambda_{\nu \mu} v^\nu \; ,
\end{equation}
while if we calculate it from the projection of the invariant derivative we find:
\begin{align}
v^\lambda_{; \mu} &= e^\lambda_B e_\mu^A \left( v^B_{\cdot A} \right)
= e^\lambda_B e_\mu^A \left( \partial_A v^B + {\omega^B}_{CA} v^C \right)
\nonumber \\
&= e^\lambda_B \partial_\mu v^B + {\omega^\lambda}_{\nu \mu} v^\nu
= \partial_\mu \left( e^\lambda_B v^B \right) - v^B \partial_\mu e^\lambda_B
+ {\omega^\lambda}_{\nu \mu} v^\nu \nonumber \\
&= \partial_\mu v^\lambda + \left( {\omega^\lambda}_{\nu \mu} 
- e^B_\nu \partial_\mu e^\lambda_B \right) v^\nu \; .
\end{align}
Equating both expressions, and using the fact that this must be valid for an
arbitrary vector $\vec{v}$, we find:
\begin{equation}
\Gamma^\lambda_{\nu \mu} = {\omega^\lambda}_{\nu \mu} 
- e^B_\nu \partial_\mu e^\lambda_B \; ,
\label{eq:gamma-omega}
\end{equation}
where we used the fact that the Christoffel symbols
are symmetric in the lower indices.  Solving for the
spin coefficients we finally find:
\begin{equation}
{\omega^\lambda}_{\nu \mu} = \Gamma^\lambda_{\nu \mu}
+ e^B_\nu \partial_\mu e^\lambda_B \; .
\label{eq:omega-gamma1}
\end{equation}

In a similar way, by considering now the derivative of a 1-form we
also find:
\begin{equation}
{\omega^\lambda}_{\nu \mu} = \Gamma^\lambda_{\nu \mu}
- e^\lambda_B \partial_\mu e_\nu^B \; .
\label{eq:omega-gamma2}
\end{equation}
Notice that the last expression can be obtained immediately from the
fact that $e_\nu^B e_B^\lambda = \delta_\nu^\lambda$.  Finally,
projecting the first two indices onto the tetrad we obtain:
\begin{equation}
\omega_{AB \mu} = e_B^\nu \left( e_{\lambda A} \Gamma^\lambda_{\mu \nu}
- \partial_\mu e_{\nu A} \right) \; .
\label{eq:omega-gamma3}
\end{equation}
Even though not obvious, the expression above is in fact antisymmetric
in the indices $A$ and $B$ as will become apparent below.  This
expression is particularly useful in order to calculate the
coefficients $\omega_{AB \mu}$ which, as we well see in the next
sections, are necessary in order to write the Dirac equation in
general relativity.

If we now substitute the definition of the Christoffel symbols in
terms of the metric tensor, and the expression of the metric in terms
of the tetrad $g_{\mu \nu} = e_{\mu A} e_\nu^A$, a straightforward
(but somewhat long) algebra allows us to find:
\begin{equation}
\omega_{ABC} = - \frac{1}{2} \left[ \left(f_{ABC} 
+ f_{ACB} + f_{CAB} \right) - A \leftrightarrow B \right] \; ,
\label{eq:omega-f}
\end{equation}
where we have projected the third index onto the tetrad, and where we
have defined the quantities:
\begin{equation}
f_{ABC} := \left( \partial_A e_{\alpha B} \right) e_C^\alpha \; .
\label{eq:f-definition}
\end{equation}
The antisymmetry in the first two indices of $\omega_{ABC}$ is now
evident.

\vspace{5mm}

To finish this section we will now write the Riemann curvature tensor
in terms of the Ricci rotation coefficients.  One finds:
\begin{equation}
R_{AB \mu \nu} = \partial_\mu \omega_{AB \nu} - \partial_\nu \omega_{AB \mu}
+ \omega_{AC \mu} \: \omega\indices{^C_B_\nu}
- \omega_{AC \nu} \: \omega\indices{^C_B_\mu} \; .
\label{eq:Riemann-omega}
\end{equation}
Notice that in this expression the Riemann tensor has the first two
indices projected onto the tetrad, so that it behaves as a rank 2
tensor (in fact a 2-form) with respect to coordinate changes.  The
previous expression can be proved by direct calculation (the algebra
is rather long) by substituting the $\omega$'s in terms of the
Christoffel symbols from~\eqref{eq:gamma-omega}, and using the usual
definition of the Riemann tensor:
\begin{equation}
{R^\alpha}_{\beta \mu \nu} := \partial_\mu \Gamma^\alpha_{\beta \nu}
- \partial_\nu \Gamma^\alpha_{\beta \mu}
+ \Gamma^\alpha_{\sigma \mu} \Gamma^\sigma_{\beta \nu}
- \Gamma^\alpha_{\sigma \nu} \Gamma^\sigma_{\beta \mu} \; .
\label{eq:Riemann}
\end{equation}


\subsection{Generally covariant form of the Dirac equation}

In order to generalize the Dirac equation to the case of a curved
spacetime we must first go back to
equation~\eqref{eq:DiracMatrices-metric} that defines the Clifford
algebra, which in the language of tetrads now takes the form:
\begin{equation}
\left\{ \gamma^A , \gamma^B \right\} = - 2 \eta^{AB} I_4 \; ,
\label{eq:DiracMatrices-metrictetrad}
\end{equation}
with $\{ , \}$ the anticommutator, and where the $\gamma^A$ matrices
are the same constant matrices we defined in the case of special
relativity.  We will now define new $\gamma^\mu$ matrices that
depend on the tetrad as:
\begin{equation}
\gamma^\mu := \gamma^A {e_A}^\mu \; .
\label{eq:gamma-mu}
\end{equation}
Notice that these new $\gamma^\mu$ are in general not constant anymore
and can change from one point to another. From this definition we find
immediately:
\begin{equation}
\left\{ \gamma^\mu , \gamma^\nu \right\}
= \left\{ \gamma^A , \gamma^B \right\} {e_A}^\mu {e_B}^\nu
= - 2 \eta^{AB} {e_A}^\mu {e_B}^\nu \; ,
\end{equation}
and using now~\eqref{eq:tetrad-contraction2} we finally obtain:
\begin{equation}
\left\{ \gamma^\mu , \gamma^\nu \right\} = - 2 g^{\mu \nu} I_4 \; .
\label{eq:DiracMatrices-metriccurve}
\end{equation}
This is the form of the Clifford algebra in general relativity. In
particular, notice that the new $\gamma^\mu$ matrices with different
indices in general do not anti-commute anymore, and only do so for the
special case of orthogonal coordinates.  From the previous result it
is easy to show that:
\begin{equation}
\gamma_\mu \gamma^\mu = \gamma_ A \gamma^A = -  4 I_4 \; .
\label{eq:gamma2-cov}
\end{equation}

The next step is to consider the transformation rule for a spinor.
For a 4D rotation of the tetrad we will have:
\begin{equation}
\psi' = S(x) \: \psi \; ,
\end{equation}
where now $S(x)$ is a general Lorentz spinor transformation as those
we saw before, but which can now depend on position. The derivative of
a spinor, however, does not transform as a spinor anymore since we
will have:
\begin{equation}
\partial_\mu \psi' = \partial_\mu \left( S \: \psi \right) =
S \: (\partial_\mu \psi ) + (\partial_\mu S ) \: \psi \; .
\end{equation}
In order to take this into account we will now define a {\em spinor
  covariant derivative}\/ as:
\begin{equation}
\mathcal{D}_\mu \psi := \partial_\mu \psi + \Gamma_\mu \psi \; ,
\label{eq:spinorderiv-psi}
\end{equation}
where the $\Gamma_\mu$ are some matrices to be determined, and are
known as the {\em spinor affine connection}\/ (not to be confused with
the spin connection coefficients we saw above), or the {\em
  Fock--Ivanenko coefficients}~\cite{Fock29a,Fock29b}. We will find
the explicit form of these coefficients in the next Section.

Consider now the adjoint of equation~\eqref{eq:spinorderiv-psi} which takes the form:
\begin{equation}
\mathcal{D}_\mu \bar{\psi} := \partial_\mu \bar{\psi} + \bar{\psi} \: \bar{\Gamma}_\mu \; ,
\end{equation}
In order to find the relation between $\bar{\Gamma}_\mu$ and $\Gamma_\mu$
we now ask for our spinor covariant derivative to obey the Leibniz rule,
and also for it to be compatible with the usual covariant derivative.  Consider
then the covariant derivative of $(\bar{\psi} \psi)$, we have:
\begin{equation}
\mathcal{D}_\mu \left( \bar{\psi} \psi \right)
= \left( \mathcal{D}_\mu \bar{\psi} \right) \psi
+ \bar{\psi} \left( \mathcal{D}_\mu \psi \right)
= \left( \partial_\mu \bar{\psi} \right) \psi
+ \bar{\psi} \left( \partial_\mu \psi \right)
+ \bar{\psi} \left( \bar{\Gamma}_\mu + \Gamma_\mu \right) \psi \; .
\end{equation}
On the other hand, since we know that $\bar{\psi} \psi$ behaves as a
scalar we must also have:
\begin{equation}
\mathcal{D}_\mu \left( \bar{\psi} \psi \right)
= \partial_\mu \left( \bar{\psi} \psi \right)
= \left( \partial_\mu \bar{\psi} \right) \psi
+ \bar{\psi} \left( \partial_\mu \psi \right) \; .
\end{equation}
Comparing now both expressions we find immediately:
\begin{equation}
\bar{\Gamma}_\mu = - \Gamma_\mu \; ,
\end{equation}
so that the spinor covariant derivative of $\bar{\psi}$ 
takes the final form:
\begin{equation}
\mathcal{D}_\mu \bar{\psi} := \partial_\mu \bar{\psi} - \bar{\psi} \: \Gamma_\mu \; .
\label{eq:spinorderiv-psi-adj}
\end{equation}

With the above definitions, the Dirac equation in a curved spacetime
can be written as:
\begin{equation}
i \gamma^\mu \mathcal{D}_\mu \psi - m \psi = 0 \; ,
\label{eq:Dirac-curved}
\end{equation}
with $\mathcal{D}_\mu \psi = \partial_\mu \psi + \Gamma_\mu \psi$.
Similarly, the adjunct equation takes the form:
\begin{equation}
i \left( \mathcal{D}_\mu \bar{\psi} \right) \gamma^\mu + m \bar{\psi} = 0 \; ,
\label{eq:Dirac-curved-adj}
\end{equation}
with $\mathcal{D}_\mu \bar{\psi} = \partial_\mu \bar{\psi} - \bar{\psi}
\Gamma_\mu$.


\subsection{The Fock--Ivanenko coefficients}

We still need to find the explicit form of the Fock--Ivanenko
coefficients $\Gamma_\mu$. In order to do this, let us first assume
that we have a matrix operator with spacetime indices $Q^\alpha$, such
that $(\bar{\psi} Q^\alpha \psi)$ transforms as a tensor with respect
to a general change of coordinates, and where $\alpha$ can represent
any combination of covariant and contravariant indices. For the spinor
derivative of $(\bar{\psi} Q^\alpha \psi)$ we will have:
\begin{align}
\mathcal{D}_\mu \left( \bar{\psi} Q^\alpha \psi \right)
&= \left( \mathcal{D}_\mu \bar{\psi} \right) Q^\alpha \psi
+ \bar{\psi} \left( \mathcal{D}_\mu Q^\alpha \right) \psi
+ \bar{\psi} Q^\alpha \left( \mathcal{D}_\mu \psi \right) \nonumber \\
&= \left( \partial_\mu \bar{\psi} \right) Q^\alpha \psi
+ \bar{\psi} Q^\alpha \left( \partial_\mu \psi \right)
+ \bar{\psi} \left( \mathcal{D}_\mu Q^\alpha - \Gamma_\mu Q^\alpha
+ Q^\alpha \Gamma_\mu \right) \psi \; .
\end{align}
On the other hand, since $(\bar{\psi} Q^\alpha \psi)$ behaves as a
tensor, and spinors should behave as scalars for the usual covariant
derivative (they only have Lorentz indices), we have:
\begin{equation}
\mathcal{D}_\mu \left( \bar{\psi} Q^\alpha \psi \right)
= \nabla_\mu \left( \bar{\psi} Q^\alpha \psi \right)
= \left( \partial_\mu \bar{\psi} \right) Q^\alpha \psi
+ \bar{\psi} Q^\alpha \left( \partial_\mu \psi \right)
+ \bar{\psi} \left( \nabla_\mu Q^\alpha \right) \psi \; .
\end{equation}
Equating both expressions we now find:
\begin{equation}
\bar{\psi} \left( \mathcal{D}_\mu Q^\alpha - \Gamma_\mu Q^\alpha
+ Q^\alpha \Gamma_\mu \right) \psi
= \bar{\psi} \left( \nabla_\mu Q^\alpha \right) \psi \; ,
\end{equation}
and since this must hold for any $\psi$ we finally obtain:
\begin{equation}
\mathcal{D}_\mu Q^\alpha = \nabla_\mu Q^\alpha + \left[ \Gamma_\mu , Q^\alpha \right] \; .
\label{eq:spinorderiv-matrix}
\end{equation}
Clearly, if we take $Q=I_4$ we find
$\mathcal{D}_\mu I_4=0$, as expected.  On the other hand, if we take our
matrix operator as $Q^{\alpha \beta} = g^{\alpha \beta} I_4$, we
immediately find $\mathcal{D}_\mu ( g^{\alpha \beta} I_4 )=0$, which
indicates that the spinor derivative must be compatible with the
metric. Going back to the Clifford algebra,
equation~\eqref{eq:DiracMatrices-metriccurve}, it is easy to see that
a sufficient condition for this to be satisfied is to ask for the
spinor derivative of the $\gamma^\mu$ matrices to vanish, that is:
\begin{equation}
\mathcal{D}_\mu \gamma^\nu = 0 \; ,
\label{eq:Dgamma}
\end{equation}
or more explicitly:
\begin{equation}
\nabla_\mu \gamma^\nu + \left[ \Gamma_\mu , \gamma^\nu \right] = 0 \; .
\end{equation}
A somewhat long algebraic calculation shows that the
previous equation will be satisfied if we take:
\begin{equation}
\Gamma_\mu = - \frac{1}{4} \: \omega_{AB \mu} \gamma^A \gamma^B
= - \frac{1}{2} \: \omega_{AB \mu} \sigma^{AB} \; ,
\label{eq:FockIvanenko}
\end{equation}
where the $\sigma^{AB}$ matrices are the same we had previously
defined in~\eqref{eq:Sigma-munu}, but now expressed in terms of the
tetrad:
\begin{equation}
\sigma^{AB} := \frac{1}{4} \left[ \gamma^A , \gamma^B \right]
= \frac{1}{2} \left( \gamma^A \gamma^B + \eta^{AB} I_4 \right) \; .
\end{equation}
Equation~\eqref{eq:FockIvanenko} gives us the explicit form of the
Fock--Ivanenko coefficients $\Gamma_\mu$.

Let us now return to equation~\eqref{eq:Dgamma}.  Since by definition
we have $\gamma^\mu = \gamma^A e^\mu_A$, and the $\gamma^A$ are constant,
the condition that the spinor derivative of $\gamma^\mu$ should vanish
implies that we must ask for:
\begin{equation}
\mathcal{D}_\mu e^\nu_A = 0 \; .
\end{equation}
The more natural way to accomplish this is to ask for the spinor derivative
of tensor objects with mixed indices (spacetime and Lorentz) to reduce
to the invariant derivative that we defined in the previous section.


\subsection{Geometric derivation of the Fock--Ivanenko coefficients}

In the previous Section we arrived at an explicit form for the
Fock--Ivanenko coefficients through a series of algebraic requirements
that might seem somewhat obscure.  Here we will show an alternative
geometric derivation that arrives at the same result (the discussion
here is based on that of~\cite{Jhangiani:1975}).

We start by considering the fact that the geometrically meaningful
derivative of a spinor $\psi$ can not simply be given by the
difference between its values at neighboring points, since the tetrad
with respect to which $\psi$ is defined will in general not be
parallelly transported between these two points.  The first step is
then to consider how the tetrad changes. Let $\vec{e}_A(x+dx)$ be the
value of the tetrad at point $(x+dx)$, and
${\vec{e}_A\!}^\parallel(x+dx)$ the value of the corresponding tetrad
that has been parallelly transported from $x$ to $x+dx$. Since both
these tetrads are now evaluated at the same point, the difference
between them must be an infinitesimal Lorentz transformation, that is:
\begin{equation}
\vec{e}_A(x+dx) - \vec{e}_A\!^\parallel(x+dx) = {\lambda_A}^B \vec{e}_B(x+dx)
\simeq {\lambda_A}^B \vec{e}_B(x) \; ,
\end{equation}
with $\lambda_{AB}$ the infinitesimal Lorentz transformation that we
introduced in Section~\ref{sec:LorentzGroup} above.  But this
difference is precisely the definition of the usual covariant
derivative, so we must have:
\begin{equation}
e^\mu_{A;\nu} dx^\nu = {\lambda_A}^B e^\mu_B \; .
\end{equation}
Contracting both sides of this equation with $e_{\mu C}$ we immediately
find:
\begin{equation}
e_{\mu C} e^\mu_{A;\nu} dx^\nu = {\lambda_A}^B e^\mu_B e_{\mu C}
\implies \lambda_{AC} = e_{\mu C} e^\mu_{A;\nu} dx^\nu \; ,
\end{equation}
or:
\begin{equation}
\lambda_{AC} = e_{\mu C} {\omega^\mu}_{A \nu} dx^\nu
= \omega_{CA \nu} dx^\nu \
= - \omega_{AC \nu} dx^\nu \; .
\end{equation}
From equation~\eqref{eq:S-inf}, the change of the spinor $\psi$ under
such a Lorentz transformation will then be:
\begin{equation}
\delta \psi = - \frac{1}{2} \: \lambda_{AB} \sigma^{AB} \psi \; ,
\end{equation}
where we already used the fact that the coefficients for the
transformation are given by $-\lambda_{AB}$ (see
equation~\eqref{eq:Lorentz-infinitesimal2}).

Now, just as before, the geometrically meaningful derivative of $\psi$
must be given by the difference between the value of $\psi$ at point
$x+dx$ and the value of $\psi^\parallel$ that has been parallelly transported
from $x$ to $x+dx$:
\begin{align}
\mathcal{D} \psi &= \psi(x+dx) - \psi^\parallel(x+dx)
= \psi(x+dx) - \left( \psi(x) + \delta \psi \right)
\nonumber \\
&= \partial_\nu \psi \: dx^\nu - \delta \psi
= \partial_\nu \psi \: dx^\nu + \frac{1}{2} \: \lambda_{AB} \sigma^{AB} \psi
\nonumber \\
&= \left( \partial_\nu \psi dx^\nu - \frac{1}{2} \: \omega_{AB \nu} \sigma^{AB} \psi \right) dx^\nu
\equiv \mathcal{D}_\nu \psi \: dx^\nu \; .
\end{align}
Comparing this to our definition for the spinor covariant derivative~\eqref{eq:spinorderiv-psi}
we find:
\begin{equation}
\Gamma_\mu = - \frac{1}{2} \: \omega_{AB \mu} \sigma^{AB} \; ,
\end{equation}
which is the same as~\eqref{eq:FockIvanenko}.


\subsection{Spinor Ricci identity}
\label{sec:RicciIndentity}

In the same way as the usual covariant derivatives, the commutator
of the spinor derivative can also be written in terms of the Riemann
tensor.  Using the expression for the Fock--Ivanenko coefficients that
we found above, as well as the expression for the Riemann tensor
in terms of the Ricci rotation coefficients~\eqref{eq:Riemann-omega}
together with equation~\eqref{eq:fourgammas}, it is not difficult to
show that:
\begin{equation}
\left[ \mathcal{D}_\mu , \mathcal{D}_\nu  \right] \psi
= - \frac{1}{2} \: R_{AB \mu \nu} \sigma^{AB} \psi \; .
\label{eq:Ricci-identity-spinor}
\end{equation}
This expression is a generalization of the Ricci identity for the case of
spinor covariant derivatives.  The previous result can also be
written in terms of the $\Gamma_\mu$ matrices as:
\begin{equation}
\partial_\mu \Gamma_\nu - \partial_\nu \Gamma_\mu + \Gamma_\mu \Gamma_\nu
- \Gamma_\nu \Gamma_\mu = - \frac{1}{2} \: R_{AB \mu \nu} \sigma^{AB} \; ,
\label{eq:Ricci-identity-spinor2}
\end{equation}
which bears an obvious resemblance to the form that the Riemann
tensor takes in terms of the usual Christoffel symbols.
In order to prove the previous expressions one must remember that
$\mathcal{D}_\mu \psi$ is both a spinor and a 1-form, so that:
\begin{align}
\mathcal{D}_\mu \mathcal{D}_\nu \psi
&= \partial_\mu \left( \mathcal{D}_\nu \psi \right)
+ \Gamma_\mu  \left( \mathcal{D}_\nu \psi \right)
- \Gamma^\alpha_{\mu \nu} \left( \mathcal{D}_\alpha \psi \right)
\nonumber \\
&= \partial_\mu \left( \partial_\nu \psi + \Gamma_\nu \psi \right)
+ \Gamma_\mu  \left( \partial_\nu \psi + \Gamma_\nu \psi \right)
- \Gamma^\alpha_{\mu \nu} \left( \partial_\alpha \psi + \Gamma_\alpha \psi \right)
\nonumber \\
&= \left( \partial_\mu \partial_\nu \psi - \Gamma^\alpha_{\mu \nu} \partial_\alpha \psi \right)
+ \Gamma_\mu \partial_\nu \psi  + \Gamma_\nu \partial_\mu \psi
+ \left( \partial_\mu \Gamma_\nu - \Gamma^\alpha_{\mu \nu} \Gamma_\alpha
+ \Gamma_\mu \Gamma_\nu \right) \psi
 \; ,
\end{align}
and finally:
\begin{equation}
\mathcal{D}_\mu \mathcal{D}_\nu \psi
= \nabla_\mu \nabla_\nu \psi
+ \Gamma_\mu \partial_\nu \psi  + \Gamma_\nu \partial_\mu \psi
+ \left( \nabla_\mu \Gamma_\nu + \Gamma_\mu \Gamma_\nu  \right) \psi ,
\label{eq:SecondSpinorDerivative}
\end{equation}
where $\nabla_\mu$ is the usual covariant derivative that acts on
$\psi$ as a scalar, and on $\Gamma_\mu$ as a 1-form.  Notice that,
even if the first three terms are clearly symmetric on $(\mu,\nu)$,
the last term breaks this symmetry since there is no reason to assume
that $\nabla_\mu \Gamma_\nu$ is symmetric, and $\Gamma_\mu \Gamma_\nu$
also isn't symmetric since the $\Gamma_\mu$ in general do not commute.


\subsection{Invariance of the spinor affine connection}
\label{sec:InvarianceSpinorConnection}

In the previous Sections we found the final form of the Fock--Ivanenko
coefficients $\Gamma_\mu$ in terms of the spin connection given by
equation~\eqref{eq:FockIvanenko}. We still need to show that the spinor
covariant derivative that we defined above does indeed transform as a
spinor.  That is, we want to show that:
\begin{equation}
\mathcal{D}'_\mu \psi' := S \: \mathcal{D}_\mu \psi \; ,
\label{eq:D-transf}
\end{equation}
where:
\begin{equation}
\mathcal{D}'_\mu \psi' = \partial_\mu \psi' + \Gamma'_\mu \psi' \; .
\end{equation}
It is not difficult to show that equation~\eqref{eq:D-transf}
will be satisfied if we ask for the $\Gamma_\mu$ matrices
to transform according to the rule:
\begin{equation}
\Gamma'_\mu = S \: \Gamma_\mu \: S^{-1} - ( \partial_\mu S ) \: S^{-1} \; .
\label{eq:Gamma-mu-transf}
\end{equation}

We will now show that the $\Gamma_\mu$ given
by~\eqref{eq:FockIvanenko} do in fact satisfy this transformation
rule. In order to do this, we consider an infinitesimal Lorentz
transformation of the form (see
equation~\eqref{eq:Lorentz-infinitesimal1}):
\begin{equation}
\Lambda\indices{^A_B} \simeq \delta^A_B + \lambda\indices{^A_B} \; , \qquad
\lambda\indices{^A_B} := \frac{1}{2} \: C_{CD} \: (M^{CD})\indices{^A_B} \; ,
\label{eq:Lorentz-infinitesimal3}
\end{equation}
with $C_{CD}=C_{CD}(x)$ now functions of position, and $|C_{CD}| \ll
1$.  The associated spinor transformation will be:
\begin{equation}
S\indices{^A_B} \simeq \delta^A_B + s\indices{^A_B} \; , \qquad
s\indices{^A_B} := \frac{1}{2} \: C_{CD} \: (\sigma^{CD})\indices{^A_B} \; .
\label{eq:Spin-infinitesimal}
\end{equation}
The inverse transformations will then have the form, to first
order in small quantities:
\begin{equation}
(\Lambda^{-1})\indices{^A_B} \simeq \delta^A_B - \lambda\indices{^A_B} \; ,
\qquad
(S^{-1})\indices{^A_B} \simeq \delta^A_B - s\indices{^A_B} \; .
\end{equation}

Substituting now the expressions for $S$ and $S^{-1}$ in the
transformation rule~\eqref{eq:Gamma-mu-transf}, and keeping to first
order in $C_{AB}$ we find, after some algebra:
\begin{equation}
\Gamma'_\mu = \Gamma_\mu + \left[ s , \Gamma_\mu \right]
- \partial_\mu s \; ,
\end{equation}
and explicitly substituting the form of $s$:
\begin{equation}
\Gamma'_\mu = \Gamma_\mu
+ \frac{1}{2} \: C_{CD} \left[ \sigma^{CD} , \Gamma_\mu \right]
- \frac{1}{2} \left( \partial_\mu C_{CD} \right) \sigma^{CD} \; .
\end{equation}

Let us now write the $\Gamma_\mu$ in the form~\eqref{eq:FockIvanenko},
but instead of using the $\omega_{AB \mu}$ as coefficients of the
$\sigma$ matrices we consider some arbitrary coefficients $B_{AB \mu}$:
\begin{equation}
\Gamma_\mu = - \frac{1}{2} \: B_{AB \mu} \sigma^{AB} \; ,
\end{equation}
where at the moment we are not assuming anything about the
coefficients $B_{AB \mu}$, except for the fact that they must be
antisymmetric in their Lorentz indices.  Substituting into the
transformation law we find:
\begin{align}
B'_{AB \mu} \sigma^{AB} &= B_{AB \mu} \sigma^{AB}
+ \frac{1}{2} \: C_{AB} \left[ \sigma^{AB} , B_{CD \mu} \sigma^{CD} \right] 
+ \frac{1}{2} \left( \partial_\mu C_{AB} \right) \sigma^{AB} \nonumber \\
&= \left( B_{AB \mu} + \partial_\mu C_{AB} \right) \sigma^{AB}
+ \frac{1}{2} \: C_{AB} B_{CD \mu} \left[ \sigma^{AB} , \sigma^{CD} \right] \; ,
\end{align}
where we used the fact that the $\sigma^{CD}$ are constant and as such
are invariant, and also that the $C$'s and $B$´s are real numbers.
Using now the commutation relations of the $\sigma$ matrices given
in~\eqref{eq:sigma-sigma} we find, after some algebra:
\begin{equation}
B'_{AB \mu} \sigma^{AB} = \left[  B_{AB \mu} 
+ \partial_\mu C_{AB} - \left( B\indices{_A_C_\mu} C\indices{_B^C}
+ B\indices{_C_B_\mu} C\indices{_A^C} \right) \right] \sigma^{AB} \; ,
\end{equation}
which implies that the $B$ coefficients must transform as:
\begin{equation}
B'_{AB \mu} = B_{AB \mu} 
+ \partial_\mu C_{AB} - \left( B\indices{_A_C_\mu} C\indices{_B^C}
+ B\indices{_C_B_\mu} C\indices{_A^C} \right) \; .
\end{equation}

On the other hand, the transformation rule for the spin connection is:
\begin{align}
\omega'_{AB \mu} &= e'^\lambda_A \omega'_{\lambda B \mu}
= e'^\lambda_A e'_{\lambda B ; \mu}
= {\Lambda_A}^C e^\lambda_C \left( {\Lambda_B}^D e_{\lambda D} \right)_{; \mu}
\nonumber \\
&= {\Lambda_A}^C {\Lambda_B}^D e^\lambda_C \: e_{\lambda D ; \mu}
+ {\Lambda_A}^C e^\lambda_C \: e_{\lambda D} \: \partial_\mu {\Lambda_B}^D \; .
\end{align}
Using now the fact that $e^\lambda_C \: e_{\lambda D} = \eta_{CD}$ we
find:
\begin{equation}
\omega'_{AB \mu} = {\Lambda_A}^C {\Lambda_B}^D \: \omega_{C D \mu}
+ {\Lambda_A}^C \partial_\mu \Lambda_{BC} \; .
\end{equation}
Assuming an infinitesimal transformation, the last expression 
reduces to first order to:
\begin{equation}
\omega'_{AB \mu} = \omega_{AB \mu} + \omega_{AC \mu} {\lambda_B}^C
+ \omega_{CB \mu} {\lambda_A}^C - \partial_\mu \lambda_{AB} \; ,
\end{equation}
where we used the fact that $\lambda_{AB}$ is antisymmetric.
Remembering now that for an infinitesimal transformation we have
$\lambda_{AB} = - C_{AB}$ (see
equation~\eqref{eq:Lorentz-infinitesimal2}), we finally find:
\begin{equation}
\omega'_{AB \mu} = \omega_{AB \mu} + \partial_\mu C_{AB}
- \left( \omega_{AC \mu} {C_B}^C + \omega_{CB \mu} {C_A}^C \right) \; .
\end{equation}
But this is precisely the transformation rule for the $B$ coefficients
we found above, so we conclude that the Fock--Ivanenko coefficients
given by~\eqref{eq:FockIvanenko} do transform in the correct way.


\section{The Schroedinger--Dirac equation}
\label{eq:SchoedingerDirac}

For the case of Minkowski spacetime we have already shown that Dirac's
equation takes us directly to the Klein--Gordon equation for each of
the spinor components.  In the case of a curved spacetime this is no
longer true, and what we find is a generalization of the Klein--Gordon
equation known as the Schroedinger--Dirac
equation~\cite{Schroedinger:1932} (see also~\cite{Fleury:2023} and
references therein).

The first step in finding this equation is to calculate the quantity
$\slashed{\mathcal{D}}^2 \psi := \gamma^\mu \mathcal{D}_\mu
(\gamma^\nu \mathcal{D}_\nu) \psi = \gamma^\mu \gamma^\nu
\mathcal{D}_\mu \mathcal{D}_\nu \psi$. We have:
\begin{align}
\slashed{\mathcal{D}}^2 \psi &= \gamma^\mu \gamma^\nu \mathcal{D}_\mu \mathcal{D}_\nu \psi
= \frac{1}{2} \left( \left\{\gamma^\mu , \gamma^\nu \right\}
+ \left[ \gamma^\mu , \gamma^\nu \right] \right) \mathcal{D}_\mu \mathcal{D}_\nu \psi
\nonumber \\
&= - g^{\mu \nu} \mathcal{D}_\mu \mathcal{D}_\nu \psi
+ \frac{1}{2} \left[ \gamma^\mu , \gamma^\nu \right] \mathcal{D}_\mu \mathcal{D}_\nu \psi
\nonumber \\
&= - \mathcal{D}^\mu \mathcal{D}_\mu \psi
+ \frac{1}{4} \left[ \gamma^\mu , \gamma^\nu \right]
\left[ \mathcal{D}_\mu , \mathcal{D}_\nu \right] \psi
\nonumber \\
&= - \mathcal{D}^\mu \mathcal{D}_\mu \psi
- \frac{1}{2} \: \sigma^{\mu \nu} R_{CD \mu \nu} \sigma^{CD} \psi
\nonumber \\
&= - \mathcal{D}^\mu \mathcal{D}_\mu \psi
- \frac{1}{2} \: R_{ABCD} \sigma^{AB} \sigma^{CD} \psi \; ,
\end{align}
where we used the definition of the $\sigma^{\mu \nu}$ matrices and
the expression for the commutator of the spinor derivatives given
by~\eqref{eq:Ricci-identity-spinor}, and the fact that the Riemann
tensor is symmetric when we exchange the first and second pairs of
indices. Using now the antisymmetry of the Riemann tensor in the two
pairs of indices, and substituting the definition of the $\sigma$
matrices, we can rewrite the previous result as:
\begin{equation}
\slashed{\mathcal{D}}^2 \psi = - \mathcal{D}^\mu \mathcal{D}_\mu \psi
- \frac{1}{8} \: R_{ABCD} \gamma^A \gamma^B \gamma^C \gamma^D \psi \: .
\end{equation}
The second term of the above result in fact turns out to be
proportional to the scalar curvature.  In fact we have:
\begin{equation}
R = - \frac{1}{2} \: R_{ABCD} \gamma^A \gamma^B \gamma^C \gamma^D \; .
\label{eq:scalar-gamma}
\end{equation}
We will leave the proof of this result to the end of this section.  The
quantity $\slashed{\mathcal{D}}^2 \psi$ then takes the final form:
\begin{equation}
\slashed{\mathcal{D}}^2 \psi = \left( - \mathcal{D}^\mu \mathcal{D}_\mu
+ \frac{R}{4} \right) \psi \; .
\label{eq:D2psi}
\end{equation}
This last expression is the natural form of the Laplace operator when
applied to spinors in a curved spacetime.  Having a contribution from
the curvature scalar is not surprising, and comes from the fact that
the covariant derivatives of spinors in general do not commute
(something similar happens in the case of vectors and 1-forms, and
the natural Laplace operator in that case is the so-called ``de Rham
Laplacian'' which also has contributions from the curvature tensor).

Having found this result, we can now go back to the Dirac equation.
Applying the operator $i \gamma^\mu \mathcal{D}_\mu$ from the left we find:
\begin{align}
& i \gamma^\mu \mathcal{D}_\mu \left( i \gamma^\mu \mathcal{D}_\mu \psi 
- m \psi \right) = 0 \nonumber \\
& \implies \quad - \slashed{\mathcal{D}}^2 \psi - i m \gamma^\mu \mathcal{D}_\mu \psi = 0
\nonumber \\
& \implies \quad \slashed{\mathcal{D}}^2 \psi + m^2 \psi = 0 \; ,
\end{align}
where in the second term of the third row we used again Dirac's
equation. Substituting now~\eqref{eq:D2psi} we finally find:
\begin{equation}
\left( \mathcal{D}^\mu \mathcal{D}_\mu - \frac{R}{4} - m^2 \right) \psi = 0 \; .
\label{eq:Schroedinger-Dirac}
\end{equation}
This is the Schroedinger--Dirac equation, and is the generalization of
the Klein--Gordon equation for spinors in a curved spacetime. In the
previous equation one should remember that the operator
$\mathcal{D}^\mu \mathcal{D}_\mu$ must be calculated as (see
equation~\eqref{eq:SecondSpinorDerivative}):
\begin{equation}
\mathcal{D}^\mu \mathcal{D}_\mu \psi = \Box \psi + 2 \Gamma^\mu \partial_\mu \psi
+ \left( \nabla_\mu \Gamma^\mu + \Gamma_\mu \Gamma^\mu \right) \psi \; ,
\end{equation}
where $\Box$ is the usual d'Alambertian applied to scalars, and
$\nabla_\mu \Gamma^\mu = \partial_\mu ( |g|^{1/2} \Gamma^\mu ) /
|g|^{1/2}$. Notice that, since the operator $\mathcal{D}^\mu
\mathcal{D}_\mu$ involves the $\Gamma_\mu$ matrices, in the
Schroedinger--Dirac equation the different components of $\psi$ are in
fact coupled, something that does not happen with the Klein--Gordon
equation.

\vspace{5mm}

We will now prove equation~\eqref{eq:scalar-gamma} that we used in
order to derive the Schroedinger--Dirac equation. The first step is
to consider the contraction of the Ricci tensor $R_{\mu \nu}$
with two $\gamma$ matrices.  using the symmetry of the Ricci
tensor we find:
\begin{equation}
R_{\mu \nu} \gamma^\mu \gamma^\nu = R_{AB} \gamma^A \gamma^B
= \frac{1}{2} \: R_{AB} \left( \gamma^A \gamma^B + \gamma^B \gamma^A \right)
= - R_{AB} \eta^{AB} = - R \; . 
\label{eq:Ricci-gammas}
\end{equation}
Next we must express  $R_{AB} \gamma^A \gamma^B$ in terms
of the Riemann tensor:
\begin{align}
R_{AB} \gamma^A \gamma^B &= R_{ACBD} \eta^{CD} \gamma^A \gamma^B
= R_{ACBD} \gamma^A \eta^{CD} \gamma^B \nonumber \\
&= - \frac{1}{2} \: R_{ACBD} \gamma^A \left( \gamma^C \gamma^D + \gamma^D \gamma^C \right) \gamma^B
\nonumber \\
&= \frac{1}{2} \left( R_{ACDB} \gamma^A \gamma^C \gamma^D \gamma^B
- R_{ACBD} \gamma^A \gamma^D \gamma^C \gamma^B \right)
\nonumber \\
&= \frac{1}{2} \left( R_{ABCD} \gamma^A \gamma^B \gamma^C \gamma^D
- R_{ACBD} \gamma^A \gamma^D \gamma^C \gamma^B \right) \; ,
\end{align}
where we used the antisymmetry of the Riemann tensor in the second pair
of indices, and in the last step we renamed indices on the first term.
In the first term above we can recognize already the contraction
$R_{ABCD} \gamma^A \gamma^B \gamma^C \gamma^D$ that we need in order
to prove~\eqref{eq:scalar-gamma}. For the second term we use the
cyclic symmetry of Riemann so that:
\begin{align}
R_{AB} \gamma^A \gamma^B &= \frac{1}{2} \left[ R_{ABCD} \gamma^A \gamma^B \gamma^C \gamma^D
+ \left( R_{ADCB} + R_{ABDC} \right) \gamma^A \gamma^D \gamma^C \gamma^B \right]
\nonumber \\
&= R_{ABCD} \gamma^A \gamma^B \gamma^C \gamma^D
+ \frac{1}{2} \: R_{ABDC} \gamma^A \gamma^D \gamma^C \gamma^B \; ,
\end{align}
where we again renamed indices to show that two of the terms are
identical. To calculate the second term of the last expression
we use the fact that the Clifford algebra implies:
\begin{equation}
\gamma^D \gamma^C \gamma^B = \gamma^B \gamma^D \gamma^C
+ 2 \eta^{BD} \gamma^C - 2 \eta^{BC} \gamma^D \; .
\end{equation}
Using this result we can show that:
\begin{align}
R_{ABDC} \gamma^A \gamma^D \gamma^C \gamma^B &=
R_{ABDC} \gamma^A \left( \gamma^B \gamma^D \gamma^C
+ 2 \eta^{BD} \gamma^C - 2 \eta^{BC} \gamma^D \right)
\nonumber \\
&= R_{ABDC} \gamma^A \gamma^B \gamma^D \gamma^C
+ 2 R\indices{_A^B_B_C} \gamma^A \gamma^C
- 2 R\indices{_A^B_D_B} \gamma^A \gamma^D
\nonumber \\
&= R_{ABCD} \gamma^A \gamma^B \gamma^C \gamma^D
- 4 R_{AB} \gamma^A \gamma^B \; ,
\end{align}
where once more we renamed indices and grouped terms. Collecting our
results we find:
\begin{equation}
R_{AB} \gamma^A \gamma^B = \frac{3}{2} \: R_{ABCD} \gamma^A \gamma^B \gamma^C \gamma^D
- 2 R_{AB} \gamma^A \gamma^B \; ,
\end{equation}
and solving for $R_{AB} \gamma^A \gamma^B$:
\begin{equation}
R_{AB} \gamma^A \gamma^B = \frac{1}{2} \: R_{ABCD} \gamma^A \gamma^B \gamma^C \gamma^D \: .
\end{equation}
Finally, using equation~\eqref{eq:Ricci-gammas} we obtain
the desired result:
\begin{equation}
R = - \frac{1}{2} \: R_{ABCD} \gamma^A \gamma^B \gamma^C \gamma^D \: .
\end{equation}


\section{Lagrangian and stress--energy tensor of the Dirac field}

\subsection{Lagrangian, first version}

The Lagrangian associated with the Dirac field should be a scalar
function that depends on the spinor $\psi$ and its derivatives.
Furthermore, since the Dirac equation is of first order, the
Lagrangian should also be of first order.  One possible expression
that satisfies all the previous conditions is:
\begin{equation}
L = \bar{\psi} \left( i \gamma^\mu \mathcal{D}_\mu - m \right) \psi \; ,
\label{eq:Lagrangian-Dirac0}
\end{equation}
with $\bar{\psi} = \psi^\dag \gamma^T$, and where here $\gamma^T$
corresponds to the constant matrix $\gamma^0$ used in Minkowski
spacetime.  The change in notation from $\gamma^0$ to $\gamma^T$ is
introduced here in order to avoid a possible confusion bewtween
$\gamma^T$ and coordinate component $\gamma^t= \gamma^A e_A^t$ when
considering the case of a curved spacetime.  From here on the
constante $\gamma$ matrices will be associated with the tetrad basis
and not the coordiante basis (we will discuss this further in
Sec.~\ref{sec:Dirac-3+1}.)

The Lagrangian~\eqref{eq:Lagrangian-Dirac0} is the one found in many
text books on quantum field theory in Minkowski spacetime (with
$\partial_\mu \psi$ instead of $\mathcal{D}_\mu \psi$). The Lagrangian
density then takes the form:
\begin{equation}
\mathcal{L} = \bar{\psi} \left( i \gamma^\mu \mathcal{D}_\mu - m \right) \psi
\: |g|^{1/2}\; ,
\label{eq:Lagrangiandenisty-Dirac0}
\end{equation}
with $g$ the determinant of the metric, and the action integral
becomes:
\begin{equation}
S = \int \mathcal{L} \: d^4 x
= \int \bar{\psi} \left( i \gamma^\mu \mathcal{D}_\mu - m \right) \psi
\: |g|^{1/2} d^4 x \; .
\label{eq:Action-Dirac}
\end{equation}
For the variation of the action above one must take $\psi$ and $\bar{\psi}$
as independent fields.  The Euler--Lagrange equations are then:
\begin{equation}
\frac{\partial}{\partial x^\mu} \left( \frac{\partial \mathcal{L}}{\partial
  (\partial_\mu \psi)} \right)
- \frac{\partial \mathcal{L}}{\partial \psi} = 0 \; ,
\label{eq:EulerLagrange}
\end{equation}
with an analogous equation for $\bar{\psi}$. In fact, in this case it
turns out to be far easier to work with $\bar{\psi}$ since the
Lagrangian~\eqref{eq:Lagrangian-Dirac0} does not depend on its
derivatives.  We find:
\begin{equation}
\frac{\partial \mathcal{L}}{\partial \bar{\psi}}
= \left( i \gamma^\mu \mathcal{D}_\mu - m \right) \psi \: |g|^{1/2} \; ,
\qquad
\frac{\partial \mathcal{L}}{\partial (\partial_\mu \bar{\psi})} = 0 \; ,
\end{equation}
so the Euler--Lagrange equation for $\bar{\psi}$ gives us immediately
Dirac's equation:
\begin{equation}
i \gamma^\mu \mathcal{D}_\mu \psi - m \psi = 0 \; .
\end{equation}

The Dirac equation for $\psi$ is a bit more subtle.  In this case it
is necessary to write the spinor covariant derivative explicitly
in the Lagrangian density:
\begin{equation}
\mathcal{L} = \left[ i \bar{\psi} \gamma^\mu \left( \partial_\mu \psi
+ \Gamma_\mu \psi \right) - m \psi \right] |g|^{1/2}\; .
\end{equation}
From here we then find:
\begin{equation}
\frac{\partial \mathcal{L}}{\partial \psi}
= \bar{\psi} \left( i \gamma^\mu \Gamma_\mu - m  \right) \: |g|^{1/2} \; , \qquad
\frac{\partial \mathcal{L}}{\partial (\partial_\mu \psi)}
= i \bar{\psi} \gamma^\mu \: |g|^{1/2} \; ,
\end{equation}
which implies:
\begin{align}
\frac{\partial}{\partial x^\mu} \left( \frac{\partial \mathcal{L}}{\partial
(\partial_\mu \psi)} \right)
&= i \partial_\mu \left( \bar{\psi} \gamma^\mu \: |g|^{1/2} \right) \nonumber \\
&= i \left[ \left( \partial_\mu \bar{\psi} \right) \gamma^\mu + \bar{\psi} \partial_\mu
\gamma^\mu + \frac{1}{2|g|} \: \bar{\psi} \gamma^\mu \partial_\mu |g| \right] |g|^{1/2} \; .
\end{align}
The Euler--Lagrange equation then takes the form:
\begin{equation}
i \left[ \left( \partial_\mu \bar{\psi} \right) \gamma^\mu + \bar{\psi} \partial_\mu
\gamma^\mu + \frac{1}{2|g|} \: \bar{\psi} \gamma^\mu \partial_\mu |g| \right]
- \bar{\psi} \left( i \gamma^\mu \Gamma_\mu - m  \right) = 0 \; ,
\end{equation}
and regrouping terms:
\begin{equation}
i \left( \mathcal{D}_\mu \bar{\psi} \right) \gamma^\mu + m \bar{\psi}
+ i \bar{\psi} \left[ \partial_\mu \gamma^\mu
+ \frac{1}{2} \: \gamma^\mu \partial_\mu \ln |g|
+ \left[ \Gamma_\mu , \gamma^\mu \right] \right] = 0 \; .
\end{equation}
Using now the fact that $\partial_\mu \ln |g| = 2 \Gamma^\nu_{\nu
  \mu}$, we can recognize that the term in square brackets is just the
spinor divergence of the $\gamma^\mu$ matrices, but this divergence
vanishes since the spinor derivative of the $\gamma^\mu$ is zero.
We then finally obtain:
\begin{equation}
i \left( \mathcal{D}_\mu \bar{\psi} \right) \gamma^\mu + m \bar{\psi} = 0 \; ,
\end{equation}
which is precisely the adjunct Dirac equation.

\subsection{Lagrangian, second version}

Even though the Lagrangian~\eqref{eq:Lagrangian-Dirac0} results
in the correct equations of motion, it has the serious disadvantage of
not being symmetric in $\psi$ and $\bar{\psi}$.  This can be easily
fixed if we define an alternative form of the Lagrangian as:
\begin{equation}
L = \frac{i}{2} \left[ \bar{\psi} \gamma^\mu \left( \mathcal{D}_\mu \psi \right)
- \left( \mathcal{D}_\mu \bar{\psi} \right) \gamma^\mu \psi \right]
- m \bar{\psi} \psi \; .
\label{eq:Lagrangian-Dirac}
\end{equation}
This is the form of the Lagrangian that we will use from now on (this
form for the Lagrangian is also well known, see for
example~\cite{Birrel:1982,Freedman:2012,Burgess:2022}). The Lagrangian
density is now:
\begin{equation}
\mathcal{L} = \left\{ \frac{i}{2} \left[ \bar{\psi} \gamma^\mu \left( \mathcal{D}_\mu \psi \right)
- \left( \mathcal{D}_\mu \bar{\psi} \right) \gamma^\mu \psi \right]
- m \bar{\psi} \psi \right\} |g|^{1/2} \; .
\label{eq:Lagrangiandenisty-Dirac}
\end{equation}
A similar procedure to the one presented above shows that this
Lagrangian density results on precisely the same Dirac equations (one
should mention the fact that the $i$ factor is frequently absorbed in
the definition of the $\gamma$ matrices, so that it does not appear in
the Lagrangian, or indeed in the Dirac equation). It is interesting to
note that both expressions for the Lagrangian in fact become zero when
we substitute the Dirac equation, that is when we evaluate them ``on
shell''.  This represents no problem, since what one is interested in
is the functional form of the Lagrangian in terms of $\psi$ and its
derivatives, and not its specific numerical value on shell.

\vspace{5mm}

The Lagrangian~\eqref{eq:Lagrangiandenisty-Dirac} is clearly invariant
under a transformation of the form:
\begin{equation}
\psi \rightarrow e^{-i \theta} \psi \; , \qquad
\bar{\psi} \rightarrow e^{+i \theta} \bar{\psi} \; , 
\end{equation}
with $\theta$ an arbitrary constant.  This implies the existence
of a conserved Noether current given by:
\begin{align}
j^\mu &= \left( \frac{\partial L}{\partial
(\partial_\mu \psi)} \right) \left( - i \psi \right)
+ \left( i \bar{\psi} \right) \left( \frac{\partial L}{\partial
(\partial_\mu \bar{\psi})} \right) 
\nonumber \\
&= \frac{i}{2} \: \bar{\psi} \gamma^\mu \left( - i \psi \right)
 - \frac{i}{2} \left( i \bar{\psi} \right) \gamma^\mu \psi
\nonumber \\
&= \bar{\psi} \gamma^\mu \psi \; ,
\end{align}
such that $\nabla_\mu j^\mu = 0$.  We can now see that this is
precisely the same conserved current that we had initially found
in equation~\eqref{eq:DiracCurrent}.

\vspace{5mm}

Even if the expression for the
Lagrangian~\eqref{eq:Lagrangiandenisty-Dirac} is correct, it is
interesting to rewrite it in a more illustrative form.  We can write
the Lagrangian as:
\begin{equation}
L = K - V \; ,
\end{equation}
where here $K$ is the so-called {\em kinetic}\/ term given by:
\begin{align}
K &= \frac{i}{2} \left[ \bar{\psi} \gamma^\mu \left( \mathcal{D}_\mu \psi \right)
- \left( \mathcal{D}_\mu \bar{\psi} \right) \gamma^\mu \psi \right] \nonumber \\
&= \frac{i}{2} \left[ \bar{\psi} \gamma^\mu \left( \partial_\mu \psi \right)
- \left( \partial_\mu \bar{\psi} \right) \gamma^\mu \psi
+ \bar{\psi} \left( \gamma^\mu \Gamma_\mu + \Gamma_\mu \gamma^\mu \right) \psi \right] \; ,
\label{eq:Kinetic1}
\end{align}
while $V$ is the {\em potential}\/ term that is simply:
\begin{equation}
V = m \bar{\psi} \psi \; .
\end{equation}
The kinetic term can in turn be written in several different forms.
If we use the expression for the $\Gamma_\mu$ in terms of the
Ricci rotation coefficients given in~\eqref{eq:FockIvanenko}
we find:
\begin{equation}
\Gamma_\mu = - \frac{1}{2} \: \omega_{AB \mu} \sigma^{AB}
= - \frac{1}{2} \: e_\mu^C \: \omega_{ABC} \: \sigma^{AB} \; .
\label{eq:Gamma-f}
\end{equation}
Using this relation, it is not difficult to show that:
\begin{equation}
\gamma^\mu \Gamma_\mu + \Gamma_\mu \gamma^\mu
= - \frac{1}{2} \: \omega_{ABC} \: \gamma^{CAB} \; ,
\end{equation}
where we have defined $\gamma^{CAB} := \{ \gamma^C, \sigma^{AB} \}$.
For what follows it is important to notice that $\gamma^{CAB}$
is totally antisymmetric:
\begin{equation}
\gamma^{CAB} = - \gamma^{CBA} = - \gamma^{ACB} = - \gamma^{BAC}\; .
\end{equation}
The kinetic term then reduces to:
\begin{equation}
K = \frac{i}{2} \left[ \bar{\psi} \gamma^\mu \left( \partial_\mu \psi \right)
- \left( \partial_\mu \bar{\psi} \right) \gamma^\mu \psi \right]
- \frac{i}{4} \: \bar{\psi} \left( \omega_{ABC} \: \gamma^{CAB} \right) \psi \; .
\label{eq:Kinetic2}
\end{equation}
Finally, if we substitute the $\omega$'s using
equation~\eqref{eq:omega-f}, and use the anticommutation property of
the $\gamma^A$ matrices, a little algebra allows us to show that the
kinetic term takes the final form:
\begin{equation}
K = \frac{i}{2} \left[ \bar{\psi} \gamma^\mu \left( \partial_\mu \psi \right)
- \left( \partial_\mu \bar{\psi} \right) \gamma^\mu \psi \right]
+ \frac{i}{4} \: \bar{\psi} \left( f_{ABC} \: \gamma^{CAB} \right) \psi \; ,
\label{eq:Kineti3}
\end{equation}
where we must remember that the $f$'s are defined as $f_{ABC} :=
(\partial_A e_{\nu B}) e^\nu_C = e_A^\mu e_C^\nu \partial_\mu e_{\nu
  B}$.


\subsection{Stress--energy tensor}

In order to find the form of the stress--energy tensor associated with
the Dirac field we start from the action integral:
\begin{equation}
S = \int L \: |g|^{1/2} d^4 x \; ,
\end{equation}
with $L$ the Lagrangian that we found above:
\begin{equation}
L = \frac{i}{2} \left[ \bar{\psi} \gamma^\mu \left( \mathcal{D}_\mu \psi \right)
- \left( \mathcal{D}_\mu \bar{\psi} \right) \gamma^\mu \psi \right]
- m \bar{\psi} \psi \; .
\end{equation}

At this point one could think of using the standard definition for the
Hilbert stress--energy tensor in terms of the Lagrangian $L$ given by:
\begin{equation}
T_{\mu \nu} =  - 2 \: \frac{\partial L}
{\partial g^{\mu \nu}} + g_{\mu \nu} L \; .
\end{equation}
However, in the case of a Dirac field this definition fails since the
Lagrangian has terms that depend directly on the tetrad and not just
on the metric. Instead, we must now define the stress--energy tensor by
considering the variation of the action with respect to the tetrad
itself:
\begin{equation}
T_{\mu \nu} = - \frac{1}{2} \left( e_{\mu D} \: \frac{\delta L}{\delta e_D^\nu}
+ e_{\nu D} \: \frac{\delta L}{\delta e_D^\mu} \right) + g_{\mu \nu} L \; ,
\label{eq:Tmunu-tetrad0}
\end{equation}
where we must remember that the metric is given in terms of the tetrad
as $g_{\mu \nu} = e_{\mu A} e_\nu^A$, and where we have included two
symmetrized terms since the final stress--energy tensor must be
symmetric. It is not difficult to see that for an action that depends
only on the metric both definitions for $T_{\mu \nu}$ are in fact
equivalent.

Using the expression above one arrives, after a somewhat lengthy
algebra, at:
\begin{equation}
T_{\mu \nu} = \frac{i}{2} \left[ \left(
\mathcal{D}_{(\mu} \bar{\psi} \right) \gamma_{\nu)} \psi -
\bar{\psi} \gamma_{(\mu} \left( \mathcal{D}_{\nu)} \psi \right) \right] \; .
\label{eq:Dirac-stressenergy}
\end{equation}
This is the stress--energy tensor for the Dirac equation (this
expression is also well known, see
e.g.~\cite{Freedman:2012,Dolan:2015,Shapiro:2022}). Details of the
derivation of this stress--energy tensor, as well as a proof that it
satisfies the conservation equations $\nabla^\mu T_{\mu \nu}=0$, can
be found in Appendix A.

\vspace{5mm}

There is an property of the stress--energy
tensor~\eqref{eq:Dirac-stressenergy} that is interesting to mention.
If we take its trace, and use the Dirac equation, one can easily show
that:
\begin{equation}
{T^\mu}_\mu = - m \bar{\psi} \psi = - m \left( \psi^\dag \gamma^T \psi \right)
= - m \left( | \psi_1 |^2 + | \psi_2 |^2 - | \psi_3 |^2 - | \psi_4 |^2 \right) \; .
\label{eq:Dirac-stressenergy-trace}
\end{equation}
We then see that the trace in fact vanishes for $m=0$.


\section{Dirac equation in the 3+1 formalism}

In the previous sections we found the general form of the Dirac
equation for a curved spacetime.  For numerical applications,
or in case one is interested in the Hamiltonian formalism,
it is interesting to find the form that the Dirac equation
takes in the 3+1 formalism of general relativity.

We assume that the spacetime is globally hyperbolic, so it can be
foliated by Cauchy hypersurfaces $\Sigma_{t}$ parametrized by a global
time function $t$. The metric can then be written in the general form
(see for example~\cite{Alcubierre08a}):
\begin{equation}
ds^2 = \left( - \alpha^2 + \beta_i \beta^i \right) dt^2 + 2 \beta_i dt dx^i
+ \gamma_{ij} dx^i dx^j \; ,
\end{equation}
where $x^i$ are spatial coordinates, $\alpha$ is the lapse function,
$\beta^i$ is the shift vector, $\gamma_{ij}$ is the spatial metric
induced on the spacelike hypersurfaces of constant $t$, and where
$\beta_i = \gamma_{ij} \beta^j$ (in general the indices of purely
spatial tensors are raised and lowered with $\gamma_{ij}$).

The next step is to choose a tetrad adapted to our spacetime foliation.
In particular we will take:
\begin{equation}
e^\mu_T = n^\mu \; , \qquad e^\mu_I = E^\mu_I \; ,
\end{equation}
with $n^\mu$ the unit vector orthogonal to the spatial hypersurfaces,
and where $E^\mu_I$ with \mbox{$I \in \{1,2,3\}$} are three purely
spatial vectors orthogonal to each other that from now on we will call
the ``spatial triad''.

Do notice that we now have four different types of indices that
require special notation: 1) Spacetime coordinate indices that take
values from 0 to 3, for which we continue to use Greek letters
$\{\alpha,\beta,\cdots\}$; 2) tetrad (Lorentz) indices that also take
values from 0 to 3, for which we continue to use upper case Latin
letters starting from $\{A,B,\cdots\}$; 3) purely spatial coordinate
indices that only take values from 1 to 3, for which we use lower case
Latin letters starting from $\{i,j,\cdots\}$; 4) purely spatial
triad indices that also only take values from 1 to 3, and for which we
will use upper case Latin indices starting from $\{I,J,\cdots\}$.

The timelike vector $e^\mu_T = n^\mu$ can now be expressed in terms of
the lapse and shift as:
\begin{equation}
e^\mu_T = \left( 1/\alpha, - \beta^i/ \alpha \right) \; , \qquad
e_{\mu T} = \left( - \alpha, 0 \right) \; ,
\end{equation}
so that we clearly have $e^\mu_T e_{\mu T} = -1$.
On the other hand, since the vectors $E^\mu_I$ are purely
spatial we must have  $e^\mu_T E_{\mu I} = e_{\mu T} E^\mu_I = 0$,
so that:
\begin{equation}
E^0_I = 0 \; , \qquad E_{0 I} = \beta^m E_{m I} \; .
\end{equation}
Notice in particular that the purely spatial indices of $E^m_I$
are raised and lowered with the spatial metric $\gamma_{mn}$,
that is: $E_{mI} =\gamma_{mn} E^n_I$, $E^m_I = \gamma^{mn} E_{nI}$.
We also have:
\begin{equation}
E_{mI} E^m_J = \delta_{IJ} \; ,
\qquad
E_m^I E_{nI} = \gamma_{mn} \; .
\end{equation}

At this point it is convenient to introduce the projection operator
onto the spatial hypersurfaces defined as:
\begin{equation}
P^\mu_\nu := \delta^\mu_\nu + n^\mu n_\nu \; .
\label{eq:projection3D}
\end{equation}
Notice that, so defined, this operator corresponds directly with the
induced metric on the spatial hypersurfaces, \mbox{$\gamma_{\mu \nu}
  \equiv P_{\mu \nu}$}.  In particular the spatial metric is
$\gamma_{ij} = P_{ij}$, as can be seen directly from the above
definition.


\subsection{Ricci rotation coefficients in 3+1 form}

In order to find the components of the Ricci rotation coefficients
in the 3+1 formalism we start from equation~\eqref{eq:omega-gamma3} 
which we rewrite here:
\begin{equation}
\omega_{AB \mu} = e_B^\nu \left( e_{\lambda A} \Gamma^\lambda_{\nu \mu}
- \partial_\mu e_{\nu A} \right) \; .
\label{eq:omega-gamma4}
\end{equation}

The expressions for the Christoffel symbols $\Gamma^\lambda_{\mu \nu}$ in terms
of the  3+1 quantities are well known and can be found, for example,
in Appendix B of reference~\cite{Alcubierre08a}.  We write them again
here for completeness:
\begin{eqnarray}
\Gamma_{00}^0 &=& \left( \partial_t\,\alpha +
\beta^m \partial_m\,\alpha - \beta^m \beta^n K_{mn} \right) / \alpha \; ,
\label{eq:Gamma-000} \\
\Gamma_{0i}^0 &=& \left( \partial_i \alpha -
\beta^m K_{im} \right) / \alpha \; , \\
\Gamma_{ij}^0 &=& - K_{ij} / \alpha \; , \\
\Gamma_{00}^l &=& \alpha \partial^l \alpha - 2 \alpha \beta^m K_m^l
- \beta^l \left( \partial_t \alpha +
\beta^m \partial_m \alpha - \beta^m \beta^n K_{mn} \right) /
\alpha + \partial_t \beta^l + \beta^m \; D_m \beta^l \; , \\
\Gamma_{m0}^l &=& - \beta^l \left( \partial_m \alpha -
\beta^n K_{mn} \right) / \alpha
- \alpha K^l_m + D_m \beta^l \; , \\
\Gamma_{ij}^l &=& {}^{(3)}\Gamma_{ij}^l + \beta^l K_{ij} / \alpha \; ,
\label{eq:Gamma-lij}
\end{eqnarray}
with $D_i$ the covariant derivative associated with the spatial metric
$\gamma_{ij}$, ${}^{(3)}\Gamma_{ij}^l$ the corresponding
three-dimensional Christoffel symbols, and where $K_{ij}$ refers to
the extrinsic curvature tensor of the spatial hypersurfaces of
constant time (also known as the second fundamental form), defined as
the spatial projection of the covariant derivative of the timelike
unit normal vector to the spatial hypersurfaces, that is $K_{\mu \nu}
:= - P^\alpha_\mu \nabla_\alpha n_\nu$.  The purely spatial components
of the extrinsic curvature can be shown to be related to the time
derivative of the spatial metric $\gamma_{ij}$ as:
\begin{equation}
\partial_t \gamma_{ij} = - 2 \alpha K_{ij} + D_i \beta_j + D_j \beta_i \; .
\label{eq:gamma-dot}
\end{equation}
Equations~\eqref{eq:Gamma-000}-\eqref{eq:Gamma-lij} are used for
deriving the results that follow.

Below I show the final results for the different components of the
Ricci rotation coefficients $\omega_{AB \mu}$.  The calculations are
straightforward and will not be presented here in detail.  Also,
remember that the $\omega$'s are antisymmetric in the first two
indices, so that in a four-dimensional spacetime they only have 24
independent components.  In what follows the index $T$ refers
specifically to the Lorentz time component, while the index $I$ refers
to Lorentz purely spatial components.

\begin{enumerate}

\item Coefficients $\omega_{TI0} = - \omega_{IT0}$ (3 coefficients):
\begin{equation}
\omega_{TI0} = - \omega_{IT0} = - E^m_I \left( \partial_m \alpha - \beta^n K_{mn} \right) \; .
\label{eq:omega_TI0}
\end{equation}

\item Coefficients $\omega_{TIm} = - \omega_{ITm}$ (9 coefficients):
\begin{equation}
\omega_{TIm} = - \omega_{ITm} = E^n_I K_{nm} \; .
\label{eq:omega_TIm}
\end{equation}

\item Coefficients $\omega_{IJ0} = - \omega_{JI0}$ (3 coefficients):
\begin{equation}
\omega_{IJ0} = - \omega_{JI0} = - E^m_J \left[ \partial_t E_{m I}
- E^n_I \left( - \alpha K_{mn} + D_m \beta_n \right) \right] \; .
\label{eq:omega_IJ0}
\end{equation}
Here it is important to mention that, even if it is not immediately
evident, when using equation~\eqref{eq:gamma-dot} for the time
derivative of the spatial metric, plus the fact that $\gamma_{mn} =
E_{mI} E^I_n$, a short calculation allows one to show that the
previous result is in fact antisymmetric in $(I,J)$.

\item Coefficients $\omega_{IJm} = - \omega_{JIm}$ (9 coefficients):
\begin{equation}
\omega_{IJm} = - E^n_J D_m E_{nI} \; .
\end{equation}
Again, using the fact that the three-dimensional covariant derivative
of the spatial metric is zero, it is not difficult to see that the
previous expression is also antisymmetric in $(I,J)$.

If we now define the three-dimensional Ricci rotation coefficients as:
\begin{equation}
\omega_{IJm}^{(3)} := E^n_I D_m E_{nJ} \; ,
\end{equation}
our result reduces simply to:
\begin{equation}
\omega_{IJm} = - \omega_{JIm}
= E^n_I D_m E_{nJ} \equiv \omega_{IJm}^{(3)} \; .
\label{eq:omega_IJm}
\end{equation}

\end{enumerate}

\vspace{5mm}

The above results provide us with 3+1 expressions for the 24
independent components of the four-dimensional Ricci rotation
coefficients $\omega_{AB \mu}$.  Using these results we can now also
project the third index of the $\omega_{AB \mu}$ onto the tetrad to
obtain $\omega_{ABC}$:

\begin{enumerate}

\item Coefficients $\omega_{TIT} = - \omega_{ITT}$ (3 coefficients):
\begin{equation}
\omega_{ITT} = - \omega_{TIT} = E^m_I \partial_m \alpha / \alpha
\equiv \partial_I \alpha / \alpha \; .
\label{eq:omega_ITT}
\end{equation}

\item Coefficients $\omega_{TIJ} = - \omega_{ITJ}$ (9 coefficients):
\begin{equation}
\omega_{TIJ} = - \omega_{ITJ} = E^m_I E^n_J K_{mn} \equiv K_{IJ} \; .
\label{eq:omega_TIJ}
\end{equation}
Notice in particular that the coefficients $\omega_{TIJ}$ turn out
to be symmetric in $(I,J)$.  On the other hand, from~\eqref{eq:gamma-dot}
we have:
\begin{equation}
K_{mn} = - \frac{1}{2 \alpha} \left( \partial_t \gamma_{mn}
- D_m \beta_n - D_n \beta_m \right) \; .
\end{equation}
Equation~\eqref{eq:omega_TIJ} can then also be written as:
\begin{equation}
\omega_{TIJ} = - \omega_{ITJ} =
- \frac{E^m_I E^n_J}{2 \alpha} \left( \partial_t \gamma_{mn}
- D_m \beta_n - D_n \beta_m \right) \; .
\label{eq:omega_TIJ-2}
\end{equation}

\item Coefficients $\omega_{IJT} = - \omega_{JIT}$ (3 coefficients):
\begin{equation}
\omega_{IJT} = - \omega_{JIT}
= - \frac{1}{\alpha} \left[ E^m_J \left( \partial_t E_{m I}
- \pounds_{\vec{\beta}}\:  E_{mI} \right) + \alpha K_{IJ} \right] \; ,
\label{eq:omega_IJT}
\end{equation}
where $\pounds_{\vec{\beta}}\: E_{mI}$ denotes the Lie derivative of
the 1-form $E_{mI}$ with respect to the shift vector $\beta^i$.

Again, even if it is not evident, one can show that the previous coefficients
are antisymmetric in $(I,J)$.  Notice that this antisymmetry implies that
only three of these coefficients are independent from each other.

\item Coefficients $\omega_{IJK} = - \omega_{JIK}$ (9 coefficients):
\begin{equation}
\omega_{IJK} = - \omega_{JIK} = E^n_I E^m_K D_m E_{n J}
\equiv \omega_{IJK}^{(3)} \; .
\label{eq:omega_IJK}
\end{equation}

\end{enumerate}

There are some interesting facts that we can notice about our results.
First, as already mentioned, there are only 24 independent
coefficients.  We can see that once we have defined a spatial triad
$E^m_I$, nine of these coefficients are given simply by the projection
of the extrinsic curvature tensor on the triad, $\omega_{TIJ} =
K_{IJ}$.  Another nine coefficients are just projections onto the
triad of the 3-covariant derivatives of the triad itself,
$\omega_{IJK} = E^n_I E^m_K D_m E_{nJ}=\omega_{IJK}^{(3)}$.  This
means that 18 of the coefficients depend only on information at a
given hypersurface.  Of the six remaining coefficients, three depend
directly on our choice of the lapse function $\alpha$, in other words
on our slicing condition, $\omega_{ITT} = \partial_I \alpha / \alpha$.

The last three coefficients correspond to $\omega_{IJT}$, and they depend
on the form in which the spatial triad propagates through time, as can
be seen from equation~\eqref{eq:omega_IJT}. Just as the choice of
lapse and shift is free, the choice of the propagation of the spatial
triad in time is also free, so it represents a new gauge degree of
freedom.  Now, since the triad must by definition be orthonormal, the
only thing it can do as it evolves is rotate rigidly in space, and
this rotation can be parametrized by the usual three Euler angles.  This
explains why these coefficients have only three degrees of freedom.


\subsection{Triad evolution and Fermi--Walker transport}
\label{sec:FermiWalker}

In principle there are many different forms to choose the evolution of
the spatial triad, but there is one particular choice that is quite
natural and can be useful in many cases.  Such a choice consists on
asking for the triad {\em not}\/ to rotate as it propagates along the
normal direction to the spatial hypersurfaces, or in other words,
asking for the triad to evolve in such a way that it always
corresponds to the natural local inertial frame associated with the
normal (Eulerian) observers as they move through time.  The condition
we need to impose to achieve this is known as {\em Fermi--Walker}\/
transport~\cite{Synge60}.

Consider the worldline of an arbitrary observer with 4-velocity
$u^\mu$, such that $u_\mu u^\mu=-1$. In the general case we say that a
vector $v^\mu$ is transported without rotation along the curve with
tangent $u^\mu$, with $v^\mu$ not necessarily normal to $u^\mu$, if we
have:
\begin{equation}
u^\mu \nabla_\mu v_\nu =  v^\mu \left( a_\mu u_\nu - a_\nu u_\mu \right) \; ,
\label{eq:FermiWalker0}
\end{equation}
where $a^\mu := u^\nu \nabla_\nu u^\mu$ is the 4-acceleration
associated with our observer (if the observer moves on a geodesic we
will clearly have $a^\mu=0$ and Fermi-Walker transport becomes
equivalent to parallel transport). The previous condition defines what is
known as the Fermi--Walker transport of the vector $v^\mu$ along the
integral lines of the vector field $u^\mu$.

In the case when we consider the transport of the spatial triad along
the worldline of the Eulerian observers we will have $u^\mu = n^\mu$
and $v^\mu = E^\mu_I$, with $n^\mu$ the unit timelike vector to the
spatial hypersurfaces. Since by construction we have $n^\mu E_{\mu I}
= 0$, the second term inside the parenthesis in
equation~\eqref{eq:FermiWalker0} vanishes and our condition reduces
to:
\begin{equation}
n^\mu \nabla_\mu E_{\nu I} =  \left( E^\mu_I a_\mu \right) n_\nu \; ,
\label{eq:FermiWalker1}
\end{equation}
where $a^\mu := n^\nu \nabla_\nu n^\mu$ is now the 4-acceleration of
the Eulerian observers. Let us now see the form that this condition
takes in 3+1 language.  The first step is to calculate the term
$E^\mu_I a_\mu$.  We have:
\begin{align}
E^\mu_I a_\mu &= E^\mu_I n^\nu \nabla_\nu n_\mu
= E^\mu_I n^\nu \left( \partial_\nu n_\mu
- \Gamma^\lambda_{\mu \nu} n_\lambda \right) \nonumber \\
&=  E^\mu_I \left[ \frac{1}{\alpha} \left( \partial_t n_\mu
- \Gamma^\lambda_{\mu 0} n_\lambda \right)
- \frac{\beta^m}{\alpha} \left( \partial_m n_\mu
- \Gamma^\lambda_{\mu m} n_\lambda \right) \right] \nonumber \\
&= - \frac{\cancelto{0}{E^0_I}}{\alpha} \left( \partial_t \alpha - \beta^m \partial_m \alpha \right)
+ E^\mu_I \left( \Gamma^0_{\mu 0} - \beta^m \Gamma^0_{\mu m} \right)
= E^n_I \left( \Gamma^0_{n 0} - \beta^m \Gamma^0_{n m} \right) \nonumber \\
&= E^n_I \left[ \frac{1}{\alpha} \left( \partial_n \alpha - \cancel{\beta^m K_{mn}} \right)
+ \cancel{\frac{\beta^m}{\alpha} \: K_{nm}} \right]
= \frac{\partial_I \alpha}{\alpha} \; .
\end{align}
The Fermi--Walker condition then becomes:
\begin{equation}
n^\mu \nabla_\mu E_{\nu I} = n_\nu \left( \partial_I \ln \alpha \right) \; .
\label{eq:FermiWalker2}
\end{equation}

To proceed, consider first the spatial components of the previous
condition by taking $\nu=m$. Since we know that $n_m=0$, the condition
now reduces to simply $n^\mu \nabla_\mu E_{m I} = 0$. Calculating the
left hand side we find, after some algebra:
\begin{equation}
n^\mu \nabla_\mu E_{mI} = \frac{1}{\alpha} \left\{ \partial_t E_{mI} - \pounds_{\vec{\beta}}\: E_{mI}
+ \alpha K_{mI} \right\} \; .
\end{equation}
The Fermi--Walker condition the takes the final form:
\begin{equation}
\partial_t E_{mI} - \pounds_{\vec{\beta}}\: E_{mI} + \alpha K_{mI} = 0 \; ,
\end{equation}
or:
\begin{equation}
\partial_t E_{mI} = \pounds_{\vec{\beta}}\: E_{mI} - \alpha K_{mI} \; .
\label{eq:FermiWalker3}
\end{equation}

We still need to consider the time component of
equation~\eqref{eq:FermiWalker2}, that is taking $\nu=0$.  We won't
do the calculation here, but it is not too difficult to show that this
component in fact adds no new information and reduces again
to~\eqref{eq:FermiWalker3}.  This equation is then the full expression
that guarantees that the spatial triad propagates without rotation as
seen by the Eulerian observers.

\vspace{5mm}

Let us now return to equation~\eqref{eq:omega_IJT} for the
coefficients $\omega_{IJT}$. This equation can be trivially rewritten
as:
\begin{equation}
\omega_{IJT} = - \omega_{JIT}
= - \frac{E^m_J}{\alpha} \left[ \partial_t E_{m I}
- \pounds_{\vec{\beta}}\:  E_{mI} + \alpha K_{m I} \right] \; .
\end{equation}
We then see that if we impose the condition that the spatial
triad should evolve under Fermi--Walker transport along
the normal direction to the spatial hypersurfaces, we
will clearly have:
\begin{equation}
\omega_{IJT} = 0 \; .
\end{equation}

We should emphasize here that this result is just a gauge choice, and
does not need to apply in the general case.  This choice simplifies
the equations, and as such can be useful in some cases.  On the other
hand, it is not difficult to show that if we impose
condition~\eqref{eq:FermiWalker3} on the spatial triad we recover the
evolution equation for the metric~\eqref{eq:gamma-dot}, so that
imposing the condition is a perfectly consistent gauge choice.

Now, even though condition~\eqref{eq:FermiWalker3} is not general,
one can always write in the general case:
\begin{equation}
\partial_t E_{mI} = \pounds_{\vec{\beta}}\: E_{mI} - \alpha K_{mI} + \alpha Q_{mI} \; ,
\label{eq:dtEmI-general}
\end{equation}
with the $Q_{mI}$ quantities to be chosen.  Notice, however, that not
all the $Q_{mI}$ can be chosen freely since we need to guarantee that
equation~\eqref{eq:gamma-dot} holds.  A little algebra allows us to
show that this requirement implies that $Q_{mn}$ with purely space
indices, or equivalently $Q_{IJ}$ with purely triad indices, must be
antisymmetric.  This is to be expected since, as mentioned above, the
only freedom we really have is a rigid rotation of the triad, which
can always be described by a $3 \times 3$ antisymmetric matrix.
Equation~\eqref{eq:dtEmI-general} then allows us to reduce all the
gauge freedom associated with the evolution of the spatial triad to
the choice of the three-dimensional tensor $Q_{mn}$.  Given
equation~\eqref{eq:dtEmI-general}, the coefficients $\omega_{IJT}$ are
then given in general by:
\begin{equation}
\omega_{IJT} = - Q_{JI} = + Q_{IJ} \; ,
\end{equation}
with $Q_{IJ}=0$ corresponding to the choice of a triad that
is Fermi--Walker transported along the normal direction to
the spatial hypersurfaces.

\vspace{5mm}

As mentioned above, Fermi--Walker transport can be a good choice in
many cases.  As we will see below, it is the natural choice in
spherical symmetry.  However, there are many situations when such a
choice might not be adequate.  For example, in situations when there
is angular momentum asking for the triad not to rotate might be a very
bad choice.  In fact, it is not difficult to show that in the case of
the Kerr spacetime the most natural choice of triad, that is the one
associated with the already orthogonal spatial coordinates in the
standard Boyer--Lindquist form of the metric, does not
satisfy~\eqref{eq:FermiWalker3}.  Another possible choice for the
triad evolution is to simply take:
\begin{equation}
\partial_t E_{mI} = \frac{1}{2} \: E^n_I \partial_t \gamma_{mn} \; ,
\label{eq:dtEmI-static1}
\end{equation}
or equivalently:
\begin{equation}
\partial_t E^m_I = - \frac{1}{2} \: E^l_I \gamma^{mn} \partial_t \gamma_{ln} \; .
\label{eq:dtEmI-static2}
\end{equation}
This last condition has the advantage of guaranteeing that for a
stationary spacetime, such as Kerr for example, the spatial triad is
also time independent. We shall call it the {\em metric driven}\/
triad gauge choice.  It is not difficult to show that this condition
is also perfectly consistent in the sense that we find:
\begin{align}
\partial_t \left( \delta_{IJ} \right) &= \partial_t \left( E^m_I E_{mJ} \right)
= E^m_I \partial_t E_{mJ} + E_{mJ} \partial_t E^m_I \nonumber \\
&= \frac{1}{2} \left( E^m_I E^n_J \partial_t \gamma_{mn}
- E_{mJ} E^l_I \gamma^{mn} \partial_t \gamma_{ln} \right) = 0 \; ,  
\end{align}
and:
\begin{align}
\partial_t \left( E^I_m E_{nI} \right) &= \delta^{IJ} \partial_t \left(E_{mJ} E_{nI} \right)
= \delta^{IJ} \left( E_{mJ} \partial_t E_{nI} + E_{nI} \partial_t E_{mJ} \right) \nonumber \\
&= \frac{1}{2} \left( E^I_m E^a_I \partial_t \gamma_{an}
+ E^I_n E^a_J \partial_t \gamma_{am} \right)
= \frac{1}{2} \left( \delta^a_m \partial_t \gamma_{an}
+ \delta^a_n \partial_t \gamma_{am} \right) \nonumber \\
&= \partial_t \gamma_{mn} \; .
\end{align}
We can now substitute condition~\eqref{eq:dtEmI-static1} in the
general equation~\eqref{eq:dtEmI-general} in order to find the 
value of $Q_{mn}$ for this gauge choice.  We find, after some
algebra:
\begin{equation}
Q_{mn} = \frac{1}{\alpha} \left[ \frac{1}{2} \left( D_n \beta_m - D_m \beta_n \right)
- \beta^l \left( E_n^I D_l E_{mI} \right) \right] \; .
\end{equation}
The first term in the last expression is clearly antisymmetric.
On the other hand, using the fact that  $E_n^I E_{mI} =
\gamma_{nm}$ it is not difficult tom show that the second term is also
antisymmetric, so that we can rewrite the expression as:
\begin{equation}
Q_{mn} = - \frac{1}{2 \alpha} \left[ \left( \partial_m \beta_n - \partial_n \beta_m \right) 
- \beta^l \left( E_m^K D_l E_{nK} - E_n^K D_l E_{mK} \right) \right] \; ,
\label{eq:Qmn-3+1}
\end{equation}
or equivalently:
\begin{equation}
Q_{IJ} = - \frac{1}{2 \alpha} \left[
E^m_I E^n_J \left( \partial_m \beta_n - \partial_n \beta_m \right) 
+ \beta^l \left( E^m_I D_l E_{mJ} - E^m_J D_l E_{mI} \right) \right] \; .
\label{eq:Qmn-3+1-2}
\end{equation}
where we used the symmetry of the Christoffel symbols on the two lower
indices to change covariant derivatives for partial derivatives in the
first term. This is the form of the tensor $Q$ when we use the metric
driven triad evolution~\eqref{eq:dtEmI-static1}.  Notice that this
form of $Q$ vanishes for the case when we have no shift vector,
$\beta^i=0$.

Of course, other gauge choices for $\partial_t E_{mI}$ might be
useful/interesting, but we will not discuss this issue further here.


\subsection{Fock--Ivanenko coefficients in 3+1 form}

The next step is to find the form of the Fock--Ivanenko coefficients
$\Gamma_\mu$  in terms of 3+1 quantities. In order to do this we start
from equation~\eqref{eq:FockIvanenko2}, which we repeat here for
completeness:
\begin{equation}
\Gamma_\mu = - \frac{1}{4} \: \omega_{AB \mu} \gamma^A \gamma^B \; .
\label{eq:FockIvanenko2}
\end{equation}
It turns out to be more convenient to work with the coefficients
projected onto the tetrad, which now take the form:
\begin{equation}
\Gamma_C = - \frac{1}{4} \: \omega_{ABC} \gamma^A \gamma^B \; ,
\label{eq:FockIvanenko3}
\end{equation}
where here we must remember that the indices $(A,B,C)$ take values
from 0 to 3, while the indices $(I,J,K)$ will only take values from 1
to 3.

Consider first the time component $\Gamma_T$, we have:
\begin{equation}
\Gamma_T = - \frac{1}{4} \: \omega_{AB T} \gamma^A \gamma^B
= - \frac{1}{4} \left[ 2 \: \omega_{TIT} \gamma^T \gamma^I
+ \omega_{IJT} \gamma^I \gamma^J \right] \; ,
\end{equation}
where we used the fact that the $\omega_{ABC}$ are antisymmetric in
the first two indices, and $\gamma^A \gamma^B = - \gamma^B \gamma^A$
for $A \neq B$ (from Clifford's algebra). Substituting now the
values of $\omega_{ITT}$ and $\omega_{IJT}$ from
equations~\eqref{eq:omega_ITT} and~\eqref{eq:omega_IJT} we find:
\begin{align}
\Gamma_T &= \frac{1}{4} \left[ 2 \left( \frac{\partial_I \alpha}{\alpha} \right) \gamma^T \gamma^I 
+ \frac{E^m_J}{\alpha} \left( \partial_t E_{m I}
- \pounds_{\vec{\beta}}\:  E_{mI} + \alpha K_{mI} \right) \gamma^I \gamma^J \right] \nonumber \\
&= \left( \frac{\partial_I \alpha}{2 \alpha} \right) \gamma^T \gamma^I
- \frac{1}{4} \: Q_{IJ} \gamma^I \gamma^J \; ,
\label{eq:Gamma_T}
\end{align}
where in the last step we used equation~\eqref{eq:dtEmI-general}.
Remember that in the last expression the antisymmetric matrix
$Q_{IJ}$ is a free gauge choice that vanishes when the triad
evolves under Fermi--Walker transport.

Before considering the spatial components of $\Gamma_A$, let us
remember that in Dirac's equation~\eqref{eq:Dirac-curved} we have the
contraction $\gamma^\mu \Gamma_\mu$ coming from the operator
$\gamma^\mu \mathcal{D}_\mu$, and we have in general $\gamma^\mu
\Gamma_\mu = \gamma^A \Gamma_A = \gamma^T \Gamma_T + \gamma^I
\Gamma_I$. At  this point we can already calculate the product $\gamma^T
\Gamma_T$:
\begin{equation}
\gamma^T \Gamma_T = \left( \frac{\partial_I \alpha}{2 \alpha} \right)
\left( \gamma^T \right)^2 \gamma^I
- \frac{1}{4} \: Q_{IJ} \gamma^T \gamma^I \gamma^J
= \left( \frac{\partial_I \alpha}{2 \alpha} \right) \gamma^I
- \frac{1}{4} \: Q_{IJ} \: \gamma^T \gamma^I \gamma^J \; .
\end{equation}
Here one should emphasize the fact that in the first term above we
have a sum over $I$, while in the second we have sums over $I$ and
$J$.  Notice that the term $\gamma^T \Gamma_T$ clearly only depends on
our gauge choices, in particular on the choice of the lapse function
$\alpha$ and the triad rotation matrix $Q_{IJ}$. For the particular
case when $Q_{IJ}=0$ we simply have:
\begin{equation}
\gamma^T \Gamma_T = \left( \frac{\partial_I \alpha}{2 \alpha} \right) \gamma^I \; .
\end{equation}

\vspace{5mm}

Let us now consider the spatial components of the Fock--Ivanenko
coefficients $\Gamma_I$.  We have:
\begin{equation}
\Gamma_I = - \frac{1}{4} \: \omega_{AB I} \gamma^A \gamma^B
= - \frac{1}{4} \left[ 2 \: \omega_{TJI} \gamma^T \gamma^J
+ \omega_{JKI} \gamma^J \gamma^K \right] \; .
\end{equation}
Substituting the values of $\omega_{TJI}$ and $\omega_{JKI}$ from
equations~\eqref{eq:omega_TIJ} and $\eqref{eq:omega_IJK}$ we find:
\begin{align}
\Gamma_I &= - \frac{1}{2} \: K_{IJ} \gamma^T \gamma^J
- \frac{1}{4} \left( E^n_J E^m_I D_m E_{nK} \right) \gamma^J \gamma^K
\nonumber \\
&= - \frac{1}{2} \: K_{IJ} \gamma^T \gamma^J
- \frac{1}{4} \: \omega_{JKI}^{(3)} \gamma^J \gamma^K
= - \frac{1}{2} \: K_{IJ} \gamma^T \gamma^J
+ \Gamma_I^{(3)} \; ,
\label{eq:Gamma_I}
\end{align}
where in the last step we defined the purely three-dimensional
Ricci rotation and Fock--Ivanenko coefficients as:
\begin{align}
\omega_{JKI}^{(3)} :=  E^n_J E^m_I D_m E_{nK} \; ,
\label{eq:omega3D}
\\
\Gamma^{(3)}_I := - \frac{1}{4} \: \omega_{JKI}^{(3)} \gamma^J \gamma^K \; .
\label{eq:FockIvanenko3D}
\end{align}

 Let us now calculate the contraction $\gamma^I \Gamma_I$:
\begin{equation}
\gamma^I \Gamma_I = - \frac{1}{2} \: K_{IJ} \gamma^I \gamma^T \gamma^J
+ \gamma^I \Gamma^{(3)}_I \; .
\end{equation}
For the first term above we find:
\begin{equation}
- \frac{1}{2} \: K_{IJ} \gamma^I \gamma^T \gamma^J
= \frac{1}{2} \left( K_{IJ} \gamma^I \gamma^J \right) \gamma^T
= \frac{1}{2} \left( \sum_I K_{II} \left( \gamma^I \right)^2
+ \sum_{I \neq J} K_{IJ} \gamma^I \gamma^J \right) \gamma^T \; .
\end{equation}
Now, since $K_{IJ}$ is symmetric, and from Clifford's algebra
we know that $\gamma^I \gamma^J = - \gamma^J \gamma^I$ for $I \neq J$,
the second term in the last equation cancels. On the other hand,
we also have $(\gamma^I)^2 = - 1$, so that we finally find:
\begin{equation}
K_{IJ} \gamma^I \gamma^J = - \sum_I K_{II} = - K \; ,
\label{eq:K-ll}
\end{equation}
with $K$ the trace of $K_{mn}$. We can then rewrite the contraction
$\gamma^I \Gamma_I$ as:
\begin{equation}
\gamma^I \Gamma_I 
= - \left( \frac{K}{2} \right) \gamma^T + \gamma^I \Gamma^{(3)}_I \; .
\label{eq:gammaI-GammaI}
\end{equation}

To finish, we can add both contributions to $\gamma^A \Gamma_A$
to obtain:
\begin{equation}
\gamma^A \Gamma_A = \gamma^\mu \Gamma_ \mu
= \left[ \left( \frac{\partial_I \alpha}{2 \alpha} \right) \gamma^I
- \frac{1}{4} \: Q_{IJ} \: \gamma^T \gamma^I \gamma^J
- \left( \frac{K}{2} \right) \gamma^T
+ \gamma^I \Gamma^{(3)}_I \right] \; .
\label{eq:gammaGamma-3+1}
\end{equation}

\vspace{5mm}

Notice that we can also write the spacetime components of the
Fock--Ivanenko coefficients as $\Gamma_\mu = e_\mu^A \Gamma_A$, from
which we find:
\begin{align}
\Gamma_t &= e^A_t \Gamma_A
= \alpha \Gamma_T + \beta^I \Gamma_I
= \left( \frac{\partial_I \alpha}{2} \right) \gamma^T \gamma^I
- \frac{\alpha}{4} \: Q_{IJ} \gamma^I \gamma^J 
- \beta^I \left( \frac{K_{IJ}}{2} \:  \gamma^T \gamma^J - \Gamma^{(3)}_I \right) \; ,
\label{eq:Gamma-t}
\\
\Gamma_m &= e^A_m \Gamma_A = E_m^I \Gamma_I
= - \frac{K_{mJ}}{2} \: \gamma^T \gamma^J + \Gamma^{(3)}_m \; .
\label{eq:Gamma-m}
\end{align}
From the above expression one can now easily verify find that:
\begin{equation}
\Gamma_t - \beta^m \Gamma_m = \left( \frac{\partial_I \alpha}{2} \right) \gamma^T \gamma^I
- \frac{\alpha}{4} \: Q_{IJ} \gamma^I \gamma^J
= \alpha \Gamma_T \; .
\label{eq:GammaT-spacetime}
\end{equation}


\subsection{Dirac equation in 3+1 form}
\label{sec:Dirac-3+1}

We are now ready to write the Dirac equation in 3+1 language.
Remember that the Dirac equation in general relativity takes the
form~\eqref{eq:Dirac-curved}:
\begin{equation}
i \gamma^\mu \mathcal{D}_\mu \psi - m \psi = 0 \; ,
\end{equation}
or equivalently:
\begin{equation}
\left( \gamma^\mu \partial_\mu + \gamma^\mu \Gamma_\mu
+ i m \right) \psi = 0 \; ,
\end{equation}
The last equation can also be written as:
\begin{equation}
\left( \gamma^t \partial_t + \gamma^m \partial_m \right) \psi
= - \left( \gamma^\mu \Gamma_\mu + i m \right) \psi \; .
\end{equation}
Notice now that (remember that $I$ takes values from 1 to 3):
\begin{align}
\gamma^t &= e^t_A \gamma^A = e^t_T \gamma^T + \cancelto{0}{e^t_I \gamma^I}
= \left( \frac{1}{\alpha} \right) \gamma^T \; , \\
\gamma^m &= e^m_A \gamma^A = e^m_T \gamma^T + e^m_I \gamma^I
= - \left( \frac{\beta^m}{\alpha} \right) \gamma^T + E^m_I \gamma^I \; .
\end{align}
Substituting these results into the Dirac equation we find:
\begin{equation}
\gamma^T \left( \partial_t - \beta^m \partial_m \right) \psi
= - \alpha \left( \lambda^m \partial_m
+ \gamma^\mu \Gamma_\mu + i m \right) \psi \; ,
\label{eq:Dirac_3+1_0}
\end{equation}
where we have defined the purely spatial Dirac matrices as $\lambda^m
:= E^m_I \gamma^I$. Notice that, so defined, $\lambda^m$ is different
from $\gamma^m = e^m_A \gamma^A$.  In fact we have:
\begin{equation}
\gamma^m = e^m_A \gamma^A = e^m_T \gamma^T + e^m_I \gamma^I
= - \left( \frac{\beta^m}{\alpha} \right) \gamma^T + E^m_I \gamma^I
= - \left( \frac{\beta^m}{\alpha} \right) \gamma^T + \lambda^m \; .
\end{equation}
The $\lambda^m$ can also be defined in a completely equivalent way by
simply projecting the $\gamma^\mu$ onto the spatial hypersurfaces:
\begin{equation}
\lambda^\mu := P^\mu_\nu \gamma^\nu \; ,
\label{eq:SpatialDiracMatrices}
\end{equation}
with $P^\mu_\nu$ the projection operator defined above
in~\eqref{eq:projection3D}.  From this definition we find immediately
$\lambda^t=0$, as expected.  Notice also that even if $\lambda^m \neq
\gamma^m$, if we now lower the indices of $\lambda^m$ using the
spatial metric, that is if we define $\lambda_m := \gamma_{mn}
\lambda^n$, then we do find that $\lambda_m = \gamma_m$.

The purely spatial Dirac matrices satisfy the three-dimensional
Clifford algebra:
\begin{equation}
\lambda^m \lambda^n + \lambda^n \lambda^m = - 2 \gamma^{mn} I_3 \; ,
\end{equation}
with $\gamma^{mn}$ the inverse spatial metric and $I_3$ the $3 \times
3$ identity matrix.  There is one important comment to make here with
respect to the spatial Dirac matrices $\lambda^m$.  When we project
these matrices back onto the triad we find $\lambda^I = E^I_m
\lambda^m = E^I_m E^m_J \gamma^J = \delta^I_J \gamma^J = \gamma^I$.
So that we have $\lambda^I = \gamma^I$, but crucially $\lambda^m \neq
\gamma^m$ whenever $\beta^m \neq 0$.  The reason for this is that we
have $\lambda^m := E^m_I \gamma^I$, while $\gamma^m := e^m_A
\gamma^A$.  In particular, for any three-dimensional tensor $T_m$ we
will have:
\begin{equation}
\gamma^m T_m = \left( \lambda^m - \frac{\beta^m}{\alpha} \right) T_m
\neq \lambda^m T_m = \lambda^I T_I = \gamma^I T_I \; ,
\end{equation}
so that in general we have $\gamma^m T_m \neq \gamma^I T_I$, while
$\lambda^m T_m = \lambda^I T_I$ .  Because of this, in order to
avoid possible confusions, it is best to always try to use the
$\lambda$'s instead of the $\gamma$'s when considering purely spatial
contractions of indices, be them coordinate or triad indices.

Multiplying now equation~\eqref{eq:Dirac_3+1_0} with $\gamma^T$ from
the left, and using the fact that \mbox{$(\gamma^T)^2 = 1$}, we find:
\begin{equation}
\left( \partial_t - \beta^m \partial_m \right) \psi =
- \alpha \gamma^T \left( \lambda^m \partial_m
+ \gamma^\mu \Gamma_\mu + i m \right) \psi \; .
\end{equation}
We can now use our result for $\gamma^\mu
\Gamma_\mu$, equation~\eqref{eq:gammaGamma-3+1}, 
to obtain:
\begin{equation}
\left( \partial_t - \beta^m \partial_m \right)  \psi =
- \alpha \gamma^T \left\{ \lambda^m \partial_m
+ \left[ \left( \frac{\partial_I \alpha}{2 \alpha} \right) \gamma^I
- \frac{1}{4} \: Q_{IJ} \: \gamma^T \gamma^I \gamma^J
- \left( \frac{K}{2} \right) \gamma^T + \gamma^I \Gamma^{(3)}_I \right]
+ i m \right\} \psi \; .
\end{equation}
The above equation can be further simplified by noticing first that:
\begin{equation}
\gamma^I \partial_I = \gamma^I \left( e_I^\mu \partial_\mu \right)
= \gamma^I E_I^m \partial_m = \lambda^m \partial_m \; .
\end{equation}
Here one must remember again that in general $\lambda^m \neq
\gamma^m$, so that $\gamma^I \partial_I = \lambda^m \partial_m \neq
\gamma^m \partial_m$.  Similarly we find $\gamma^I \Gamma^{(3)}_I =
\lambda^m \Gamma^{(3)}_m $. The Dirac equation then takes the form:
\begin{equation}
\left( \partial_t - \beta^m \partial_m \right) \psi =
\alpha \left[ - \gamma^T \lambda^m \left( \partial_m
+ {}^{(3)} \Gamma_m + \frac{\partial_m \alpha}{2 \alpha} \right)
+ \left( \frac{K}{2} - i m \gamma^T \right)
+ \frac{1}{4} \: Q_{IJ} \lambda^I \lambda^J \right] \psi \: .
\end{equation}
Finally, if we define the three-dimensional spinorial
covariant derivative as:
\begin{equation}
D_m \psi :=  \partial_m \psi + \Gamma^{(3)}_m \psi \; ,
\end{equation}
the Dirac equation then becomes:
\begin{equation}
\left( \partial_t - \beta^m \partial_m \right) \psi
= - \alpha \gamma^T \left[ \lambda^m \left( D_m
+ \frac{\partial_m \alpha}{2 \alpha} \right)
+ i m \right] \psi + \alpha \left( \frac{K}{2}
+ \frac{1}{4} \: Q_{mn} \lambda^m \lambda^n \right) \psi \; .
\end{equation}
The above equation can be written in a somewhat more compact form as:
\begin{equation}
\left( \partial_t - \beta^m \partial_m \right) \psi
= - \alpha \gamma^T \left( \lambda^m D_m \psi + i m \psi \right)  + \alpha \left( \frac{K}{2}
- \Gamma_T \right) \psi \; ,
\label{eq:Dirac_3+1}
\end{equation}
where $\Gamma_T$ is given by (confront equation~\eqref{eq:Gamma_T}):
\begin{equation}
\Gamma_T = \gamma^T \lambda^m \left( \frac{\partial_m \alpha}{2 \alpha} \right)
- \frac{1}{4} \: Q_{mn} \lambda^m \lambda^n \; .
\label{eq:Gamma_T-2}
\end{equation}
Equation~\eqref{eq:Dirac_3+1} is the final form of the Dirac equation
in the 3+1 formalism.

The expression for $\Gamma_T$ given above is in principle valid for
any arbitrary choice of the triad evolution gauge represented by the
matrix $Q_{mn}$.  In the particular case when we choose a triad that
evolves via Fermi--Walker transport we have $Q_{mn}=0$, so the
$\Gamma_T$ reduces simply to:
\begin{equation}
\Gamma_T = \gamma^T \lambda^m \left( \frac{\partial_m \alpha}{2
  \alpha} \right) \; .
\end{equation}

Notice also that in the Dirac equation there is an explicit dependence
on the Dirac matrix $\gamma^T$.  Since $\gamma^T$ is associated to the
time direction in a local inertial frame, from now on we will simple
take $\gamma^T = \gamma^0$, with $\gamma^0$ the usual Dirac matrix
from Minkowski spacetime, so that:
\begin{equation}
\gamma^T = \left(
\begin{array}{cc}
I_2 & 0 \\
0 & -I_2 
\end{array}
\right) \; .
\end{equation}
However, when we resurrect the spacetime index we will now have:
\begin{equation}
\gamma^t = e^t_A \gamma^A
= n^t_T \gamma_T + \cancelto{0}{E^t_I} \gamma_I = \frac{\gamma^T}{\alpha} \; .
\end{equation}
We then see that the matrix $\gamma^t$ is not just the $\gamma^0$ used
in Minkowski spacetime, so we must be somewhat careful with the
notation.  Therefore, from here on we will always write $\gamma^T$
instead of $\gamma^0$ when we refer to the time component of the Dirac
matrices in a local inertial frame.  In particular, the above result
implies that when we project the gamma matrices onto the normal
direction to the spatial hypersurfaces we find:
\begin{equation}
n_\mu \gamma^\mu = - \alpha \gamma^t = - \gamma^T \; .
\end{equation}

From the expression for $\Gamma_T$ above it is not difficult to show
that $\gamma^T {\Gamma_T}^\dag \gamma^T = - \Gamma_T$.  Using this
result, a little algebra allows us to find the adjunct
Dirac equation in 3+1 form:
\begin{equation}
\left( \partial_t - \beta^m \partial_m \right) \bar{\psi}
= - \alpha \left( (D_m \bar{\psi}) \lambda^m - i m \bar{\psi} \right) \gamma^T
+ \alpha \bar{\psi} \left( \frac{K}{2} + \Gamma_T \right) \; .
\label{eq:Dirac_3+1-adjunct}
\end{equation}


\subsection{Conserved current and stress--energy tensor in 3+1 form}

Let us now consider the conserved current in 3+1 form.  To do this we
must first find the adjoint spinor $\bar{\psi} = \psi^\dag \gamma^T$:
\begin{equation}
\bar{\psi} = \left( \psi_1^* , \psi_2^* , - \psi_3^* , - \psi_4^* \right) \; .
\end{equation}

As we have shown before, the conserved current will now be given by
$j^\mu = \bar{\psi} \gamma^\mu \psi$.  We now define the particle
density measured by the Eulerian observers as $\rho_p := - n_\mu
j^\mu$.  We the find:
\begin{equation}
\rho_p := - n_\mu j^\mu = \alpha j^t = \alpha \bar{\psi} \gamma^t \psi
= \bar{\psi} \gamma^T \psi
= | \psi_1 |^2 + | \psi_2 |^2 + | \psi_3 |^2 + | \psi_4 |^2 \; ,
\label{eq:ParticleDensity-3+1}
\end{equation}
which is, of course, what we would have expected.

On the other hand, the particle flux measured by the Eulerian
observers is defined as $f^i := P^i_\nu j^\nu$, with $P^\mu_\nu$ the
projection operator onto the spatial hypersurfaces.  We now find:
\begin{align}
f^i = P^i_\nu j^\nu = P^i_\nu \left( \bar{\psi} \gamma^\nu \psi \right)
= \bar{\psi} \left( P^i_\nu \lambda^\nu \right) \psi
= \bar{\psi} \lambda^i \psi \; ,
\label{eq:ParticleCurrent-3+1}
\end{align}
where we used the definition of the purely spatial Dirac
matrices~\eqref{eq:SpatialDiracMatrices}.

\vspace{5mm}

Let us now consider the stress--energy tensor expressed in 3+1 terms.
For this, it turns out to be convenient to define $\Pi := n^\mu
\mathcal{D}_\mu \psi$.  A little algebra then allows us to show that:
\begin{equation}
\Pi = \frac{1}{\alpha} \left( \partial_t \psi - \beta^i \partial_i \psi
\right) + \Gamma_T \psi \; .
\label{eq:Pi}
\end{equation}
From this definition one can see that $\Pi$ represents the geometric
change of the spinor along the normal direction to the spatial
hypersurfaces.  In an analogous way we also define $\bar{\Pi} := n^\mu
\mathcal{D}_\mu \bar{\psi}$, so that:
\begin{equation}
\bar{\Pi} = \frac{1}{\alpha} \left( \partial_t \bar{\psi} - \beta^i \partial_i \bar{\psi}
\right) - \bar{\psi} \Gamma_T \; .
\end{equation}

The energy density measured by the Eulerian observers is now defined
as $\rho_E := n^\mu n^\nu T_{\mu \nu}$. Using the expression for the
stress--energy tensor of the Dirac field,
equation~\eqref{eq:Dirac-stressenergy}, we find:
\begin{equation}
\rho_E = \frac{i}{2} \: n^\mu n^\mu \left[
\left( \mathcal{D}_{(\mu} \bar{\psi} \right) \gamma_{\nu)} \psi
- \bar{\psi} \gamma_{(\mu} \left( \mathcal{D}_{\nu)} \psi \right) \right]
= \frac{i}{2} \left[ \bar{\psi} \gamma^T \Pi - \bar{\Pi} \gamma^T \psi \right] \; ,
\end{equation}
where we used the fact that $n^\mu \gamma_\mu = - \gamma^T$. We can
simplify this further by noticing first that $\bar{\psi} \gamma^T =
\psi^\dag$. Also, from the transpose of the definition of $\Pi$ we
clearly have:
\begin{equation}
\Pi^\dag = \frac{1}{\alpha} \left( \partial_t \psi^\dag - \beta^i \partial_i \psi^\dag
\right) + \psi^\dag {\Gamma_T}^\dag \; .
\end{equation}
Multiplying this from the right with $\gamma^T$ we find:
\begin{equation}
\Pi^\dag \gamma^T = \frac{1}{\alpha} \left( \partial_t \bar{\psi} -
\beta^i \partial_i \bar{\psi} \right) + \psi^\dag {\Gamma_T}^\dag \gamma^T
= \frac{1}{\alpha} \left( \partial_t \bar{\psi} -
\beta^i \partial_i \bar{\psi} \right) - \bar{\psi} \Gamma_T \; ,
\end{equation}
which implies that $\Pi^\dag \gamma^T = \bar{\Pi}$, or
equivalently $\bar{\Pi} \gamma^T = \Pi^\dag$.  The energy density
then reduces to:
\begin{equation}
\rho_E = \frac{i}{2} \left( \psi^\dag \Pi - \Pi^\dag \psi \right) \; .
\label{eq:EnergyDensity-3+1}
\end{equation}
It is clear from this expression that the energy density is not
positive definite, as already mentioned.  In terms of the components
of the spinor we will have:
\begin{equation}
\rho_E = \frac{i}{2} \left[ \left( \psi_1^* \Pi_1 + \psi_2^* \Pi_2
+ \psi_3^* \Pi_3 + \psi_4^* \Pi_4 \right) - c.c. \right] \; ,
\end{equation}
where $c.c.$ denotes the complex conjugate of the previous expression.

There is an interesting observation to be made with respect to our
final expression for the energy density,
equation~\eqref{eq:EnergyDensity-3+1}.  If we now define:
\begin{equation}
\tilde{\Pi} := n^\mu \partial_\mu \psi = \frac{1}{\alpha} \left( \partial_t \psi
- \beta^i \partial_i \psi \right) \; ,
\end{equation}
we will clearly have $\Pi = \tilde{\Pi} + \Gamma_T \psi$.  In the same
way, if we define $\tilde{\Pi}^\dag := n^\mu \partial_\mu \psi^\dag$ we find
$\Pi^\dag = \tilde{\Pi}^\dag + \psi^\dag {\Gamma_T}^\dag$. The energy density
then becomes:
\begin{equation}
\rho_E = \frac{i}{2} \left[ \left( \tilde{\psi}^\dag \Pi - \tilde{\Pi}^\dag \psi \right)
+ \psi^\dag \left( \Gamma_T - {\Gamma_T}^\dag \right) \psi \right] \; .
\end{equation}
Using now the expression for $\Gamma_T$ given by equation~\eqref{eq:Gamma_T-2}
it is not difficult to show that:
\begin{equation}
\Gamma_T - {\Gamma_T}^\dag = - \frac{1}{2} \: Q_{mn} \lambda^m \lambda^n \: .
\end{equation}
The energy density then becomes:
\begin{equation}
\rho_E = \frac{i}{2} \left[ \left( \psi^\dag \tilde{\Pi} - \tilde{\Pi}^\dag \psi \right)
- \frac{1}{2} \: \psi^\dag \left( Q_{mn} \lambda^m \lambda^n \right) \psi \right] \; .
\end{equation}
The interesting fact about this last expression is that all
dependencies coming from the gradient of the lapse that appear in
$\Gamma_T$ have cancelled, and we are only left with a dependency
on the $Q_{mn}$.  For a triad that evolves via Fermi--Walker transport
we have $Q_{mn}=0$, and the energy density reduces simply to:
\begin{equation}
\rho_E = \frac{i}{2} \left( \psi^\dag \tilde{\Pi} - \tilde{\Pi}^\dag \psi \right) \; .
\label{eq:EnergyDensity-3+1-FW}
\end{equation}

\vspace{5mm}

Let us now consider the momentum density measured by the Eulerian
observers, which is defined as \mbox{$J_i := - n^\mu P_i^\nu T_{\mu
    \nu}$}. We will then have:
\begin{align}
J_i &= - \frac{i}{2} \: n^\mu P_i^\nu \left[
\left( \mathcal{D}_{(\mu} \bar{\psi} \right) \gamma_{\nu)} \psi
- \bar{\psi} \gamma_{(\mu} \left( \mathcal{D}_{\nu)} \psi \right) \right]
\nonumber \\
&= - \frac{i}{4} \: n^\mu P_i^\nu \left[
\left\{ \left( \mathcal{D}_{\mu} \bar{\psi} \right) \gamma_{\nu}
+ \left( \mathcal{D}_{\nu} \bar{\psi} \right) \gamma_{\mu} \right] \psi
- \bar{\psi} \left[ \gamma_{\mu} \left( \mathcal{D}_{\nu} \psi \right)
+ \gamma_{\nu} \left( \mathcal{D}_{\mu} \psi \right) \right\} \right]
\nonumber \\
&= - \frac{i}{4} \left[ \bar{\Pi} \lambda_i \psi - \bar{\psi} \lambda_i \Pi
+ \bar{\psi} \gamma^T \left( P_i^\nu  \mathcal{D}_{\nu} \psi \right)
- \left( P_i^\nu  \mathcal{D}_{\nu} \bar{\psi} \right) \gamma^T \psi
 \right] \; ,
\end{align}
where we used the fact that $n^\mu \gamma_\mu = - \gamma^T$ and
$P^\mu_i \gamma_\mu = \lambda_i$. We now need to calculate the
projections of the spinor derivatives $P_i^\nu \mathcal{D}_{\nu} \psi$
and $P_i^\nu \mathcal{D}_{\nu} \bar{\psi}$. In order to do this notice
first that, since the covariant components of the normal vector
vanish, we will have $P_i^\mu = \delta_i^\mu$
(confront~\eqref{eq:projection3D}).  This implies:
\begin{equation}
P_i^\nu \mathcal{D}_{\nu} \psi = \mathcal{D}_i \psi
= \partial_i \psi + \Gamma_i \psi
= \partial_i \psi + \left( \Gamma^{(3)}_i
- \frac{K_{im}}{2} \: \gamma^T \lambda^m \right) \psi
= D_i \psi - \frac{K_{im}}{2} \: \gamma^T \lambda^m \psi \; ,
\end{equation}
with $D_i$ the three-dimensional covariant derivative we define above,
and where we used equation~\eqref{eq:Gamma-m}. Similarly:
\begin{equation}
P_i^\nu \mathcal{D}_{\nu} \bar{\psi} = \mathcal{D}_i \bar{\psi}
= D_i \bar{\psi} + \frac{K_{im}}{2} \: \bar{\psi} \gamma^T \lambda^m \; .
\end{equation}
We then find:
\begin{align}
\bar{\psi} \gamma^T \left( P_i^\nu  \mathcal{D}_{\nu} \psi \right)
- \left( P_i^\nu  \mathcal{D}_{\nu} \bar{\psi} \right) \gamma^T \psi
&= \bar{\psi} \gamma^T \left( D_i \psi \right) - \left( D_i \bar{\psi} \right) \gamma^T \psi
- \frac{K_{im}}{2} \: \bar{\psi} \left( \lambda^m
+ \gamma^T \lambda^m \gamma^T \right) \psi
\nonumber \\
&= \psi^\dag \left( D_i \psi \right) - \left( D_i \psi^\dag \right) \psi \; ,
\end{align}
where we used the fact that $(\gamma^T)^2=1$ and $\gamma^T \lambda^m
\gamma^T = E^m_I \gamma^T \gamma^I \gamma^T = - E^m_I \gamma^I
(\gamma^T)^2 = - \lambda^m$. The momentum density then takes the final form:
\begin{equation}
J_i = - \frac{i}{4} \left[ \bar{\Pi} \lambda_i \psi - \bar{\psi} \lambda_i \Pi
+ \psi^\dag \left( D_i \psi \right) - \left( D_i \psi^\dag \right) \psi
\right] \; .
\label{eq:MomentumDensity-3+1}
\end{equation}

\vspace{5mm}

Finally, the spatial stress tensor is defined as defined as $S_{ij} :=
P_i^\mu P_j^\nu T_{\mu \nu} = \delta_i^\mu \delta_j^\nu T_{\mu \nu} =
T_{ij}$, where we used the fact that $P^\mu_i = \delta^\mu_i$ since
the covariant spatial components of the normal vector vanish, $n_i=0$
(confront~\eqref{eq:projection3D}). We now find:
\begin{align}
S_{ij} &= \frac{i}{2}
\left[ \left( \mathcal{D}_{(i} \bar{\psi} \right) \gamma_{j)} \psi
- \bar{\psi} \gamma_{(i} \left( \mathcal{D}_{j)} \psi \right) \right]
\nonumber \\
&= \frac{i}{2}
\left[ \left( D_{(i} \bar{\psi} \right) \lambda_{j)} \psi
- \bar{\psi} \lambda_{(i} \left( D_{j)} \psi \right) 
+ \frac{1}{2} \: \bar{\psi} \left( K_{m(i} \gamma^T \lambda^m \lambda_{j)}
+ \lambda_{(i} K_{j)m} \gamma^T \lambda^m \right) \psi \right]
\nonumber \\
&= \frac{i}{2}
\left[ \left( D_{(i} \bar{\psi} \right) \lambda_{j)} \psi
- \bar{\psi} \lambda_{(i} \left( D_{j)} \psi \right)
+ \frac{1}{2} \: \bar{\psi} K_{m(i} \left( \gamma^T \lambda^m \lambda_{j)}
+ \lambda_{j)} \gamma^T \lambda^m \right) \psi \right]
\nonumber \\
&= \frac{i}{2}
\left[ \left( D_{(i} \bar{\psi} \right) \lambda_{j)} \psi
- \bar{\psi} \lambda_{(i} \left( D_{j)} \psi \right)
+ \frac{1}{2} \left( \bar{\psi} \gamma^T \right) K_{m(i} \left( \lambda^m \lambda_{j)}
- \lambda_{j)} \lambda^m \right) \psi \right] \; .
\end{align}
The last expression can be further simplified noticing first that
$\bar{\psi} \gamma^T = \psi^\dag$, and:
\begin{align}
K_{mi} \left( \lambda^m \lambda_j - \lambda_j \lambda^m \right)
&= K_{mi} \gamma_{jn} \left( \lambda^m \lambda^n - \lambda^n \lambda^m \right)
= - 2 K_{mi} \gamma_{jn} \left( \lambda^n \lambda^m + \gamma^{nm} \right)
\nonumber \\
&= - 2 \left( K_{ji} + K_{mi} \lambda_j \lambda^m \right) \; ,
\end{align}
where we used the spatial Clifford algebra. We finally find for the
spatial stress tensor:
\begin{equation}
S_{ij} = \frac{i}{2} \left[ \left( D_{(i} \bar{\psi} \right) \lambda_{j)} \psi
- \bar{\psi} \lambda_{(i} \left( D_{j)} \psi \right)
- \psi^\dag \left( K_{ij} + K_{m(i} \lambda_{j)} \lambda^m \right) \psi \right] \; .
\label{eq:StressTensor-3+1}
\end{equation}

It is interesting at this point to calculate the trace of this spatial
stress tensor.  We have:
\begin{equation}
S \equiv {S^m}_m = \frac{i}{2} \left[ \left( D_m \bar{\psi} \right) \lambda^m \psi
- \bar{\psi} \lambda^m \left( D_m \psi \right)
- \psi^\dag \left( K + K_{mn} \lambda^m \lambda^n \right) \psi \right] \; .
\end{equation}
But, from equation~\eqref{eq:K-ll}, we know that $K_{mn} \lambda^m \lambda^n = - K$,
so that $S$ reduces to:
\begin{equation}
S = \frac{i}{2} \left[ \left( D_m \bar{\psi} \right) \lambda^m \psi
- \bar{\psi} \lambda^m \left( D_m \psi \right) \right] \; .
\end{equation}
The previous expression can be further simplified by using the Dirac
equation~\eqref{eq:Dirac_3+1} and its
adjunct~\eqref{eq:Dirac_3+1-adjunct}, which can be written in terms
of $\Pi$ and $\bar{\Pi}$ as:
\begin{align}
\Pi &= - \gamma^T \left( \lambda^m (D_m \psi) + i m \psi \right) + \frac{K}{2} \: \psi \; , \\
\bar{\Pi} &= - \left( (D_m \bar{\psi}) \lambda^m - i m \bar{\psi} \right) \gamma^T
+ \frac{K}{2} \: \bar{\psi} \; .
\end{align}
Solving for $\lambda^m (D_m \psi)$ and $(D_m \bar{\psi}) \lambda^m$
from the above equations, and substituting into $S$ we find, after
some algebra:
\begin{equation} 
S = \frac{i}{2} \left[ \psi^\dag \Pi - \Pi^\dag \psi \right] - m \bar{\psi} \psi \; .
\end{equation}
By comparing this with the expression for the energy density above
it is easy to see that:
\begin{equation}
S = \rho_E - m \bar{\psi} \psi \; .
\end{equation}
But this result is to be expected since the trace of the full
stress--energy tensor can be written in 3+1 terms as ${T^\mu}_\mu = S
- \rho_E$, so that we have ${T^\mu}_\mu = - m \bar{\psi} \psi$, which
is precisely the result we found before in
equation~\eqref{eq:Dirac-stressenergy-trace}.


\section{Example: Spherically symmetric spacetime}

As an example of the Dirac equation in a curved spacetime, we will
consider the particular case of a spacetime with spherical symmetry.
We will first only consider the Dirac field as a test field in a
background spherically-symmetric spacetime, and only later we will
consider the self-gravitating case.

We start by writing the metric of a spherically-symmetric spacetime in
spherical coordinates $(t,r,\theta,\varphi)$ as:
\begin{equation}
ds^2 = (- \alpha^2 + \beta_r \beta^r ) dt^2 + 2 \beta_r dr dt 
+ a^2 dr^2 + b^2 r^2 d \Omega^2 \; ,
\label{eq:3+1metric}
\end{equation}
where $d \Omega^2 = d\theta^2 + \sin^2 (\theta) \: d\varphi^2$ is
the standard solid angle element, $\alpha=\alpha(r,t)$ is the lapse
function, $\beta^r=\beta^r(r,t)$ the shift vector which in this case
only has a non-zero radial component, and where $a=a(r,t)$ and
$b=b(r,t)$ are the spatial metric components.  In particular we have
$\beta_r := \gamma_{rr} \beta^r = a^2 \beta^r$.  Notice that we recover the
Minkowski metric in spherical coordinates if we take $\alpha=a=b=1$,
$\beta^r=0$.

\subsection{Dirac equation}

The first step in order to write the Dirac equation is to choose our
tetrad.  As we have already mentioned when discussing the Dirac
equation in the 3+1 formalism, for the timelike vector we take the
unit normal vector to the spatial hypersurfaces:
\begin{equation}
e^\mu_T = n^\mu = \left( 1/\alpha, - \beta^r / \alpha ,0,0 \right) \; .
\label{eq:tetrad-CurvSphere-0}
\end{equation}
Notice that for $\beta^r \neq 0$ this vector has a non-trivial radial
component.  For the spatial vectors we take as a natural choice the
unit vectors along the radial and angular directions, which now take
the form:
\begin{equation}
e^\mu_{R} = \left( 0,\frac{1}{a},0,0 \right) \; , \quad
e^\mu_{\Theta} = \left( 0,0,\frac{1}{rb},0 \right) \; , \quad
e^\mu_{\Phi} = \left( 0,0,0,\frac{1}{rb \sin \theta} \right) \; .
\label{eq:tetrad-CurvSphere-1}
\end{equation}
Is is clear that these three vectors are already orthogonal to each
other.

In order to avoid confusions, from here on we will always denote the
spacetime indices by $(t,r,\theta,\varphi)$, and their associated
Lorentz indices by $(T,R,\Theta,\Phi)$.  Using now the
metric~\eqref{eq:3+1metric} one can now show that the associated
1-forms (the co-tetrad) are:
\begin{equation}
e_{\mu T} = \left( -\alpha,0,0,0 \right) , \:
e_{\mu R} = \left( a \beta^r,a,0,0 \right) , \:
e_{\mu \Theta} = \left( 0,0,rb,0 \right) , \:
e_{\mu \Phi} = \left( 0,0,0,rb \sin \theta \right) .
\label{eq:tetrad-CurvSphere-2}
\end{equation}
Notice again that for $\beta^r \neq 0$ the radial 1-form has a
non-zero time component.  When thinking only of the spatial triad we
will have:
\begin{equation}
E^i_{R} = \left(\frac{1}{a},0,0 \right) \; , \quad
E^i_{\Theta} = \left(0,\frac{1}{rb},0 \right) \; , \quad
E^i_{\Phi} = \left( 0,0,\frac{1}{rb \sin \theta} \right) \; ,
\label{eq:triad-CurvSphere-1}
\end{equation}
and:
\begin{equation}
E_{i R} = \left(a,0,0 \right) \; , \quad
E_{i \Theta} = \left(0,rb,0 \right) \; , \quad
E_{i \Phi} = \left( 0,0,rb \sin \theta \right) \; ,
\label{eq:triad-CurvSphere-2}
\end{equation}

Next, we need to construct the Dirac matrices. Since spherical
coordinates are already orthogonal, the natural choice is to associate
the $\gamma$ matrices directly to the coordinate directions.  We will
start by defining the $\gamma^A$ matrices with Lorentz indices since
they correspond to a local inertial frame, and can therefore be
constructed directly from the usual Dirac matrices in Minkowski
spacetime.  As already mentioned, we will take $\gamma^T=\gamma^0$
along the time direction, but we now have to ask ourselves in what
order should be associate the $\gamma^i$ to the spatial coordinates
$(r,\theta,\phi)$. An obvious choice (used frequently) is to associate
$\gamma^1$ to $r$, $\gamma^2$ to $\theta$ and $\gamma^3$ to $\varphi$.
However, it turns out to be more convenient when separating variables
(see below) to make a different choice and associate instead
$\gamma^3$ to the radial coordinate $r$, and $\gamma^2$ and $\gamma^1$
to the angular coordinates $\theta$ and $\varphi$ respectively.  This
is the choice we will follow here. Notice that the different choices
only correspond to changing the order of the coordinates and should be
completely equivalent physically.  We will then take:
\begin{equation}
\gamma^T = \gamma^0 \; , \qquad
\gamma^R = \gamma^3 \; , \qquad
\gamma^\Theta = \gamma^2 \; , \qquad
\gamma^\Phi = \gamma^1 \; .
\end{equation}
The spacetime components of the $\gamma$ matrices are then defined as
$\gamma^\mu = e^\mu_A \gamma^A$. Using our choice of tetrad above we now
find:
\begin{equation}
\gamma^t = \frac{\gamma^T}{\alpha} \; , \qquad
\gamma^r = \frac{\gamma^R}{a} - \frac{\beta^r \gamma^T}{\alpha} \; , \qquad
\gamma^\theta = \frac{\gamma^\Theta}{rb} \; , \qquad
\gamma^\varphi = \frac{\gamma^\Phi}{rb \sin \theta} \; .
\end{equation}
Notice that when $\beta^r \neq 0$ the matrix $\gamma^r$ now has a
contribution from $\gamma^T$.  It is not difficult to verify that the
above matrices satisfy the Clifford algebra:
\begin{equation}
\gamma^\mu \gamma^\nu + \gamma^\nu \gamma^\mu = - 2 g^{\mu \nu} \; .
\end{equation}
On the other hand, for the purely spatial Dirac matrices $\lambda$ we
have:
\begin{equation}
\lambda^R = \gamma^3 \; , \qquad
\lambda^\Theta = \gamma^2 \; , \qquad
\lambda^\Phi = \gamma^1 \; ,
\end{equation}
and:
\begin{equation}
\lambda^r = \frac{\gamma^R}{a} \; , \qquad
\lambda^\theta = \frac{\gamma^\Theta}{rb} \; , \qquad
\lambda^\varphi = \frac{\gamma^\Phi}{rb \sin \theta} \; .
\end{equation}
Notice in particular that when $\beta^r$ is not zero we clearly have
$\lambda^r \neq \gamma^r$.
 
The $\lambda^m$ now clearly satisfy the spatial Clifford algebra:
\begin{equation}
\lambda^m \lambda^n + \lambda^n \lambda^m = - 2 \gamma^{mn} I_3 \; ,
\end{equation}
with $\gamma^{mn}$.

We now need to calculate the three-dimensional Ricci rotation
coefficients, since they are necessary in order to obtain the
Fock--Ivanenko coefficients. In order to do this we first need to find
the 3D Christoffel symbols $^{(3)}\Gamma^i_{jk}$ and
use~\eqref{eq:omega3D}. The calculation is not particularly
illuminating so we will not write it in detail here. In particular,
for the $\omega^{(3)}_{ABC}$ we find that the only non-zero components
are:
\begin{equation}
\omega^{(3)}_{R \Theta \Theta} = \omega^{(3)}_{R \Phi \Phi} = - \frac{1}{a} \left(
\frac{1}{r} + \frac{\partial_r b}{b} \right) \; , \qquad
\omega^{(3)}_{\Theta \Phi \Phi} = - \frac{\cot \theta}{rb} \; .
\label{eq:omega-CurvSphe-1}
\end{equation}

The next step is to find the matrices $\sigma^{IJ} =
[\gamma^I,\gamma^J]/4$. Given equation~\eqref{eq:FockIvanenko}, plus
the fact that we only have three non-zero Ricci rotation coefficients,
it is not difficult to see that we only need the three matrices
$\sigma^{R \Theta}$, $\sigma^{R \Phi}$ and $\sigma^{\Theta \Phi}$.
Using now equation~\eqref{eq:sigma_ij} we find:
\begin{align}
\sigma^{R \Theta} &= - \sigma^{23} =
+ \frac{i}{2}
\left(
\begin{array}{cc}
\sigma_1 & 0
\\
0 & \sigma_1 
\end{array}
\right) \; ,
\\
\sigma^{R \Phi} &= - \sigma^{13} =
- \frac{i}{2} \left(
\begin{array}{cc}
\sigma_2 & 0
\\
0 & \sigma_2 
\end{array}
\right) \; ,
\\
\sigma^{\Theta \Phi} &= - \sigma^{12} =
+ \frac{i}{2}
\left(
\begin{array}{cc}
\sigma_3 & 0
\\
0 & \sigma_3 
\end{array}
\right) \; ,
\end{align}
with $\sigma_i$ the usual Pauli matrices (see
Eq.~\eqref{eq:PauliMatrices}).  We can now use the previous results to
calculate the three-dimensional Fock--Ivanenko coefficients from
equation~\eqref{eq:FockIvanenko3D}. We find:
\begin{align}
\Gamma^{(3)}_R &= 0 \; , \\
\Gamma^{(3)}_\Theta &= - \frac{i}{2} \: \omega_{R \Theta \Theta}
\left(
\begin{array}{cc}
\sigma_1 & 0 \\
0 & \sigma_1
\end{array}
\right) \; ,\\
\Gamma^{(3)}_\Phi &= + \frac{i}{2} \: \omega_{R \Phi \Phi}
\left(
\begin{array}{cc}
\sigma_2 & 0 \\
0 & \sigma_2
\end{array}
\right)
- \frac{i}{2} \: \omega_{\Theta \Phi \Phi}
\left(
\begin{array}{cc}
\sigma_3 & 0 \\
0 & \sigma_3
\end{array}
\right) \; ,
\end{align}
and for the coordinate components $\Gamma^{(3)}_r = 0$,
$\Gamma^{(3)}_\theta = rb \: \Gamma^{(3)}_\Theta$, $\Gamma^{(3)}_\varphi
= rb \sin \theta \: \Gamma^{(3)}_\Phi$.

We can now calculate the contraction $\lambda^I \Gamma^{(3)}_I$ that
appears in Dirac's equation.  A little algebra yields:
\begin{equation}
\lambda^I \Gamma^{(3)}_I = \lambda^m \Gamma^{(3)}_m
= \frac{1}{2} \left(
\begin{array}{cccc}
0     & 0     & + M_1 & + M_2 \\
0     & 0     & - M_2 & - M_1 \\
- M_1 & - M_2 & 0     & 0 \\
+ M_2 & + M_1 & 0     & 0 
\end{array}
\right) \; ,
\end{equation}
where:
\begin{equation}
M_1 := \frac{2}{a} \left( \frac{1}{r} + \frac{\partial_r b}{b} \right) \; ,
\qquad
M_2 := - i \: \frac{\cot \theta}{rb} \; .
\end{equation}

\vspace{5mm}

We now have all the necessary ingredients to write down the Dirac
equation. Notice first that in spherical symmetry our spatial triad
does not rotate, so we have $Q_{IJ}=0$ and the Dirac equation reduces
to:
\begin{equation}
\left( \partial_t - \beta^r \partial_r \right) \psi
= - \alpha \gamma^T \left[ \lambda^r \left( \partial_r
+ \frac{\partial_r \alpha}{2 \alpha} \right) + \lambda^m \Gamma^{(3)}_m
+ i m \right] \psi + \left( \frac{\alpha K}{2} \right) \psi \: ,
\label{eq:Dirac_3+1-Sphe}
\end{equation}
or explicitly for each of the spinor components:
\begin{align}
\partial_t \psi_1 - \beta^r \partial_r \psi_1 &= \alpha \left[
- \frac{1}{a} \left( \partial_r  + \frac{\partial_r \alpha}{2 \alpha}
+ \frac{\partial_r b}{b} + \frac{1}{r} \right) \psi_3
+ \frac{i}{rb} \left( \partial_\theta 
+ \frac{i}{\sin \theta} \: \partial_\varphi + \frac{\cot \theta}{2} \right) \psi_4
+ \left( \frac{K}{2} - i m \right) \psi_1 \right] \; ,
\label{eq:DiracCurvSphe1} \\
\partial_t \psi_2 - \beta^r \partial_r \psi_2 &= \alpha \left[
+ \frac{1}{a} \left( \partial_r + \frac{\partial_r \alpha}{2 \alpha}
+ \frac{\partial_r b}{b} + \frac{1}{r} \right) \psi_4
- \frac{i}{rb} \left( \partial_\theta 
- \frac{i}{\sin \theta} \: \partial_\varphi + \frac{\cot \theta}{2} \right) \psi_3
+ \left( \frac{K}{2} - i m \right) \psi_2 \right] \; ,
\label{eq:DiracCurvSphe2} \\
\partial_t \psi_3 - \beta^r \partial_r \psi_3 &= \alpha \left[
- \frac{1}{a} \left( \partial_r + \frac{\partial_r \alpha}{2 \alpha}
+ \frac{\partial_r b}{b} + \frac{1}{r} \right) \psi_1
+ \frac{i}{rb} \left( \partial_\theta 
+ \frac{i}{\sin \theta} \: \partial_\varphi + \frac{\cot \theta}{2} \right) \psi_2
+ \left( \frac{K}{2} + i m \right) \psi_3 \right] \; ,
\label{eq:DiracCurvSphe3} \\
\partial_t \psi_4 - \beta^r \partial_r \psi_4 &= \alpha \left[
+ \frac{1}{a} \left( \partial_r + \frac{\partial_r \alpha}{2 \alpha}
+ \frac{\partial_r b}{b} + \frac{1}{r} \right) \psi_2
- \frac{i}{rb} \left( \partial_\theta 
- \frac{i}{\sin \theta} \: \partial_\varphi + \frac{\cot \theta}{2} \right) \psi_1
+ \left( \frac{K}{2} + i m \right) \psi_4 \right] \; .
\label{eq:DiracCurvSphe4}
\end{align}
These are the Dirac equations in a general spherically symmetric
spacetime.

In particular, by taking $\beta^r=K=0$ and $\alpha=a=b=1$ we will have
the explicit form of the Dirac equation for the case of Minkowski
spacetime in spherical coordinates. If, on the other hand, we take
$\beta^r=K=0$, \mbox{$\alpha=(1-2M/r)^{1/2}$}, \mbox{$a=1/(1-2M/r)$}
and $b=1$ we will have the Dirac equation in a Schwarzschild spacetime
in the standard coordinates.  Alternatively, by taking
\mbox{$\alpha=1/(1+2M/r)^{1/2}$}, \mbox{$\beta^r=2M/(r+2M)$},
\mbox{$a=(1+2M/r)$}, $b=1$, and
\mbox{$K=(2M/r^2)(1+3M/r)/(1+2M/r)^{3/2}$} we will have the Dirac
equation in a Schwarzschild spacetime in horizon penetrating
coordinates of Kerr--Schild type.

At this point is it important to mention the fact that, even though we
arrived at the previous equations using the 3+1 form of the Dirac
equation, we would have obtained precisely the same result starting
from the fully covariant four-dimensional version.

The equations we just found can be written in a more compact form
if we define:
\begin{equation}
\psi^I_\pm := \psi_1 \pm \psi_3 \; , \qquad
\psi^{II}_\pm := \psi_4 \mp \psi_2 \; .
\end{equation}
The Dirac equations then reduce to:
\begin{align}
\partial_t \psi^I_\pm - \beta^r \partial_r \psi^I_\pm
&= \alpha \left[ \mp \frac{1}{a} \left( \partial_r + \frac{\partial_r \alpha}{2 \alpha}
+ \frac{\partial_r b}{b} + \frac{1}{r} \right) \psi^I_\pm
+ \frac{i}{rb} \left( \partial_\theta 
+ \frac{i}{\sin \theta} \: \partial_\varphi + \frac{\cot \theta}{2} \right) \psi^{II}_\mp 
+ \frac{K}{2} \: \psi^I_\pm - i m \psi^I_\mp \right] ,
\\
\! \! \partial_t \psi^{II}_\pm - \beta^r \partial_r \psi^{II}_\pm
&= \alpha \left[ \mp \frac{1}{a} \left( \partial_r + \frac{\partial_r \alpha}{2 \alpha}
+ \frac{\partial_r b}{b} + \frac{1}{r} \right) \psi^{II}_\pm
- \frac{i}{rb} \left( \partial_\theta 
- \frac{i}{\sin \theta} \: \partial_\varphi + \frac{\cot \theta}{2} \right) \psi^{I}_\mp
+ \frac{K}{2} \: \psi^{II}_\pm + i m \psi^{II}_\mp \right] .
\end{align}

An interesting property of the previous system of equations is the
fact that we cannot have a spinor with spherical symmetry, which
makes perfect sense since spinors represent spin $1/2$ particles.  To
see this notice that if even if we start with initial data such that
all the different spinor components are functions only of the radial
coordinate $r$, the terms with $\cot \theta$ above will immediately
introduce a dependence on the angular coordinate $\theta$ during
evolution.


\subsection{Conserved current and stress--energy tensor}

We can now calculate the particle density and its associated flux. For
the particle density we simply have, from
equation~\eqref{eq:ParticleDensity-3+1}:
\begin{equation}
\rho_p = | \psi_1 |^2 + | \psi_2 |^2 + | \psi_3 |^2 + | \psi_4 |^2 \; .
\label{eq:rhop-CurvSpherical}
\end{equation}

On the other hand, the spatial components of the current which give
us the particle flux, can be found using
equation~\eqref{eq:ParticleCurrent-3+1}. For the radial flux we find:
\begin{equation}
f_r = \bar{\psi} \lambda_r \psi = a \left( \psi^\dag \gamma^T \lambda^R \right) \psi
= a \left[ \left( \psi_1 \psi_3^* - \psi_2 \psi_4^* \right) + c.c. \right] \; ,
\label{eq:fr-CurvSpherical}
\end{equation}
and for the angular components:
\begin{align}
f_\theta &= \bar{\psi} \lambda_\theta \psi
= rb \left( \psi^\dag \gamma^T \lambda^\Theta \psi \right)
= i rb \left[ \left( \psi_1 \psi_4^* - \psi_2 \psi_3^* \right)
- c.c. \right] \; ,
\label{eq:jT-CurvSpherical}
\\
f_\varphi &= \bar{\psi} \lambda_\varphi \psi
= rb \sin \theta \left( \psi^\dag \gamma^T \lambda^\Phi \psi \right)
= rb \sin \theta \left[ \left( \psi_1 \psi_4^* + \psi_2 \psi_3^* \right)
+ c.c. \right] \; .
\label{eq:jP-CurvSpherical}
\end{align}

\vspace{5mm}

The next step is to find the 3+1 components of the stress--energy
tensor.  These components involve the quantity $\Pi$ defined
in~\eqref{eq:Pi}, so it is convenient at this point to find an
expression for the Fock--Ivanenko coefficient $\Gamma_T$ that appears
in the definition of $\Pi$ for the case of spherical symmetry. First,
notice that in spherical symmetry our spatial triad does not rotate so
we clearly have $Q_{mn}=0$.  Also, the lapse function only depends on
the radial coordinate $r$, so the coefficient $\Gamma_T$ reduces in
this case to:
\begin{equation}
\Gamma_T = \gamma^T \lambda^r \left( \frac{\partial_r \alpha}{2 \alpha} \right) \; .
\end{equation}

Consider now the energy density. Since, as we just mentioned, in
spherical symmetry we have $Q_{mn}=0$, we can use
equation~\eqref{eq:EnergyDensity-3+1-FW} for the energy density in
terms of $\tilde{\Pi}$ instead of $\Pi$.  We find:
\begin{equation}
\rho_E = \frac{i}{2} \left[ \psi^\dag \tilde{\Pi} - \tilde{\Pi}^\dag \psi \right]
= \frac{i}{2} \left[ \left( \psi_1^* \tilde{\Pi}_1 + \psi_2^* \tilde{\Pi}_2
+ \psi_3^* \tilde{\Pi}_3 + \psi_4^* \tilde{\Pi}_4 \right) - c.c. \right] \; ,
\label{eq:rhoE-CurvGen}
\end{equation}
where now:
\begin{equation}
\tilde{\Pi}_i = \frac{1}{\alpha} \: \left( \partial_t \psi_i - \beta^r
\partial_r \psi_i \right) \; .
\end{equation}

The momentum density $J_i$ is given by
equation~\eqref{eq:MomentumDensity-3+1}. In particular, for the radial
component we find, using the fact that $\Gamma^{(3)}_r=0$:
\begin{equation}
J_r = - \frac{i}{4} \left[ \bar{\Pi} \lambda_r \psi - \bar{\psi} \lambda_r \Pi 
+ \psi^\dag (\partial_r \psi) - (\partial_r \psi^\dag) \psi
\right] \; .
\end{equation}
On the other hand, substituting $\Pi = \tilde{\Pi} + \Gamma_T \psi$, we
have:
\begin{equation}
\bar{\Pi} \lambda_r \psi - \bar{\psi} \lambda_r \Pi
= \Pi^\dag \gamma^T \lambda_r \psi - \psi^\dag \gamma^T \lambda_r \Pi
= \tilde{\Pi}^\dag \gamma^T \lambda_r \psi - \psi^\dag \gamma^T \lambda_r \tilde{\Pi}
+ \psi^\dag \left( \Gamma_T^\dag \gamma^T \lambda_r - \gamma^T \lambda_r \Gamma_T \right) \psi \; .
\end{equation}
Using now our expression for $\Gamma_T$ above, plus the fact that
$(\gamma^T)^\dag = \gamma^T$, $(\lambda^m)^\dag = - \lambda^m$,
$(\gamma^T)^2=1$, and the spatial Clifford algebra, one can show that
in the case of spherical symmetry we have:
\begin{equation}
\Gamma_T^\dag \gamma^T \lambda_i - \gamma^T \lambda_i \Gamma_T
= \frac{\partial_r \alpha}{\alpha} \: \left( \lambda_i \lambda^r + \delta_i^r \right) \; .
\end{equation}
For $i=r$ the Clifford algebra implies that $\lambda_r \lambda^r =
-1$, so the above term vanishes. The radial component of the
momentum density can then be written using $\tilde{\Pi}$ instead of
$\Pi$ as:
\begin{equation}
J_r = - \frac{i}{4} \left[
\tilde{\Pi}^\dag \gamma^T \lambda_r \psi - \psi^\dag \gamma^T \lambda_r \tilde{\Pi}
+ \psi^\dag (\partial_r \psi) - (\partial_r \psi^\dag) \psi \right] \; ,
\end{equation}
or in terms of the spinor components:
\begin{align}
J_r &= - \frac{i}{4} \left[ a \left( \psi_1 \tilde{\Pi}_3^* - \psi_2 \tilde{\Pi}_4^*
+ \psi_3 \tilde{\Pi}_1^* - \psi_4 \tilde{\Pi}_2^* \right) \right. \nonumber \\
&+ \left. \left( \psi_1^* \partial_r \psi_1 + \psi_2^* \partial_r \psi_2
\label{eq:Jr-CurvGen}
+ \psi_3^* \partial_r \psi_3 + \psi_4^* \partial_r \psi_4 \right) - c.c. \right] \; .
\end{align}
The angular components of the momentum density can also be calculated
in a straightforward way, but they turn out to be rather long
expressions that are not particularly interesting and I will not write
them here.  But it is important to stress the fact that these angular
components do not vanish in general since, even if here we are
considering a spherically symmetric spacetime, we are not asking for a
spherically symmetric $\psi$ (but see Sec.~\ref{sec:sphere-solutions}
below).

Finally, for the spatial stress tensor $S_{ij}$ we use
equation~\eqref{eq:StressTensor-3+1}. Let us consider the diagonal
components $S_{ii}$.  Since in spherical symmetry the metric is
diagonal, the Clifford algebra implies that (no sum) $\lambda_i
\lambda^i = -1$. Moreover, in spherical symmetry the extrinsic
curvature tensor $K_{ij}$ is also diagonal.  One can then easily see
that the contributions from the extrinsic curvature in
equation~\eqref{eq:StressTensor-3+1} cancel out for $i=j$. The
diagonal components of the spatial stress tensor then become:
\begin{equation}
S_{ii} = \frac{i}{2} \left[ \left( D_i \bar{\psi} \right) \lambda_i \psi
- \bar{\psi} \lambda_i \left( D_i \psi \right) \right]
=  \frac{i}{2} \left[ \left( \partial_i \bar{\psi} \right) \lambda_i \psi
- \bar{\psi} \lambda_i \left( \partial_i \psi \right) 
- \bar{\psi} \left(\Gamma^{(3)}_i \lambda_i + \lambda_i \Gamma^{(3)}_i \right) \psi \right] \; .
\end{equation}
We can simplify this even further.  Using now the expressions for the
$\Gamma^{(3)}_i$ and $\lambda_i$ found above, plus the fact that the
Pauli matrices anti-commute with each other, it is not difficult to
show that in our case we have $\Gamma^{(3)}_i \lambda_i + \lambda_i
\Gamma^{(3)}_i = 0$ for all three possible values of $i$. The diagonal
components of the spatial stress tensor then reduce simply to:
\begin{equation}
S_{ii} = \frac{i}{2} \left[ \left( \partial_i \bar{\psi} \right) \lambda_i \psi
- \bar{\psi} \lambda_i \left( \partial_i \psi \right) \right] \; ,
\end{equation}
or more explicitly:
\begin{align}
S_{rr} &= \frac{i}{2} \left[ \left( \partial_r \bar{\psi} \right) \lambda_r \psi
- \bar{\psi} \lambda_r \left( \partial_r \psi \right) \right]
= \frac{i a}{2} \left[ \left( \psi_1 \partial_r \psi_3^*
- \psi_2 \partial_r \psi_4^* + \psi_3 \partial_r \psi_3^* - \psi_4
\partial_r \psi_2^* \right) - c.c. \right] \; ,
\label{eq:Srr-CurvGen} \\
S_{\theta \theta} &= \frac{i}{2} \left[ \left( \partial_\theta \bar{\psi} \right) \lambda_\theta \psi
- \bar{\psi} \lambda_\theta \left( \partial_\theta \psi \right) \right]
= \frac{r b}{2} \left[ \left( \psi_2 \partial_\theta \psi_3^*
  + \psi_4 \partial_\theta \psi_1^* - \psi_1 \partial_\theta \psi_4^*
  - \psi_3 \partial_\theta \psi_2^* \right) + c.c. \right] \; ,
\label{eq:Stt-CurvGen} \\
S_{\varphi \varphi} &= \frac{i}{2} \left[ \left( \partial_\varphi \bar{\psi} \right) \lambda_\varphi \psi
- \bar{\psi} \lambda_\varphi \left( \partial_\varphi \psi \right) \right]
= \frac{i r b \sin \theta}{2}
\left[ \left( \psi_1 \partial_\varphi \psi_4^*
+ \psi_2 \partial_\varphi \psi_3^* + \psi_3 \partial_\varphi \psi_2^*
+ \psi_4 \partial_\varphi \psi_1^*\right) - c.c. \right] \; .
\label{eq:Spp-CurvGen}
\end{align}

The mixed components of $S_{ij}$ for $i \neq j$ again turn out to be
rather long expressions that I will not write here.


\subsection{Separation of variables}

We are be interested in finding solutions to Dirac's equation that
are compatible with a spherically symmetric spacetime. For the moment
we will still consider the Dirac field as a test field on a fixed
background spacetime, and only later consider the self-gravitating
case.

The first step in looking for solutions is to use the method of
separation of variables.  We then propose an ansatz of the form (the
discussion here is based in part in that of~\cite{Daka:2019}):
\begin{equation}
\psi_i = R_i(t,r) T_i(\theta,\varphi) \; ,
\end{equation}
where here $i=1,2,3,4$ denotes spinor components, and $R_i$ and $T_i$
are complex functions to be determined.  Substituting this in
equations~\eqref{eq:DiracCurvSphe1}-\eqref{eq:DiracCurvSphe4}, and
regrouping terms we find, after some algebra:
\begin{align}
\frac{rb}{R_4} \left[ \frac{T_1}{T_3} \left( \frac{1}{\alpha}
\left( \partial_t - \beta^r \partial_r \right)
+ i m - \frac{K}{2} \right) R_1
+ \frac{1}{a} \left( \partial_r + \frac{\partial_r \alpha}{2 \alpha}
+ \frac{\partial_r b}{b} + \frac{1}{r} \right) R_3 \right]
&= + \frac{i}{T_3} \left[ \partial_\theta
+ \frac{i}{\sin \theta} \: \partial_\varphi + \frac{\cot \theta}{2} \right] T_4  ,
\label{eq:SeparationCurv1} \\
\frac{rb}{R_3} \left[ \frac{T_2}{T_4} \left( \frac{1}{\alpha}
\left( \partial_t - \beta^r \partial_r \right)
+ im - \frac{K}{2} \right) R_2
- \frac{1}{a} \left( \partial_r + \frac{\partial_r \alpha}{2 \alpha}
+ \frac{\partial_r b}{b} + \frac{1}{r} \right) R_4 \right]
&= - \frac{i}{T_4} \left[ \partial_\theta
- \frac{i}{\sin \theta} \: \partial_\varphi + \frac{\cot \theta}{2} \right] T_3 ,
\label{eq:SeparationCurv2} \\
\frac{rb}{R_2} \left[ \frac{T_3}{T_1} \left( \frac{1}{\alpha}
\left( \partial_t - \beta^r \partial_r \right)
- im - \frac{K}{2} \right) R_3
+ \frac{1}{a} \left( \partial_r + \frac{\partial_r \alpha}{2 \alpha}
+ \frac{\partial_r b}{b} + \frac{1}{r} \right) R_1 \right]
&= + \frac{i}{T_1} \left[ \partial_\theta
+ \frac{i}{\sin \theta} \: \partial_\varphi + \frac{\cot \theta}{2} \right] T_2 , 
\label{eq:SeparationCurv3} \\
\frac{rb}{R_1} \left[ \frac{T_4}{T_2} \left[(\frac{1}{\alpha}
\left( \partial_t - \beta^r \partial_r \right)
- im - \frac{K}{2} \right) R_4
- \frac{1}{a} \left( \partial_r+ \frac{\partial_r \alpha}{2 \alpha}
+ \frac{\partial_r b}{b} + \frac{1}{r} \right) R_2 \right]
&= - \frac{i}{T_2} \left[ \partial_\theta
- \frac{i}{\sin \theta} \: \partial_\varphi + \frac{\cot \theta}{2} \right] T_1 .
\label{eq:SeparationCurv4}
\end{align}

The right hand side of the previous equations is now only a function
of the angular coordinates $(\theta,\varphi)$, but the separation of
variables is not complete since we still have angular functions on the
left hand side. This can be fixed if we ask for $T_3 = a T_1$ and $T_4
= b T_2$, with $(a,b)$ some proportionality constants. In that case
the left hand side of the above equations will now be only a function
of $(t,r)$.

The constants $(a,b)$ are in principle arbitrary, but a convenient
choice is $a=1$ and $b=-1$. With this choice we can see that, except
for a sign, the right hand side of
equations~\eqref{eq:SeparationCurv1} and~\eqref{eq:SeparationCurv3} is
now the same, and also the right hand side of
equations~\eqref{eq:SeparationCurv2} and~\eqref{eq:SeparationCurv4}.
This implies that we must now have:
\begin{align}
& \frac{rb}{R_4} \left[ \left( \frac{1}{\alpha}
\left( \partial_t - \beta^r \partial_r \right)
+ i m - \frac{K}{2} \right) R_1
+ \frac{1}{a} \left( \partial_r + \frac{\partial_r \alpha}{2 \alpha}
+ \frac{\partial_r b}{b} + \frac{1}{r} \right) R_3 \right]
\nonumber \\
& \hspace{20mm} = - \frac{rb}{R_2} \left[ \left( \frac{1}{\alpha}
\left( \partial_t - \beta^r \partial_r \right)
- im - \frac{K}{2} \right) R_3
+ \frac{1}{a} \left( \partial_r + \frac{\partial_r \alpha}{2 \alpha}
+ \frac{\partial_r b}{b} + \frac{1}{r} \right) R_1 \right] \; ,
\\
& \frac{rb}{R_3} \left[ \left( \frac{1}{\alpha}
\left( \partial_t - \beta^r \partial_r \right)
+ im - \frac{K}{2} \right) R_2
- \frac{1}{a} \left( \partial_r + \frac{\partial_r \alpha}{2 \alpha}
+ \frac{\partial_r b}{b} + \frac{1}{r} \right) R_4 \right]
\nonumber \\
& \hspace{20mm} = - \frac{rb}{R_1} \left[ \left( \frac{1}{\alpha}
\left( \partial_t - \beta^r \partial_r \right)
- im - \frac{K}{2} \right) R_4
- \frac{1}{a} \left( \partial_r + \frac{\partial_r \alpha}{2 \alpha}
+ \frac{\partial_r b}{b} + \frac{1}{r} \right) R_2 \right] \; .
\end{align}
We can now reduce these two equations to just one if
we take  $R_2 = c R_1$ and $R_4 = c R_3$, with $c$
a new constant. Again, the value of $c$ is arbitrary,
but a convenient choice is $c=i$. With these
choices we now have:
\begin{equation}
R_2 = i R_1 \; , \qquad R_4 = i R_3 \; , \qquad
T_3 = T_1 \; , \qquad T_4 = - T_2 \; ,
\end{equation}
and our system of equation reduces to:
\begin{align}
\frac{rb}{R_3} \left[ \left( \frac{1}{\alpha}
\left( \partial_t - \beta^r \partial_r \right)
+ i m - \frac{K}{2} \right) R_1
+ \frac{1}{a} \left( \partial_r  + \frac{\partial_r \alpha}{2 \alpha}
+ \frac{\partial_r b}{b} + \frac{1}{r} \right) R_3 \right]
&= + \frac{1}{T_1} \left[ \partial_\theta
+ \frac{i}{\sin \theta} \: \partial_\varphi + \frac{\cot \theta}{2} \right] T_2 \; ,
\\
\frac{rb}{R_3} \left[ \left( \frac{1}{\alpha}
\left( \partial_t - \beta^r \partial_r \right)
+ im - \frac{K}{2} \right) R_1
+ \frac{1}{a} \left( \partial_r + \frac{\partial_r \alpha}{2 \alpha}
+ \frac{\partial_r b}{b} + \frac{1}{r} \right) R_3 \right]
&= - \frac{1}{T_2} \left[ \partial_\theta
- \frac{i}{\sin \theta} \: \partial_\varphi + \frac{\cot \theta}{2} \right] T_1 \; ,
\\
\frac{rb}{R_1} \left[ \left( \frac{1}{\alpha}
\left( \partial_t - \beta^r \partial_r \right)
- im - \frac{K}{2} \right) R_3
+ \frac{1}{a} \left( \partial_r + \frac{\partial_r \alpha}{2 \alpha}
+ \frac{\partial_r b}{b} + \frac{1}{r} \right) R_1 \right]
&= - \frac{1}{T_1} \left[ \partial_\theta
+ \frac{i}{\sin \theta} \: \partial_\varphi + \frac{\cot \theta}{2} \right] T_2 \; , 
\\
\frac{rb}{R_1} \left[ \left( \frac{1}{\alpha}
\left( \partial_t - \beta^r \partial_r \right)
- im - \frac{K}{2} \right) R_3
+ \frac{1}{a} \left( \partial_r + \frac{\partial_r \alpha}{2 \alpha}
+ \frac{\partial_r b}{b} + \frac{1}{r} \right) R_1 \right]
&= + \frac{1}{T_2} \left[ \partial_\theta
- \frac{i}{\sin \theta} \: \partial_\varphi + \frac{\cot \theta}{2} \right] T_1 \; .
\end{align}
The previous equations now have the following structure:
\begin{align}
f_1(t,r) &= + g_1(\theta,\varphi) \; , \\
f_1(t,r) &= - g_2(\theta,\varphi) \; , \\
f_2(t,r) &= - g_1(\theta,\varphi) \; , \\
f_2(t,r) &= + g_2(\theta,\varphi) \; .
\end{align}
Since one side of these equations depends only on $(r,t)$, and the
other only on $(\theta,\varphi)$, we conclude that both sides must be
equal to the same constant.  Also, given the above structure we must
have $f_1=g_1=-f_2=-g_2=k$, with $k$ a separation constant to be
determined. We will then have the following two radial equations:
\begin{align}
\left( \frac{1}{\alpha} \left( \partial_t - \beta^r \partial_r \right)
+ im - \frac{K}{2} \right) R_1
+ \frac{1}{a} \left( \partial_r + \frac{\partial_r \alpha}{2 \alpha}
+ \frac{\partial_r b}{b} + \frac{1}{r} \right) R_3
&= + \frac{k R_3}{rb} \; ,
\label{eq:RadCurv1} \\
\left( \frac{1}{\alpha} \left( \partial_t - \beta^r \partial_r \right)
- im - \frac{K}{2} \right) R_3
+ \frac{1}{a} \left( \partial_r + \frac{\partial_r \alpha}{2 \alpha}
+ \frac{\partial_r b}{b} + \frac{1}{r} \right) R_1
&= - \frac{k R_1}{rb} \; ,
\label{eq:RadCurv2}
\end{align}
and two angular equations:
\begin{align}
\left( \partial_\theta - \frac{i}{\sin \theta} \: \partial_\varphi
+ \frac{\cot \theta}{2} \right) T_1 &= - k T_2 \; ,
\label{eq:AngCurv1}
\\
\left( \partial_\theta + \frac{i}{\sin \theta} \: \partial_\varphi
+ \frac{\cot \theta}{2} \right) T_2 &= + k T_1 \; .
\label{eq:AngCurv2}
\end{align}

The angular equations are particularly interesting.  In order to see
this let us first define the operators:
\begin{align}
\slashed{\partial}_s^+ &:= - \partial_\theta
- \frac{i}{\sin \theta} \: \partial_\varphi + s \cot \theta \; ,
\label{eq:SpinRaiseDef}
\\
\slashed{\partial}_s^- &:= - \partial_\theta
+ \frac{i}{\sin \theta} \: \partial_\varphi - s \cot \theta \; ,
\label{eq:SpinLowerDef}
\end{align}
with $s$ an integer or half-integer constant.  The above operators are
known as the raising and lowering spin operators respectively, and are
associated with the spherical harmonics with spin weight $s$ first
introduced by Newman and Penrose in~\cite{Newman-Penrose-1966} (see
also Appendix D of reference~\cite{Alcubierre08a}). For an integer $s$
such that $|s|<l$, the spin weighted spherical harmonics are defined
in terms of the usual spherical harmonics $Y^{l,m}(\theta,\varphi)$
as:
\begin{equation}
{}_sY^{l,m} := \left\{
\begin{array}{lc}
\left[ \frac{(l-s)!}{(l+s)!} \right]^{1/2}
\slashed{\partial}_{s-1}^+ \cdots \slashed{\partial}_0^+ \left( Y^{l,m} \right) \; ,
& +l \geq s \geq 0 \: , \\
(-1)^s \left[ \frac{(l-|s|)!}{(l+|s|)!} \right]^{1/2}
\slashed{\partial}_{s-1}^- \cdots \slashed{\partial}_0^- \left( Y^{l,m} \right) \; ,
\hspace{10mm} & -l \leq s \leq 0 \; , \\
0 \; , & |s| > l \; .
\end{array}
\right.
\label{eq:SpinHarmonics}
\end{equation}
We also define ${}_0Y^{l,m} := Y^{l,m}$. For example, we find:
\begin{align}
{}_{\pm 1} Y^{l,m} &= \pm \left[ \frac{(l-1)!}{(l+1)!} \right]^{1/2} \slashed{\partial}_0^\pm Y^{l,m}
\nonumber \\
&= \mp \left[ \frac{(l-1)!}{(l+1)!} \right]^{1/2} \left( \partial_\theta
\pm \frac{i}{\sin \theta} \: \partial_\varphi \right) Y^{l,m} \; ,
\\
{}_{\pm 2} Y^{l,m} &= \left[ \frac{(l-2)!}{(l+2)!} \right]^{1/2}
\slashed{\partial}_1^\pm \slashed{\partial}_0^\pm Y^{l,m} \; ,
\nonumber \\
&= \left[ \frac{(l-2)!}{(l+2)!} \right]^{1/2} \left( \partial^2_\theta
- \cot \theta \partial_\theta \pm \frac{2i}{\sin \theta}
\left( \partial_\theta - \cot \theta \right) \partial_\varphi
- \frac{1}{\sin^2 \theta} \: \partial^2 _\varphi \right) Y^{l,m} \; .
\end{align}
On the other hand, the above definitions imply that:
\begin{align}
\slashed{\partial}_s^+ \left( {}_sY^{l,m} \right)
&= + \left[ (l-s) (l+s+1) \right]^{1/2} {}_{s+1}Y^{l,m} \; ,
\label{eq:SpinRaiseAction}
\\
\slashed{\partial}_s^- \left( {}_sY^{l,m} \right)
&= - \left[ (l+s) (l-s+1) \right]^{1/2} {}_{s-1}Y^{l,m} \; ,
\label{eq:SpinLowerAction}
\end{align}
which explains why they are called spin raising and lowering
operators.  These relations also allow us to show that:
\begin{align}
\slashed{\partial}_{s+1}^- \slashed{\partial}_s^+ \left( {}_sY^{l,m} \right)
&= - \left[ l(l+1) - s (s+1) \right] {}_sY^{l,m} \; , \\
\slashed{\partial}_{s-1}^+ \slashed{\partial}_s^- \left( {}_sY^{l,m} \right)
&= - \left[ l(l+1) - s (s-1) \right] {}_sY^{l,m} \; ,
\end{align}
that is, the ${}_sY^{l,m}$ are eigenfunctions of the operators
$\slashed{\partial}_{s+1}^- \slashed{\partial}_s^+$ and
$\slashed{\partial}_{s-1}^+ \slashed{\partial}_s^-$.
In particular, for a function $f$ with spin weight $s=0$
we have $\slashed{\partial}_{1}^-
\slashed{\partial}_0^+ = \slashed{\partial}_{-1}^+
\slashed{\partial}_0^- = L^2$, with $L^2$ the usual
angular Laplacian operator:
\begin{align}
L^2 f &= \frac{1}{\sin \theta} \: \partial_\theta \left( \sin \theta \partial_\theta f \right)
+ \frac{1}{\sin^2 \theta} \: \partial^2_\varphi f
= \partial^2_\theta f + \cot \theta \: \partial_\theta f
+ \frac{1}{\sin^2 \theta} \: \partial^2_\varphi f \; .  
\end{align}
Notice now that in our equations~\eqref{eq:AngCurv1}
and~\eqref{eq:AngCurv2} we in fact have the raising and lowering
operators with spin $s = \pm 1/2$. In that case the
definition~\eqref{eq:SpinHarmonics} can not be used since we have a
half-integer value for $s$. However, we can define the functions
${}_{\pm 1/2}Y^{l,m}$ simply as the eigenfunctions of the
corresponding operators $\slashed{\partial}_{s+1}^-
\slashed{\partial}_s^+$ and $\slashed{\partial}_{s-1}^+
\slashed{\partial}_s^-$ with $s=\pm 1/2$.  Notice that in that case we
must also have $l$ and $m$ as half-integers, with $m=-l,\cdots,l$.
For $s=+1/2$ we find:
\begin{align}
\slashed{\partial}_{3/2}^- \slashed{\partial}_{1/2}^+ f
&= \partial^2 _\theta f + \cot \theta \: \partial_\theta f
+ \frac{1}{\sin^2 \theta} \left( \partial^2_\varphi f
+ i \cos \theta \: \partial_\varphi f \right)
- \frac{1}{4} \left( \frac{1}{\sin^2 \theta} - 3 \right) f \; ,
\\
\slashed{\partial}_{-1/2}^+ \slashed{\partial}_{1/2}^-
&= \partial^2 _\theta f + \cot \theta \: \partial_\theta f
+ \frac{1}{\sin^2 \theta} \left( \partial^2_\varphi f
+ i \cos \theta \: \partial_\varphi f \right)
- \frac{1}{4} \left( \frac{1}{\sin^2 \theta} + 1 \right) f \; ,
\end{align}
while for $s=-1/2$ we find:
\begin{align}
\slashed{\partial}_{1/2}^- \slashed{\partial}_{-1/2}^+ f
&= \partial^2 _\theta f + \cot \theta \: \partial_\theta f
+ \frac{1}{\sin^2 \theta} \left( \partial^2_\varphi f
- i \cos \theta \: \partial_\varphi f \right)
- \frac{1}{4} \left( \frac{1}{\sin^2 \theta} + 1 \right) f \; ,
\\
\slashed{\partial}_{-3/2}^+ \slashed{\partial}_{-1/2}^-
&= \partial^2 _\theta f + \cot \theta \: \partial_\theta f
+ \frac{1}{\sin^2 \theta} \left( \partial^2_\varphi f
- i \cos \theta \: \partial_\varphi f \right)
- \frac{1}{4} \left( \frac{1}{\sin^2 \theta} - 3 \right) f \; .
\end{align}
In terms of our raising and lowering operators for spin $1/2$ the
angular equations~\eqref{eq:AngCurv1} y and~\eqref{eq:AngCurv2} can
now be written as:
\begin{equation}
\slashed{\partial}_{+1/2}^- T_1 = k T_2 \; , \qquad
\slashed{\partial}_{-1/2}^+ T_2 = - k T_1 \; .
\end{equation}
Comparing this with equations~\eqref{eq:SpinRaiseAction}
and~\eqref{eq:SpinLowerAction} it is clear that we can take as
solutions:
\begin{equation}
T_1 = {}_{+1/2}Y^{l,m} \; , \qquad T_2 = {}_{-1/2}Y^{l,m} \; ,
\end{equation}
with the separation constant given by $k=-(l+1/2)$.

General expressions for the ${}_sY^{l,m}$ with both integer and
half-integer indices are well known, and can be found in terms of the
Wigner rotation matrices commonly used in quantum mechanics. Here we
will just consider the cases with $l=1/2$.  For $l=1/2$, $s=\pm
1/2$, $m=1/2$ we have, in the standard normalization:
\begin{equation}
{}_{\pm 1/2}Y^{1/2,1/2} = \left( \frac{1}{\sqrt{4 \pi}} \right)
e^{+i \varphi / 2} y_\pm (\theta) \; ,
\end{equation}
where $y_+(\theta) = \sin (\theta/2)$ and $y_- = \cos (\theta/2)$. On the other hand, 
for $l=1/2$, $s=\pm 1/2$, $m=-1/2$ we have:
\begin{equation}
{}_{\pm 1/2}Y^{1/2,-1/2} = \pm \left( \frac{1}{\sqrt{4 \pi}} \right)
e^{-i \varphi / 2} y_\mp (\theta) \; .
\end{equation}
Notice that the functions ${}_{\pm 1/2}Y^{1/2,\pm1/2}$ are
discontinuous at $\varphi=0$.  This is an indication of the fact that
spinors change sign under a full rotation, as mentioned when we
discussed the spinor representation of the Lorentz group in
Sec.~\ref{sec:Lorentz-spinor}.

It is now not difficult to show that taking either
$T_1={}_{+1/2}Y^{1/2,1/2}$ and $T_2={}_{-1/2}Y^{1/2,1/2}$, or
alternatively \mbox{$T_1={}_{+1/2}Y^{1/2,-1/2}$} and
\mbox{$T_2={}_{-1/2}Y^{1/2,-1/2}$}, we will have two independent
solutions for our angular equations with $k=-1$.  Of course, we can
take higher half-integer values of $(l,m)$ while keeping $s=1/2$ in
order to find more solutions, but here we will only consider these two
cases.  In that case the radial equations~\eqref{eq:RadCurv1}
and~\eqref{eq:RadCurv1} become:
\begin{align}
\left( \frac{1}{\alpha} \left( \partial_t - \beta^r \partial_r \right)
+ im - \frac{K}{2} \right) R_1
+ \frac{1}{a} \left( \partial_r + \frac{\partial_r \alpha}{2 \alpha}
+ \frac{\partial_r b}{b} + \frac{1}{r} \right) R_3
&= - \frac{R_3}{rb} \; ,
\\
\left( \frac{1}{\alpha} \left( \partial_t - \beta^r \partial_r \right)
- im - \frac{K}{2} \right) R_3
+ \frac{1}{a} \left( \partial_r + \frac{\partial_r \alpha}{2 \alpha}
+ \frac{\partial_r b}{b} + \frac{1}{r} \right) R_1
&= + \frac{R_1}{rb} \; ,
\end{align}
If we now take $R_1(t,r) \equiv F(t,r)$ and $R_3(t,r) \equiv G(t,r)$
these equations can be written as:
\begin{align}
\partial_t F &= \beta^r \partial_r F
- \frac{\alpha}{a} \left[ \partial_r + \frac{\partial_r \alpha}{2 \alpha}
+ \frac{\partial_r b}{b} + \frac{1}{r} \left( 1 + \frac{a}{b} \right) \right] G
+ \alpha \left( \frac{K}{2} - im \right) F \; ,
\label{eq:Fdot-Curv} \\
\partial_t G &= \beta^r \partial_r G
- \frac{\alpha}{a} \left[ \partial_r + \frac{\partial_r \alpha}{2 \alpha}
+ \frac{\partial_r b}{b} + \frac{1}{r} \left( 1 - \frac{a}{b} \right) \right] F
+ \alpha \left( \frac{K}{2} + i m \right) G \; ,
\label{eq:Gdot-Curv}
\end{align}
These are now evolution equations for the functions $F$ and $G$, and
can be solved either numerically given some adequate initial data, or
considering some particular ansatz.  The spinors associated with our
two solutions will then be given by:
\begin{equation}
\psi_\pm = \frac{e^{\pm i \varphi / 2}}{(4 \pi)^{1/2}}
\left(
\begin{array}{r}
F(t,r) \: y_\pm (\theta) \\
\pm i F(t,r) \: y_\mp (\theta) \\
G(t,r) \: y_\pm (\theta) \\
\mp i G(t,r) \: y_\mp (\theta)
\end{array}
\right) \; ,
\label{eq:TrialSolutionsCurv}
\end{equation}
with $y_+(\theta) = \sin (\theta/2)$ and $y_- = \cos (\theta/2)$.
Equations~\eqref{eq:Fdot-Curv} and~\eqref{eq:Fdot-Curv} can be used,
for example, to study the evolution of our spinors in a general curved
spherically symmetric spacetime, such as Schwarzschild for example,
given some adequate initial data for both $F$ and $G$.

\vspace{5mm}

To finish this section, it is important to find how the functions $F$
and $G$ behave near the origin $r=0$. In order to do this we take $r
\ll 1$, and expand our functions as powers of $r$:
\begin{equation}
F = \sum_{n=0}^\infty F_n(t) r^n \; , \qquad
G = \sum_{n=0}^\infty G_n(t) r^n \; .
\end{equation}
On the other hand, regularity of the metric at the origin implies that
$a$, $b$ and $\alpha$ must be even functions of $r$, while $\beta^r$
must be odd:
\begin{align}
a &\simeq a_0(t) + a_2(t) r^2 + {\mathcal O}(r^4)  \; , \\
b &\simeq b_0(t) + b_2(t) r^2 + + {\mathcal O}(r^4)  \; , \\
\alpha &\simeq \alpha_0(t) + \alpha_2(t) r^2 + {\mathcal O}(r^4) \; , \\
\beta^r &\simeq \beta_1(t) r + {\mathcal O}(r^3)  \; .
\end{align}
Furthermore, in order for the metric to be locally flat at $r=0$ we
must ask for $b_0(t) = a_0(t)$.  Finally, from the definition of the
extrinsic curvature we find that we must also have $K \simeq K_0(t) +
K_2(t) r^2 + {\cal O} (r^4)$. When looking at
equations~\eqref{eq:Fdot-Curv} and~\eqref{eq:Gdot-Curv} there are two
terms that appear to be singular at $r=0$.  Their behavior close to
the origin is:
\begin{equation}
\frac{G}{r} \left( 1 + \frac{a}{b} \right) \simeq  \frac{2G_0}{r} \; ,
\qquad \qquad
\frac{F}{r} \left( 1 - \frac{a}{b} \right)
= \frac{F}{rb} \left( b - a \right) \simeq \frac{rF}{b_0} \left( b_2 - a_2 \right) \; ,
\end{equation}
where we used the fact that $b_0=a_0$.  The second term is now clearly
regular at $r=0$.  In order for the first term to be also regular we
must now ask for $G_0=0$, so the function $G(r)$ must vanish at the
origin.

We can go further in the analysis, but it is easier to work in the
case of Minkowski spacetime for which we take $\alpha=a=b=1$,
$\beta^r=K=0$ (the general case is more complicated but the
conclusions are the same).  Substituting the expansions for $F$
and $G$ into equations~\eqref{eq:Fdot-Curv} y~\eqref{eq:Gdot-Curv} we
now find:
\begin{align}
\sum_{n=0}^\infty \left( \dot{F}_n + im F_n \right) r^n
+ \sum_{n=0}^\infty \left( n + 2 \right) G_n r^{n-1} &= 0 \; , \\
\sum_{n=0}^\infty \left( \dot{G}_n - im G_n \right) r^n
+ \sum_{n=0}^\infty n F_n r^{n-1} &= 0 \; .
\end{align}
Again, from the first equation it is clear that we must have $G_0=0$.
Moreover, in the second term of the second equation the sum can be
taken from $n=1$ since the $n=0$ term vanishes.  We then have:
\begin{align}
\sum_{n=0}^\infty \left( \dot{F}_n + im F_n \right) r^n
+ \sum_{n=1}^\infty \left( n + 2 \right) G_n r^{n-1} &= 0 \; , \\
\sum_{n=0}^\infty \left( \dot{G}_n - im G_n \right) r^n
+ \sum_{n=1}^\infty n F_n r^{n-1} &= 0 \; .
\end{align}
Taking now $n \rightarrow n+1$ in the second term of both
equations we can rewrite them as:
\begin{align}
\sum_{n=0}^\infty \left( \dot{F}_n + im F_n
+ (n+3) G_{n+1} \right) r^n &= 0 \; , \\
\sum_{n=0}^\infty \left( \dot{G}_n - im G_n
+ (n+1) F_{n+1} \right) r^n &= 0 \; .
\end{align}
Finally, cancelling each power of $r$ separately we find:
\begin{equation}
G_{n+1} = - \frac{\dot{F}_n + im F_n}{n+3} \; , \qquad
F_{n+1} = - \frac{\dot{G}_n - im G_n}{n+1} \; .
\end{equation}
Since we must have $G_0=0$, the above result implies that $F_1=0$,
which in turn implies $G_2=0$, which now implies $F_3=0$, etc.  We
finally find that $F$ must be an even function of $r$, while $G$ must
be odd:
\begin{equation}
F = F_0(t) + F_2(t) r^2 + \ldots \; , \qquad
G = G_1(t) r + G_3(t) r^3 + \ldots \; .
\end{equation}
This behavior must be taken into account if one wishes to construct
initial data for $F$ and $G$.


\subsection{Spherically symmetric solutions}
\label{sec:sphere-solutions}

As we have already mentioned, it is not possible to have spherically
symmetric solutions with the Dirac equation.  However, we will now
show that one can have solutions that are compatible with spherical
symmetry, in the sense of having both a conserved current and a
stress--energy tensor that are spherically symmetric, if we add the
two particular solutions that we found in the last section with the
same amplitude, but considering them as independent fields.  We start
from the conserved current, which is given by:
\begin{equation}
j_\mu^{\rm Tot} = (j_\mu)_{+} + (j_\mu)_{-} = \bar{\psi}_{+} \gamma_\mu \psi_{+}
+ \bar{\psi}_{-} \gamma_\mu \psi_{-} \; ,
\end{equation}
with $\psi_\pm$ the spinors given by~\eqref{eq:TrialSolutionsCurv}.
For the particle density we find, using~\eqref{eq:rhop-CurvSpherical}:
\begin{equation}
\rho_p^{\rm Tot} = \frac{1}{2 \pi} \: \left( |F|^2 + |G|^2 \right) \; .
\label{eq:ParticleDensity-CurvFinal}
\end{equation}
On the other hand, the flux of particles in the radial direction now
takes the form, from~\eqref{eq:fr-CurvSpherical}:
\begin{equation}
f_r^{\rm Tot} = \frac{a}{2 \pi} \: \left( F G^* + G F^* \right) \; .
\label{eq:ParticleFlux-CurvFinal}
\end{equation}
Since both $F$ and $G$ depend only on $(t,r)$ we see that $\rho_p^{\rm
  Tot}$ and $f_r^{\rm Tot}$ are clearly spherically symmetric.  Notice
in particular that from the expansions for small $r$ we found above
for $F$ and $G$, it is clear that we will also have $f_r \sim r$ close
to the origin.

Consider now the particle flux in the $\theta$ direction.  It is not
difficult to see that equation~\eqref{eq:jT-CurvSpherical} immediately
implies ${j_\theta}_{+} = {j_\theta}_{-}=0$, so that we clearly have
$f_\theta^{\rm Tot} = j_\theta^{\rm Tot} = 0$.  Finally, for the flux
in the $\varphi$ direction we use
equation~\eqref{eq:jP-CurvSpherical}.  We now find that
${j_\varphi}_{+} \neq 0$ and ${j_\varphi}_{-} \neq 0$, but crucially
${j_\varphi}_{-} = - {j_\varphi}_{+}$ so that we have $f_\varphi^{\rm
  Tot} = j_\varphi^{\rm Tot} = 0$.  We then see that the conserved
current has both angular components equal to zero, so that it is
indeed compatible with spherical symmetry.

\vspace{5mm}

Let us now consider the total stress--energy tensor:
\begin {equation}
T_{\mu \nu}^{\rm Tot} = {T_{\mu \nu}}_+ + {T_{\mu \nu}}_- \; .
\end{equation}
For the energy density we find, using~\eqref{eq:rhoE-CurvGen}:
\begin{equation}
\rho_E^{\rm Tot} = \frac{i}{4 \pi} \left( F^* \tilde{\Pi}_F + G^* \tilde{\Pi}_G - c.c. \right) \; ,
\label{eq:rhoE-CurvFinal0}
\end{equation}
with:
\begin{equation}
\tilde{\Pi}_F := \frac{1}{\alpha} \left( \partial_t F - \beta^r \partial_r F \right) \; , \qquad
\tilde{\Pi}_G := \frac{1}{\alpha} \left( \partial_t G - \beta^r \partial_r G \right) \; .
\end{equation}
The previous result for the total energy density is written in a very
compact form, but it is convenient to rewrite it using the definitions
of $\tilde{\Pi}_F$ and $\tilde{\Pi}_G$ and the evolution
equations~\eqref{eq:Fdot-Curv} y~\eqref{eq:Gdot-Curv}. Doing this we
obtain an equivalent, thought somewhat larger expression, that does
not involve time derivatives:
\begin{equation}
\rho_E^{\rm Tot} = \frac{1}{2 \pi} \left[ {\rm Im} \left( \frac{1}{a}
\left( F^* \partial_r G + G^* \partial_r F \right)
+ \frac{2}{rb} \: F^* G \right) + m \left( |F|^2 - |G|^2 \right) \right] \; ,
\label{eq:rhoE-CurvFinal}
\end{equation}
where here ${\rm Im}(q)$ indicates the imaginary part of $q$ in the
sense that, if $q=a + ib$ with both $a$ and $b$ real, then ${\rm Im}(q)
\equiv b$.  The energy density is then purely real, as expected.

On the other hand, for the total momentum density in the radial
direction we find, from~\eqref{eq:Jr-CurvGen}:
\begin{align}
J_r^{\rm Tot} &= \frac{1}{4 \pi} \: {\rm Im} \left[  F^* \partial_r F + G^* \partial_r G
  - a \left( F^* \tilde{\Pi}_G + G^* \tilde{\Pi}_F \right) \right] \nonumber \\
&= \frac{1}{2 \pi} \: {\rm Im} \left[ F^* \partial_r F + G^* \partial_r G \right] \; ,
\label{eq:Jr-CurvFinal}
\end{align}
where in the last step we substituted the definitions of
$\tilde{\Pi}_F$ and $\tilde{\Pi}_G$, and used again the evolutions
equations~\eqref{eq:Fdot-Curv} and~\eqref{eq:Gdot-Curv}.  The
calculation for the angular components of the momentum density is
longer, but after some algebra one finds that both vanish,
$J_\theta^{\rm Tot}=J_\varphi^{\rm Tot}=0$. Again, it interesting to
notice that $J_\theta$ in fact vanishes for both individual solutions,
\mbox{${J_\theta}_+ = {J_\theta}_- = 0$}, while $J_\varphi$ is
non-zero for each individual solution but the sum vanishes.

Finally, for the diagonal components of the spatial stress tensor
$S_{ij}$ we find, from
equations~\eqref{eq:Srr-CurvGen}-\eqref{eq:Spp-CurvGen}:
\begin{align}
S_{rr}^{\rm Tot} &= \frac{a}{2 \pi} \: {\rm Im} \left[ F^* \partial_r G
  + G^* \partial_r F \right] \; , \label{eq:S11-CurvFinal} \\
S_{\theta \theta}^{\rm Tot} &= \frac{r b}{2 \pi} \: {\rm Im} \left[ F^* G \right] \; ,
\label{eq:S22-CurvFinal} \\
S_{\varphi \varphi}^{\rm Tot} &= \left( \sin^2 \theta \right) S_{\theta \theta}^{\rm Tot} \; .
\label{eq:S33-CurvFinal}
\end{align}
All off-diagonal components of $S_{ij}^{\rm Tot}$ now vanish. We then
conclude that the total stress-energy tensor is indeed compatible with
spherical symmetry.

In particular, the total trace of $T_{\mu \nu}$ turns out to be:
\begin{equation}
({T^\mu}_\mu)^{\rm Tot} = ({S^i}_i)^{\rm Tot} - \rho_E^{\rm Tot}
= ({S^r}_r)^{\rm Tot} + 2 \: ({S^\theta}_\theta)^{\rm Tot} - \rho_E^{\rm Tot}
= - \frac{m}{2 \pi} \left( |F|^2 - |G|^2 \right) \; ,
\end{equation}
in complete agreement with
equation~\eqref{eq:Dirac-stressenergy-trace}.  The extra factor of
$1/2 \pi$ comes from our normalization of the spinors (see
equation~\eqref{eq:TrialSolutionsCurv}), and from the fact that we now
have two spinors with equal amplitude.


\subsection{Dirac stars}

As a particular example of a Dirac field in spherical symmetry we will
consider the so-called Dirac stars, which are self-gravitating
stationary solutions of the Einstein--Dirac equations, analogous to
the usual boson stars for the case of the Klein--Gordon field (see
reference~\cite{Liebling:2012fv} for a very complete review of boson
stars and their relatives).  Dirac stars have been previously studied
in some detail for example
in~\cite{Herdeiro:2017,Sun2024a,Liang2024a}. Because of this, here we
will only consider the basic equations describing the system and will
not discuss any particular family of solutions.

We start from a spacetime metric in spherical symmetry of the form:
\begin{equation}
ds^2 = - \alpha^2 dt^2 + a^2 dr^2 + r^2 d \Omega^2 \; ,
\label{eq:metricDiracStar}
\end{equation}
where now $\alpha$ and $a$ and only functions of the radial coordinate
$r$.  In terms of our general spherically symmetric
metric~\eqref{eq:3+1metric} we are the taking $\beta^r=0$ and $b=1$,
so that we assume that our radial coordinate is the areal radius.

For the Dirac field we will use the spherically symmetric formalism we
developed in the previous sections, so that we take a solution of the
form $\psi = \psi_+ + \psi_-$, with the spinors $\psi_\pm$ given by
equation~\eqref{eq:TrialSolutionsCurv}:
\begin{equation}
\psi_\pm = \frac{e^{\pm i \varphi / 2}}{(4 \pi)^{1/2}}
\left(
\begin{array}{r}
F(t,r) \: y_\pm (\theta) \\
\pm i F(t,r) \: y_\mp (\theta) \\
G(t,r) \: y_\pm (\theta) \\
\mp i G(t,r) \: y_\mp (\theta)
\end{array}
\right) \; .
\end{equation}

We have already shown that the total spinor $\psi$ is compatible with
spherical symmetry in the sense that both the total conserved current
$j^\mu$ and the total stress-energy tensor $T_{\mu \nu}$ maintain that
symmetry.  But if we now want to have a static solution we must also
ask for $j^\mu$ and $T_{\mu \nu}$ to be time independent, and for the
associated flux of particles and momentum density to vanish.  In order
to achieve this we introduce an ansatz with a harmonic time dependence
for the functions $F$ and $G$ that define our spinors:
\begin{equation}
F(r,t) = f(r) e^{-i \omega t} \; , \qquad
G(r,t) = i g(r) e^{-i \omega t} \; ,
\label{eq:harmonic-ansatz}
\end{equation}
where now both $f(r)$ and $g(r)$ are purely real functions. It is now
easy to see that with this ansatz both the conserved current and the
stress-energy tensor are time independent.  The minus sign in the
exponential comes from the fact that the energy operator is given by
$\hat{E} = i \partial_t$ (remember that we are working in Planck
units), so the sign guarantees that we will have positive energy
solutions for $\omega > 0$ (see below).

For the particle density and flux we find, from
equations~\eqref{eq:ParticleDensity-CurvFinal}
and~\eqref{eq:ParticleFlux-CurvFinal}:
\begin{equation}
\rho_p = \frac{1}{2 \pi} \: \left( f^2 + g^2 \right) \; , \qquad
f_r = 0 \; .  
\end{equation}
Notice that the particle flux vanishes, as expected for a static
solution.

On the other hand for the different components of the stress-energy
tensor we find, from equations~\eqref{eq:rhoE-CurvFinal},
\eqref{eq:Jr-CurvFinal}, \eqref{eq:S11-CurvFinal}
and~\eqref{eq:S22-CurvFinal}:
\begin{align}
\rho_E &= \frac{1}{2 \pi} \left[ \frac{1}{a} \left( f g' - f' g \right)
+ \frac{2 f g}{r} + m \left( f^2 - g^2 \right) \right] \; ,
\label{eq:DiracStar-rhoE} \\
J_r &= 0 \; , \\
{S^r}_r &= \frac{1}{2 \pi a} \left( f g' - f' g \right) \; ,
\label{eq:DiracStar-S11} \\
{S^\theta}_\theta &= {S^\varphi}_\varphi = \frac{fg}{2 \pi r} \; ,
\label{eq:DiracStar-S22}
\end{align}
where the prime denotes derivatives with respect to $r$. Notice again
how the momentum density $J_r$ vanishes, as expected for a static
solution.  We also find for the trace pf $T_{\mu \nu}$:
\begin{equation}
{T^\mu}_\mu = - \rho_E + {S^r}_r + 2 {S^\theta}_\theta
= - \frac{m}{2 \pi} \left( f^2 - g^2 \right) \; ,
\end{equation}
consistent with equation~\eqref{eq:Dirac-stressenergy-trace}.

Here one should notice that, although the expression for $\rho_E$ above
is correct, we can in fact find an equivalent more compact expression
using equation~\eqref{eq:rhoE-CurvFinal0}, where in this case we have
from our ansatz $\Pi_F = \partial_t F / \alpha$ and $\Pi_G =
\partial_t G / \alpha$.  We then find:
\begin{equation}
\rho_E = \frac{\omega}{2 \pi \alpha} \left( f^2 + g^2 \right) \; .
\label{eq:DiracStar-rhoE-2}
\end{equation}
Notice that we will clearly have $\rho_E > 0$ for $\omega >0$.

\vspace{5mm}

The next step is to find the equations that must be satisfied by the
stationary solution. Substituting our ansatz for the metric and the
functions $F$ and $G$ into equations~\eqref{eq:Fdot-Curv}
and~\eqref{eq:Gdot-Curv} we find:
\begin{align}
\omega f &= + \frac{\alpha}{a} \left[ g' + g \left( \frac{\alpha'}{2 \alpha}
+ \frac{1}{r} (1 + a) \right) \right] + \alpha m f \; , \\
\omega g &= - \frac{\alpha}{a} \left[ f' + f \left( \frac{\alpha'}{2 \alpha}
+ \frac{1}{r} (1 - a) \right) \right] - \alpha m g \; ,
\end{align}
where we also used the fact that for a static spacetime the extrinsic
curvature vanishes, so that $K=0$. Solving for $f'$ and $g'$ we
obtain:
\begin{align}
f' &= - f \left( \frac{\alpha'}{2 \alpha} + \frac{1}{r} (1 - a) \right)
- a g \left( m + \frac{\omega}{\alpha} \right) \; , \label{eq:f-prime} \\
g' &= - g \left( \frac{\alpha'}{2 \alpha} + \frac{1}{r} (1 + a) \right)
- a f \left( m - \frac{\omega}{\alpha} \right) \; . \label{eq:g-prime}
\end{align}

We also need equations for the metric functions $a$ and $\alpha$.  The
equation for the radial metric $a$ is obtained directly from the
Hamiltonian constraint. On the other hand, the equation for the lapse
function $\alpha$ is obtained from the so-called polar-areal gauge,
which corresponds to asking for the time derivative of the angular
component of the extrinsic curvature $K_{\theta \theta}$ to vanish.
We will not write down the general expressions for the Hamiltonian
constraint and the polar-areal gauge condition here since they are
well known and can be found in text books (see
e.g.~\cite{Alcubierre08a}).  In our case these two conditions reduce
to:
\begin{equation}
a' = \frac{a}{2} \left( \frac{1-a^2}{r} + 8 \pi r a^2 \rho_E \right) \; .
\label{eq:a-prime}
\end{equation}
and:
\begin{equation}
\alpha' = \alpha \left( \frac{a^2-1}{2r} + 4 \pi r a^2 {S^r}_r \right) \; .
\label{eq:alpha-prime}
\end{equation}

The final system of equations to be solved for the functions $(a,\alpha,f,g)$
is then:
\begin{align}
\partial_r a &= \frac{a}{2} \left( \frac{1-a^2}{r} + 8 \pi r a^2 \rho_E \right) \; ,
\label{eq:dr-a} \\
\partial_r \alpha &= \alpha \left( \frac{a^2-1}{2r} + 4 \pi r a^2 {S^r}_r \right) \; ,
\label{eq:dr-alpha} \\
\partial_r f &= - f \left( \frac{\partial_r \alpha}{2 \alpha} + \frac{1}{r} (1 - a) \right)
- a g \left( m + \frac{\omega}{\alpha} \right) \; , \label{eq:dr-f} \\
\partial_r g &= - g \left( \frac{\partial_r \alpha}{2 \alpha} + \frac{1}{r} (1 + a) \right)
- a f \left( m - \frac{\omega}{\alpha} \right) \; , \label{eq:dr-g}
\end{align}
with $\rho_E$ given by~\eqref{eq:DiracStar-rhoE-2}, and  ${S^r}_r$ given by:
\begin{equation}
{S^r}_r = \frac{1}{2 \pi a} \left( f g' - f' g \right)
= \rho_E - \frac{1}{\pi} \left( \frac{f g}{r} + \frac{m}{2} \left( f^2 - g^2 \right) \right) \; .
\end{equation}
Notice that in the equations for $\partial_r f$ and $\partial_r g$ above
there are derivatives of the lapse on the right hand side, but these can
be eliminated using~\eqref{eq:dr-alpha}.

It is important to consider the behavior of solutions of our system
of equations both at infinity and at the origin. Consider first the
limit $r \rightarrow \infty$. For asymptotically flat solutions we can
assume our spacetime is Minkowski far away, so that we must have $a
\simeq 1$, $\alpha \simeq 1$ and $1/r \rightarrow 0$. The equations for
$f$ and $g$ then reduce to:
\begin{equation}
\partial_r f \simeq - g \left( m + \omega \right) \; , \qquad
\partial_r g \simeq - f \left( m - \omega \right) \; .
\end{equation}
Taking a second derivative of the first equation, and substituting the
result in the second, we find:
\begin{equation}
\partial_r^2 f \simeq f \left( m^2 - \omega^2 \right) \; .
\end{equation}
It is now clear that if we want to have exponentially decaying
solutions at infinity we must have $m^2 > \omega^2$.  Of course, in
principle we will also have solutions that grow exponentially, which
is incompatible with having an asymptotically flat spacetime.  We will
only have decaying solutions for specific values of $\omega$, so that
we must solve an eigenvalue problem.

Consider now the behaviour of the solutions near the origin $r=0$.
Since spacetime must be locally flat there we must ask for the radial
metric component $a$ to behave as:
\begin{equation}
a \simeq 1 + {\cal O} (r^2) \; .
\end{equation}
Similarly, for the lapse function $\alpha$ we will have:
\begin{equation}
\alpha \simeq \alpha_0 + {\cal O} (r^2) \; ,
\end{equation}
with $\alpha_0$ some constant.  In principle we don't know the value
of $\alpha_0$, but notice that our system of
equations~\eqref{eq:dr-a}-\eqref{eq:dr-g} is invariant under
the rescaling:
\begin{equation}
\alpha \rightarrow k \alpha \; , \qquad \omega \rightarrow k \omega \; , 
\end{equation}
with $k$ an arbitrary constant. This means that we can simply take
$\alpha_0=1$, solve the system, and then rescale $\alpha$ and $\omega$
so that we have $\alpha \rightarrow 1$ at infinity.

On the other hand, we have already shown above that $f$ must be an even
function of $r$, while $g$ must be odd, so that we will have:
\begin{equation}
f \simeq f_0 + {\cal O} (r^2) \; , \qquad
g \simeq g_1 r + {\cal O} (r^3) \; ,
\end{equation}
with $f_0$ and $g_1$ some constants. Substituting our expansions into
the system of equations we find that at the origin we must have:
\begin{equation}
\left. \partial_r a \: \right|_{r=0} = 0 \; , \qquad
\left. \partial_r \alpha \: \right|_{r=0} = 0 \; , \qquad
\left. \partial_r f \: \right|_{r=0} = 0 \; .
\end{equation}
The condition for $g$ is more interesting due to the presence of the
term $(1+a)/r$ in its equation, which might seem to be singular at
$r=0$.  The equation, however, is in fact regular since this factor is
multiplied with $g$ which goes as $\sim r$ close to the origin. When
we substitute our expansions for small $r$ we now find:
\begin{equation}
\left. \partial_r g \: \right|_{r=0} = g_1 \; ,
\end{equation}
with $g_1 = f_0 (\omega/\alpha_0 - m)/3$, so that $g_1$ is not
independent of $f_0$.

To solve the full system of equations~\eqref{eq:dr-a}-\eqref{eq:dr-g}
one can then choose $f_0$ as our only free parameter (taking
$\alpha_0=1$), and look for solutions for which $f$ and $g$ decay
exponentially at infinity in order to find the eigenvalue $\omega$,
using a variety of numerical techniques.  For example, for a given
value of $f_0$ one can choose a trial value of $\omega$ and integrate
outward from the origin with some standard ODE integrator (for example
fourth order Runge--Kutta), and use a shooting algorithm to modify the
value of $\omega$ until one finds exponentially decaying solutions at
infinity.

By changing the value of $f_0$ one can construct a whole family of
solutions for the Dirac stars.  As mentioned above, we will not
discuss the family of solutions here, as this has already been done
before in some detail in
references~\cite{Herdeiro:2017,Sun2024a,Liang2024a}.


\section{Final remarks}

The Dirac equation is one of the most fundamental equations in
physics.  It describes the behavior of fermions such as leptons and
quarks, and is at the heart of the standard model of particles and
fields. On the other hand, general relativity is our modern theory of
gravity, and describes with great success astrophysical phenomena that
go from the structure of neutron stars, to the formation of black
holes, the emission of gravitational waves, and the evolution of the
Universe as a whole.

Though currently we do not have a successful theory of quantum gravity
(though we certainly have candidates in the form of string theory,
loop quantum gravity, dynamical causal triangulations, etc.), it is
nevertheless very important to be able to study the evolution of
quantum fields in a curved spacetime.  This was, for example, what led
Hawking to the discovery that black holes in fact radiate energy.

For the case of scalar or tensor fields, such as the Higgs or
electromagnetic fields, the generalization to a curved spacetime is
rather straightforward and follows directly from the equivalence
principle.  However, in the case of spinor fields this generalization
is not that simple, and requires the introduction of the Lorentz group
and the tetrad formalism.

Here, I have presented a pedagogical review of the Dirac equation in
the case of general relativity, starting from first principles.  Even
though I have ignored the quantization of the Dirac field and have
treated it as a purely ``classical'' field, I believe that this review
can be useful to researchers in general relativity who might not be
used to working with spinor fields.  In the last sections I have also
derived expressions for the Dirac equation and its associated
stress--energy tensor in the 3+1 formalism, and shown how this can be
applied to the special case of spherical symmetry. To my knowledge,
these last sections include new material which can be very useful for
the study of the evolution of the Dirac field in a dynamical
spacetime.


\acknowledgments

I would like to thank Axel Rangel for many useful discussions and
comments, and Eduardo Nahmad for a beautiful course on general
relativity and the tetrad formalism many years ago.  This work was
partially supported by CONAHCYT Network Projects No. 376127 and
No. 304001, and DGAPA-UNAM project IN100523.


\appendix

\section{Derivation of the stress--energy tensor for the Dirac equation}
\label{app:A}

We start by considering the variation of the action with respect to
the spacetime metric $g_{\mu \nu}$:
\begin{equation}
\delta_g S = \delta \int L \: |g|^{1/2} d^4 x \; .
\end{equation}
As we know, the Dirac Lagrangian has terms that depend directly on the
tetrad and not on the metric, so we now need to consider variations
of the tetrad itself. Notice first that from the expression for the
metric in terms of the tetrad, $g_{\mu \nu} = e_{\mu A} e_\nu^A$, we
immediately find:
\begin{equation}
\delta g_{\mu \nu} = \eta_{AB} \left( e_\mu^A \delta e_\nu^B
+ e_\nu^B \delta e_\mu^A \right) \; .
\end{equation}
The variation of the tetrad $\delta e_{\mu A}$ can be naturally
decomposed into two parts, a ``symmetric part'' $\delta^+ e_{\mu A}$
that induces variations of the metric, and an ``antisymmetric part''
$\delta^- e_{\mu A}$ that leaves the metric
unchanged~\cite{Forger:2004}:
\begin{equation}
\delta^\pm e_\mu^A = \frac{1}{2} \: \left( \delta e_\mu^A
\mp \eta^{AB} g_{\mu \nu} \delta e^\nu_B \right) \; , \qquad
\delta e_\mu^A = \delta^+ e_\mu^A + \delta^- e_\mu^A \; .
\end{equation}
From these definitions we find:
\begin{align}
\delta^\pm g_{\mu \nu} &=  \eta_{AB} \left( e_\mu^A \delta^\pm e_\nu^B
+ e_\nu^B \delta^\pm e_\mu^A \right) 
= \frac{\eta_{AB}}{2} \: \left[ e_\mu^A \left( \delta e_\nu^B
\mp \eta^{BC} g_{\nu \lambda} \delta e^\lambda_C \right)
+ e_\nu^B \left( \delta e_\mu^A \mp \eta^{AC} g_{\mu \lambda} \delta e^\lambda_C \right) \right]
\nonumber \\
&= \frac{\eta_{AB}}{2} \: \left[ e_\mu^A \delta e_\nu^B + e_\nu^B \delta e_\mu^A \right]
\mp \frac{1}{2} \: \left[ g_{\nu \lambda} e_\mu^A \delta e^\lambda_A
+ g_{\mu \lambda} e_\nu^A  \delta e^\lambda_A \right] \nonumber \\
&= \frac{\delta g_{\mu \nu}}{2} \pm \frac{1}{2} \: \left[ g_{\nu \lambda} e^\lambda_A \delta e_\mu^A
+ g_{\mu \lambda} e^\lambda_A \delta e_\nu^A \right] \nonumber \\
&= \frac{\delta g_{\mu \nu}}{2} \pm \frac{1}{2} \: \left[ e_{\nu A} \delta e_\mu^A
+ e_{\mu A} \delta e_\nu^A \right]
= \frac{\delta g_{\mu \nu}}{2} \pm \frac{\delta g_{\mu \nu}}{2} \: ,
\end{align}
where we used the fact that $e_\mu^A e^\nu_A = \delta_\mu^\nu$ implies
$e_\mu^A \delta e^\nu_A = - e^\nu_A \delta e_\mu^A$.  We then find:
\begin{equation}
\delta^+ g_{\mu \nu} = \delta g_{\mu \nu} \; , \qquad \delta^- g_{\mu \nu} = 0 \; .
\end{equation}
We can also show that:
\begin{align}
\eta^{AB} e^\nu_B \delta g_{\mu \nu}
&= \eta^{AB} e^\nu_B \left( e_\nu^C \delta e_{\mu C} + e_{\mu C} \delta e_\nu^C \right)
= \delta e_\mu^A + \eta^{AB} e_{\mu C} e^\nu_B \delta e_\nu^C \nonumber \\
&= \delta e_\mu^A - \eta^{AB} e_{\mu C} e_\nu^C \delta e^\nu_B
= \delta e_\mu^A - \eta^{AB} g_{\mu \nu} \delta e^\nu_B = 2 \delta^+ e_\mu^A \; ,
\end{align}
so that we can express $\delta^+ e_\mu^A$ entirely in terms of $\delta
g_{\mu \nu}$ as:
\begin{equation}
\delta^+ e_\mu^A = \frac{1}{2} \: \eta^{AB} e^\nu_B \delta g_{\mu \nu}
= \frac{1}{2} \: e^{\nu A}  \delta g_{\mu \nu} \; .
\end{equation}
And similarly:
\begin{equation}
\delta^+ e^\mu_A = \frac{1}{2} \: e_{\nu A} \delta g^{\mu \nu}
= - \frac{1}{2} \: g^{\mu \alpha} e^\beta_A  \delta g_{\alpha \beta} \; .
\end{equation}

Now, the stress-energy tensor is defined in terms of the variation of
the action integral with respect to changes in the metric as:
\begin{equation}
\delta_g S = \frac{1}{2} \int T^{\alpha \beta} \delta g_{\alpha \beta} |g|^{1/2} d^4 x \; .
\label{eq:Tab-definition}
\end{equation}
This means that if the Lagrangian is expressed in terms of the tetrad,
as is the case of the Dirac Lagrangian, we must only consider the
changes in the tetrad that modify the metric, that is we should only
consider $\delta^+ e^\mu_A$.  The variation in the action will then
take the form:
\begin{equation}
\delta_g S = \int \left[ |g|^{1/2} \left( \frac{\delta L}{\delta e^\mu_A}
\right) \delta^+ e^\mu _A + L \: \delta |g|^{1/2} \right] d^4 x \; .
\end{equation}
For the second term we have:
\begin{equation}
L \: \delta |g|^{1/2} = \frac{L}{2 |g|^{1/2}} \: \delta |g|
= \frac{L}{2 |g|^{1/2}} \: |g| g^{\alpha \beta} \delta g_{\alpha \beta}
= \frac{L}{2} \: |g|^{1/2} g^{\alpha \beta} \delta g_{\alpha \beta} \; .
\end{equation}
On the other hand, for the first term we find:
\begin{equation}
|g|^{1/2} \left( \frac{\delta L}{\delta e^\mu_A} \right) \delta^+ e^\mu _A
= - \frac{1}{2} \: |g|^{1/2} \left( \frac{\delta L}{\delta e^\mu_A} \right)
g^{\mu \alpha} e^\beta_A \delta g_{\alpha \beta}
= - \frac{1}{4} \: |g|^{1/2} \left( g^{\mu \alpha} e^\beta_A \frac{\delta L}{\delta e^\mu_A}
+  g^{\mu \beta} e^\alpha_A \frac{\delta L}{\delta e^\mu_A} \right) \delta g_{\alpha \beta} \; .
\end{equation}
The variation of the action then becomes:
\begin{equation}
\delta_g S = \frac{1}{2} \: \int \left[ - \frac{1}{2} \:
\left( g^{\mu \alpha} e^\beta_A \frac{\delta L}{\delta e^\mu_A}
+  g^{\mu \beta} e^\alpha_A \frac{\delta L}{\delta e^\mu_A} \right)
+ g^{\alpha \beta} L \right] \delta g_{\alpha \beta} \: |g|^{1/2} d^4 x \; .
\end{equation}
Comparing this result with~\eqref{eq:Tab-definition} we find for the
stress--energy tensor:
\begin{equation}
T^{\alpha \beta} = - \frac{1}{2} \:
\left( g^{\mu \alpha} e^\beta_A \frac{\delta L}{\delta e^\mu_A}
+  g^{\mu \beta} e^\alpha_A \frac{\delta L}{\delta e^\mu_A} \right)
+ g^{\alpha \beta} L \; ,
\end{equation}
and lowering the indices we recover the expression for the
stress-energy tensor given in~\eqref{eq:Tmunu-tetrad0}:
\begin{equation}
T_{\mu \nu} = - \frac{1}{2} \left( e_{\mu D} \: \frac{\delta L}{\delta e_D^\nu}
+ e_{\nu D} \: \frac{\delta L}{\delta e_D^\mu} \right) + g_{\mu \nu} L \; .
\end{equation}

One should stress the fact that at this point we are still not
replacing the variations with partial derivatives since the Lagrangian
can depend also on derivatives of the tetrad (see below).

Notice now that, since the Dirac
Lagrangian~\eqref{eq:Lagrangian-Dirac} vanishes on shell, for case of
the Dirac field we can in fact ignore the last term in the above
expression. Moreover, the mass term that appears in the Dirac
Lagrangian is independent of the tetrad, so the stress-energy tensor
reduces to:
\begin{equation}
T_{\mu \nu} = - \frac{1}{2} \left( e_{\mu D} \: \frac{\delta K}{\delta e_D^\nu}
+ e_{\nu D} \: \frac{\delta K}{\delta e_D^\mu} \right) \; ,
\label{eq:Tmunu-tetrad2}
\end{equation}
with $K$ the kinetic term in the Lagrangian. The kinetic term can further
be split into two parts, $K = K_1 + K_2$, given by:
\begin{align}
K_1 &= \frac{i}{2} \left[ e^\beta_A \: \bar{\psi} \gamma^A
\left( \partial_\beta \psi \right) - e^\alpha_A \left( \partial_\alpha \bar{\psi} \right)
\gamma^A \psi \right] \; , \\
K_2 & = \frac{i}{4} \: \bar{\psi} \left( e_A^\alpha e_C^\beta \: \partial_\alpha e_{\beta B}
\right) \gamma^{CAB} \psi \; .
\end{align}

Now, the variations that appear in~\eqref{eq:Tmunu-tetrad2} can be
substituted with partial derivatives when $K$ depends only on the
tetrad and not its derivatives.  This is clearly the case for $K_1$,
so its contribution is not difficult to find and turns out to be:
\begin{equation}
({T_{\mu \nu}})_1 = \frac{i}{2} \left[ \left( \partial_{(\mu} \bar{\psi} \right) \gamma_{\nu)} \psi
- \bar{\psi} \gamma_{(\mu} \left( \partial_{\nu)} \psi \right) \right] \; .
\end{equation}

The contribution from $K_2$ is somewhat more difficult to find since
it depends not only on the tetrad, but also on its derivatives. The
variation is then given by the so-called Euler derivative (this can be
shown by following the same procedure used to find the Euler--Lagrange
equations through integration by parts starting from the action
integral):
\begin{equation}
\frac{\delta K_2}{\delta e_D^\mu} = \frac{\partial K_2}{\partial e_D^\mu}
- \partial_\lambda \left( \frac{\partial K_2}{\partial ( \partial_\lambda e_D^\mu )} \right) \; .
\end{equation}

From the definition of $K_2$ we can see that the term we are
interested in is:
\begin{equation}
f_{ABC} = e_A^\alpha e_C^\beta \: \partial_\alpha e_{\beta B} \; .
\end{equation}
Using now the fact that $\partial_\alpha (e^\beta_A e_{\beta B}) =
\partial_\alpha \eta_{AB} = 0$, one can show that we have
\mbox{$\partial_\alpha e_{\beta B} = - e_\beta^D e_{\sigma B}
  \partial_\alpha e_D^\sigma$}, so that we can rewrite $f_{ABC}$ as:
\begin{equation}
f_{ABC} = - e_A^\alpha e_C^\beta \: e_\beta^D
e_{\sigma B} \partial_\alpha e_D^\sigma
= - e_{\sigma B} e_A^\alpha \partial_\alpha e^\sigma_C \; .
\end{equation}

Consider first the derivative of $f_{ABC}$ with respect
to the tetrad $e^\nu_D$:
\begin{align}
\frac{\partial f_{ABC}}{\partial e^\nu_D}
&= - \frac{\partial ( e_{\sigma B} e_A^\alpha)}{\partial e^\nu_D} \: \partial_\alpha e^\sigma_C
= - \left( e_{\sigma B} \: \frac{\partial e_A^\alpha}{\partial e^\nu_D}
+ e_A^\alpha \: \frac{\partial e_{\sigma B}}{\partial e^\nu_D} \right) \partial_\alpha e^\sigma_C
\nonumber \\
&= - \left( e_{\sigma B} \delta^\alpha_\nu \delta^D_A 
- e_A^\alpha e^D_\sigma e_{\nu B} \right) \partial_\alpha e^\sigma_C
= - e_{\sigma B} \delta^D_A \partial_\nu e^\sigma_C
+ e_\sigma^D e_{\nu B} \partial_A e^\sigma_C \; ,
\end{align}
where we used the fact that $\partial e_{\sigma B} / \partial e^\nu_D
= - e_\sigma^D e_{\nu B}$, which can be easily shown from 
$\partial (e_A^\beta e_{\beta_B}) /\partial e^\nu_D = \partial \eta_{AB}
/\partial e^\nu_D = 0$. The contribution to the stress-energy tensor
coming from this term will then be:
\begin{align}
P_{\mu \nu} &:= \frac{i}{8} \:  \bar{\psi}
\left[ e_{\mu D} \left( e_{\sigma B} \delta^D_A \partial_\nu e^\sigma_C
- e_\sigma^D e_{\nu B} \partial_A e^\sigma_C \right)
+ \mu \leftrightarrow \nu \right] \gamma^{CAB} \psi \nonumber \\
&= \frac{i}{8} \: \bar{\psi} \left[ e_{\mu A} e_{\sigma B} \partial_\nu e^\sigma_C
- g_{\mu \sigma} e_{\nu B} \partial_A e^\sigma_C + \mu \leftrightarrow \nu \right]
\gamma^{CAB} \psi \: .
\end{align}
In fact, it is more convenient to project this result onto the tetrad
to find:
\begin{align}
P_{IJ} = e^\mu_I e^\nu_J \: P_{\mu \nu}
&= \frac{i}{8} \: \bar{\psi} \left[ \eta_{IA} e_{\sigma B} \partial_J e^\sigma_C
- \eta_{JB} e_{\sigma I} \partial_A e^\sigma_C + I \leftrightarrow J \right] \gamma^{CAB} \psi
\nonumber \\
&= \frac{i}{8} \: \bar{\psi} \left[ - \eta_{IA} f_{JBC} + \eta_{JB} f_{AIC}
+ I \leftrightarrow J \right] \gamma^{CAB} \psi \nonumber \\
&= - \frac{i}{8} \: \bar{\psi} \left[ \left( f_{JAB} + f_{AJB} \right) {\gamma_I}^{AB}
+ \left( f_{IAB} + f_{AIB} \right) {\gamma_J}^{AB} \right] \psi \; ,
\end{align}
where in the last step we used the symmetry properties of the
$\gamma^{ABC}$ and renamed indices.

Consider now the term associated with the derivatives of $f_{ABC}$
with respect to the derivatives of the tetrad:
\begin{align}
- \partial_\lambda \left( \frac{\partial f_{ABC}}{\partial (\partial_\lambda e^\nu_D)} \right)
&= - \partial_\lambda \left( \frac{\partial (e_{\sigma B} e^\alpha_A
\partial_\alpha e^\sigma_C )}{\partial (\partial_\lambda e^\nu_D)} \right)
= - \partial_\lambda \left( e_{\sigma B} e^\alpha_A \delta^\lambda_\alpha \delta^\sigma_\nu 
\delta^D_C \right) \nonumber \\
&= - \delta_C^D \: \partial_\lambda \left( e^\lambda_A e_{\nu B} \right)
= - \delta_C^D \left( e_A^\lambda \partial_\lambda e_{\nu B}
+ e_{\nu B} \partial_\lambda e^\lambda_A \right) \; .
\end{align}
The contribution to the stress-energy tensor coming from this term is then:
\begin{equation}
Q_{\mu \nu} := - \frac{i}{8} \: \bar{\psi}
\left[ e_{\mu C} \left( \partial_A e_{\nu B}
+ e_{\nu B} \partial_\lambda e^\lambda_A \right)
+ \mu \leftrightarrow \nu \right] \gamma^{CAB} \psi \; .
\end{equation}
Again, it is convenient to project this result onto the tetrad
to find:
\begin{align}
Q_{IJ} &= - e^\mu_I e^\nu_J \: Q_{\mu \nu} \nonumber \\
&= - \frac{i}{8} \: \bar{\psi}
\left[ \left( \eta_{IC} e^\nu_J + \eta_{JC} e^\nu_I \right) \partial_A e_{\nu B}
+ \left( \eta_{IC} \eta_{JB} + \eta_{JC} \eta_{IB} \right) \partial_\lambda e^\lambda_A
\right] \gamma^{CAB} \psi \; .
\end{align}
Notice now that the term proportional to the divergence $\partial_\lambda e^\lambda_A$
is symmetric in $C$ and $B$, but is contracted with  $\gamma^{CAB}$ which is antisymmetric,
so it cancels. We finally find:
\begin{align}
Q_{IJ} &= - \frac{i}{8} \: \bar{\psi}
\left[ \eta_{IC} e^\nu_J \partial_A e_{\nu B}
+ \eta_{JC} e^\nu_I \partial_A e_{\nu B} \right] \gamma^{CAB} \psi
\nonumber \\
&= - \frac{i}{8} \:  \bar{\psi}
\left[ e^\nu_J \partial_A e_{\nu B} \: {\gamma_I}^{AB}
+ e^\nu_I \partial_A e_{\nu B} \: {\gamma_J}^{AB} \right] \psi \nonumber \\
&= - \frac{i}{8} \: \bar{\psi} \left[ f_{ABJ} \: {\gamma_I}^{AB}
+ f_{ABI} \: {\gamma_J}^{AB} \right] \psi \; .
\end{align}
Adding both contributions coming from $K_2$ we obtain:
\begin{align}
P_{IJ} + Q_{IJ}
&= - \frac{i}{8} \: \bar{\psi} \left[ \left( f_{ABJ} + f_{JAB} + f_{AJB} \right) {\gamma_I}^{AB}
+ I \leftrightarrow J \right] \psi \nonumber \\
&= \frac{i}{8} \: \bar{\psi} \left[ \omega_{ABJ} {\gamma_I}^{AB}
+ \omega_{ABI} {\gamma_J}^{AB} \right] \psi \; .
\end{align}
And going back to spacetime indices:
\begin{align}
({T_{\mu \nu}})_2
&= \frac{i}{8} \: \bar{\psi} \left[ \omega_{AB \mu} {\gamma_\nu}^{AB}
+ \omega_{AB \nu} {\gamma_\mu}^{AB} \right] \psi \nonumber \\
&= \frac{i}{8} \: \bar{\psi} \left[ \omega_{AB \mu} \left\{ \gamma_\nu , \sigma^{AB} \right\}
+ \omega_{AB \nu} \left\{ \gamma_\mu , \sigma^{AB} \right\} \right] \psi \nonumber \\
&= \frac{i}{8} \: \bar{\psi} \left[ \left\{ \gamma_\nu , \omega_{AB \mu} \sigma^{AB} \right\}
+ \left\{ \gamma_\mu , \omega_{AB \nu} \sigma^{AB} \right\} \right] \psi \nonumber \\
&= - \frac{i}{4} \: \bar{\psi} \left[ \left\{ \gamma_\nu , \Gamma_\mu \right\}
+ \left\{ \gamma_\mu , \Gamma_\nu \right\} \right] \psi 
= - \frac{i}{2} \: \bar{\psi} \left\{ \gamma_{(\mu} , \Gamma_{\nu)} \right\} \psi \; .
\end{align}

Adding now the contributions to the stress-energy tensor from $K_1$
and $K_2$ we find:
\begin{align}
T_{\mu \nu} = ({T_{\mu \nu}})_1 + ({T_{\mu \nu}})_2
&= \frac{i}{2} \: \left[ \left( \partial_{(\mu} \bar{\psi} \right) \gamma_{\nu)} \psi
- \bar{\psi} \gamma_{(\mu} \left( \partial_{\nu)} \psi \right)
- \bar{\psi} \left\{ \gamma_{(\mu} , \Gamma_{\nu)} \right\} \psi \right] \nonumber \\
&=  \frac{i}{2} \: \left[ \left( \partial_{(\mu} \bar{\psi}
- \bar{\psi} \Gamma_{(\mu}\right) \gamma_{\nu)} \psi
- \bar{\psi} \gamma_{(\mu} \left( \partial_{\nu)} \psi + \Gamma_{\nu)} \psi \right)
\right]  \; ,
\end{align}
and finally:
\begin{equation}
T_{\mu \nu}
= \frac{i}{2} \left[ \left( \mathcal{D}_{(\mu} \bar{\psi} \right) \gamma_{\nu)} \psi
- \bar{\psi} \gamma_{(\mu} \left( \mathcal{D}_{\nu)} \psi \right) \right] \; .
\label{eq:Dirac-stressenergy-app}
\end{equation}
This is the final form of the stress-energy tensor for the Dirac
field.  Notice that there is no explicit contribution from the mass
term in this tensor, which might seem strange at first glance, but
such a contribution is implicitly there since $\psi$ must satisfy
Dirac's equation.

\vspace{5mm}

We still need to show that the stress-energy tensor we just found does
in fact satisfy the conservation equations $\nabla^\mu T_{\mu \nu} =
0$. Now, since this tensor involves spinors, and for tensors the
spinor derivative reduces to the covariant derivative, what we must
show is that we have $\mathcal{D}^\mu T_{\mu \nu} = 0$.  Substituting
the expression for $T_{\mu \nu}$ given
in~\eqref{eq:Dirac-stressenergy}, and ignoring constant factors, we
have:
\begin{align}
\mathcal{D}^\mu T_{\mu \nu} &\propto
\mathcal{D}^\mu \left[ \left( \mathcal{D}_\mu \bar{\psi} \right) \gamma_\nu \psi
+ \left( \mathcal{D}_\nu \bar{\psi} \right) \gamma_\mu \psi
- \bar{\psi} \gamma_\mu \left( \mathcal{D}_\nu \psi \right)
- \bar{\psi} \gamma_\nu \left( \mathcal{D}_\mu \psi \right) \right] \nonumber \\
&= \left( \mathcal{D}^\mu \mathcal{D}_\mu \bar{\psi} \right) \gamma_\nu \psi
+ \left( \mathcal{D}_\mu \bar{\psi} \right) \gamma_\nu \left( \mathcal{D}^\mu \psi \right)
+ \left( \mathcal{D}^\mu \mathcal{D}_\nu \bar{\psi} \right) \gamma_\mu \psi
+ \left( \mathcal{D}_\nu \bar{\psi} \right) \gamma_\mu \left( \mathcal{D}^\mu \psi \right)
\nonumber \\
& \quad - \left( \mathcal{D}^\mu \bar{\psi} \right) \gamma_\mu \left( \mathcal{D}_\nu \psi \right)
- \bar{\psi} \gamma_\mu \left( \mathcal{D}^\mu \mathcal{D}_\nu \psi \right)
- \left( \mathcal{D}^\mu \bar{\psi} \right) \gamma_\nu \left( \mathcal{D}_\mu \psi \right)
- \bar{\psi} \gamma_\nu \left( \mathcal{D}^\mu \mathcal{D}_\mu \psi \right)
\nonumber \\
&= \left( \mathcal{D}^\mu \mathcal{D}_\mu \bar{\psi} \right) \gamma_\nu \psi
+ \left( \mathcal{D}^\mu \mathcal{D}_\nu \bar{\psi} \right) \gamma_\mu \psi
+ \left( \mathcal{D}_\nu \bar{\psi} \right) \gamma_\mu \left( \mathcal{D}^\mu \psi \right)
\nonumber \\
& \quad - \left( \mathcal{D}^\mu \bar{\psi} \right) \gamma_\mu \left( \mathcal{D}_\nu \psi \right)
- \bar{\psi} \gamma_\mu \left( \mathcal{D}^\mu \mathcal{D}_\nu \psi \right)
- \bar{\psi} \gamma_\nu \left( \mathcal{D}^\mu \mathcal{D}_\mu \psi \right) \; ,
\end{align}
where we used the fact that $\mathcal{D}_\mu \gamma^\nu = 0$, and in
the last step we cancelled two terms that were clearly equal. In the
first and last terms of the previous expression we can now use the
Schroedinger--Dirac equation~\eqref{eq:Schroedinger-Dirac} for $\psi$
and $\bar{\psi}$ to show that those two terms again cancel.  On the
other hand, the third and fourth terms can be simplified using the
Dirac equation to find:
\begin{align}
\left( \mathcal{D}_\nu \bar{\psi} \right) \gamma_\mu \left( \mathcal{D}^\mu \psi \right)
- \left( \mathcal{D}^\mu \bar{\psi} \right) \gamma_\mu \left( \mathcal{D}_\nu \psi \right)
&= - i m \left[ \left( \mathcal{D}_\nu \bar{\psi} \right) \psi 
+ \bar{\psi} \left( \mathcal{D}_\nu \psi \right) \right] \nonumber \\
&= - i m \mathcal{D}_\nu \left( \bar{\psi} \psi \right) .
\end{align}
We then have:
\begin{equation}
\mathcal{D}^\mu T_{\mu \nu} \propto
\left( \mathcal{D}^\mu \mathcal{D}_\nu \bar{\psi} \right) \gamma_\mu \psi
- \bar{\psi} \gamma_\mu \left( \mathcal{D}^\mu \mathcal{D}_\nu \psi \right)
- i m \mathcal{D}_\nu \left( \bar{\psi} \psi \right) \; .
\end{equation}
For the first two terms in the previous expression we can now use
the commutation relation for the spinor derivatives~\eqref{eq:Ricci-identity-spinor}:
\begin{align}
&\left( \mathcal{D}^\mu \mathcal{D}_\nu \bar{\psi} \right) \gamma_\mu \psi
- \bar{\psi} \gamma_\mu \left( \mathcal{D}^\mu \mathcal{D}_\nu \psi \right)
= g^{\mu \lambda} \left[ \left( \mathcal{D}_\lambda \mathcal{D}_\nu \bar{\psi} \right) \gamma_\mu \psi
- \bar{\psi} \gamma_\mu \left( \mathcal{D}_\lambda \mathcal{D}_\nu \psi \right) \right] \nonumber \\
& \qquad = \left( \mathcal{D}_\nu \mathcal{D}_\lambda \bar{\psi}
- \frac{1}{2} \: R_{AB \lambda \nu} \sigma^{AB} \bar{\psi} \right) \gamma^\lambda \psi
- \bar{\psi} \gamma^\lambda \left( \mathcal{D}_\nu \mathcal{D}_\lambda \psi
- \frac{1}{2} \: R_{AB \lambda \nu} \sigma^{AB} \psi \right) \nonumber \\
& \qquad = \mathcal{D}_\nu \left[ \left( \mathcal{D}_\lambda \bar{\psi} \right) \gamma^\lambda \right] \psi
- \bar{\psi} \mathcal{D}_\nu \left( \gamma^\lambda \mathcal{D}_\lambda \psi \right) \nonumber \\
& \qquad= im \left[ \left( \mathcal{D}_\nu \bar{\psi} \right) \psi
+ \bar{\psi} \left( \mathcal{D}_\nu \psi \right) \right] \nonumber \\
& \qquad = im \mathcal{D}_\nu \left( \bar{\psi} \psi \right) \; ,
\end{align}
where we again used Dirac's equation. We finally find:
\begin{equation}
\mathcal{D}^\mu T_{\mu \nu} = 0 \; .
\end{equation}
We then see that the stress-energy tensor~\eqref{eq:Dirac-stressenergy}
does indeed satisfy the conservation laws.


\bibliographystyle{apsrev4-2}
\bibliography{bibtex/referencias}

\begin{thebibliography}{31}%
\makeatletter
\providecommand \@ifxundefined [1]{%
 \@ifx{#1\undefined}
}%
\providecommand \@ifnum [1]{%
 \ifnum #1\expandafter \@firstoftwo
 \else \expandafter \@secondoftwo
 \fi
}%
\providecommand \@ifx [1]{%
 \ifx #1\expandafter \@firstoftwo
 \else \expandafter \@secondoftwo
 \fi
}%
\providecommand \natexlab [1]{#1}%
\providecommand \enquote  [1]{``#1''}%
\providecommand \bibnamefont  [1]{#1}%
\providecommand \bibfnamefont [1]{#1}%
\providecommand \citenamefont [1]{#1}%
\providecommand \href@noop [0]{\@secondoftwo}%
\providecommand \href [0]{\begingroup \@sanitize@url \@href}%
\providecommand \@href[1]{\@@startlink{#1}\@@href}%
\providecommand \@@href[1]{\endgroup#1\@@endlink}%
\providecommand \@sanitize@url [0]{\catcode `\\12\catcode `\$12\catcode
  `\&12\catcode `\#12\catcode `\^12\catcode `\_12\catcode `\%12\relax}%
\providecommand \@@startlink[1]{}%
\providecommand \@@endlink[0]{}%
\providecommand \url  [0]{\begingroup\@sanitize@url \@url }%
\providecommand \@url [1]{\endgroup\@href {#1}{\urlprefix }}%
\providecommand \urlprefix  [0]{URL }%
\providecommand \Eprint [0]{\href }%
\providecommand \doibase [0]{https://doi.org/}%
\providecommand \selectlanguage [0]{\@gobble}%
\providecommand \bibinfo  [0]{\@secondoftwo}%
\providecommand \bibfield  [0]{\@secondoftwo}%
\providecommand \translation [1]{[#1]}%
\providecommand \BibitemOpen [0]{}%
\providecommand \bibitemStop [0]{}%
\providecommand \bibitemNoStop [0]{.\EOS\space}%
\providecommand \EOS [0]{\spacefactor3000\relax}%
\providecommand \BibitemShut  [1]{\csname bibitem#1\endcsname}%
\let\auto@bib@innerbib\@empty
\bibitem [{\citenamefont {Dirac}(1928)}]{Dirac:1928}%
  \BibitemOpen
  \bibfield  {author} {\bibinfo {author} {\bibfnamefont {P.~A.~M.}\
  \bibnamefont {Dirac}},\ }\href@noop {} {\bibfield  {journal} {\bibinfo
  {journal} {Proceedings of the Royal Society of London}\ }\textbf {\bibinfo
  {volume} {117}},\ \bibinfo {pages} {610} (\bibinfo {year}
  {1928})}\BibitemShut {NoStop}%
\bibitem [{\citenamefont {Fock}(1929)}]{Fock29a}%
  \BibitemOpen
  \bibfield  {author} {\bibinfo {author} {\bibfnamefont {L.~H.}\ \bibnamefont
  {Fock}},\ }\href@noop {} {\bibfield  {journal} {\bibinfo  {journal} {Z.
  Phys.}\ }\textbf {\bibinfo {volume} {57}},\ \bibinfo {pages} {261} (\bibinfo
  {year} {1929})}\BibitemShut {NoStop}%
\bibitem [{\citenamefont {Fock}\ and\ \citenamefont
  {Ivanenko}(1929)}]{Fock29b}%
  \BibitemOpen
  \bibfield  {author} {\bibinfo {author} {\bibfnamefont {L.~H.}\ \bibnamefont
  {Fock}}\ and\ \bibinfo {author} {\bibfnamefont {D.}~\bibnamefont
  {Ivanenko}},\ }\href@noop {} {\bibfield  {journal} {\bibinfo  {journal} {Z.
  Phys.}\ }\textbf {\bibinfo {volume} {54}},\ \bibinfo {pages} {798} (\bibinfo
  {year} {1929})}\BibitemShut {NoStop}%
\bibitem [{\citenamefont {Bargmann}(1932)}]{Bargmann:1932}%
  \BibitemOpen
  \bibfield  {author} {\bibinfo {author} {\bibfnamefont {V.}~\bibnamefont
  {Bargmann}},\ }\href@noop {} {\bibfield  {journal} {\bibinfo  {journal}
  {Sitzsungsber. Preuss. Akad. Wiss. Phys.-Math. Kl.}\ }\textbf {\bibinfo
  {volume} {28}},\ \bibinfo {pages} {346} (\bibinfo {year} {1932})}\BibitemShut
  {NoStop}%
\bibitem [{\citenamefont {Schrödinger}(1932)}]{Schroedinger:1932}%
  \BibitemOpen
  \bibfield  {author} {\bibinfo {author} {\bibfnamefont {E.}~\bibnamefont
  {Schrödinger}},\ }\href@noop {} {\bibfield  {journal} {\bibinfo  {journal}
  {Sitzsungsber. Preuss. Akad. Wiss. Phys.-Math. Kl.}\ ,\ \bibinfo {pages}
  {105}} (\bibinfo {year} {1932})}\BibitemShut {NoStop}%
\bibitem [{\citenamefont {Birrel}\ and\ \citenamefont
  {Davies}(1982)}]{Birrel:1982}%
  \BibitemOpen
  \bibfield  {author} {\bibinfo {author} {\bibfnamefont {N.~D.}\ \bibnamefont
  {Birrel}}\ and\ \bibinfo {author} {\bibfnamefont {P.~C.~W.}\ \bibnamefont
  {Davies}},\ }\href@noop {} {\emph {\bibinfo {title} {Quantum fields in curved
  space}}}\ (\bibinfo  {publisher} {Cambridge University Press},\ \bibinfo
  {address} {U.K.},\ \bibinfo {year} {1982})\BibitemShut {NoStop}%
\bibitem [{\citenamefont {Freedman}\ and\ \citenamefont
  {Proeyen}(2012)}]{Freedman:2012}%
  \BibitemOpen
  \bibfield  {author} {\bibinfo {author} {\bibfnamefont {D.~Z.}\ \bibnamefont
  {Freedman}}\ and\ \bibinfo {author} {\bibfnamefont {A.~V.}\ \bibnamefont
  {Proeyen}},\ }\href@noop {} {\emph {\bibinfo {title} {Supergravity}}}\
  (\bibinfo  {publisher} {Cambridge University Press},\ \bibinfo {address}
  {U.K.},\ \bibinfo {year} {2012})\BibitemShut {NoStop}%
\bibitem [{\citenamefont {Kaup}(1968)}]{Kaup68}%
  \BibitemOpen
  \bibfield  {author} {\bibinfo {author} {\bibfnamefont {D.~J.}\ \bibnamefont
  {Kaup}},\ }\href@noop {} {\bibfield  {journal} {\bibinfo  {journal} {Phys.
  Rev.}\ }\textbf {\bibinfo {volume} {172}},\ \bibinfo {pages} {1331} (\bibinfo
  {year} {1968})}\BibitemShut {NoStop}%
\bibitem [{\citenamefont {Ruffini}\ and\ \citenamefont
  {Bonazzola}(1969)}]{Ruffini69}%
  \BibitemOpen
  \bibfield  {author} {\bibinfo {author} {\bibfnamefont {R.}~\bibnamefont
  {Ruffini}}\ and\ \bibinfo {author} {\bibfnamefont {S.}~\bibnamefont
  {Bonazzola}},\ }\href@noop {} {\bibfield  {journal} {\bibinfo  {journal}
  {Phys. Rev.}\ }\textbf {\bibinfo {volume} {187}},\ \bibinfo {pages} {1767}
  (\bibinfo {year} {1969})}\BibitemShut {NoStop}%
\bibitem [{\citenamefont {Liebling}\ and\ \citenamefont
  {Palenzuela}(2023)}]{Liebling:2012fv}%
  \BibitemOpen
  \bibfield  {author} {\bibinfo {author} {\bibfnamefont {S.~L.}\ \bibnamefont
  {Liebling}}\ and\ \bibinfo {author} {\bibfnamefont {C.}~\bibnamefont
  {Palenzuela}},\ }\href
  {https://doi.org/https://doi.org/10.1007/s41114-023-00043-4} {\bibfield
  {journal} {\bibinfo  {journal} {Living Rev. Relativity}\ }\textbf {\bibinfo
  {volume} {26}},\ \bibinfo {pages} {1} (\bibinfo {year} {2023})},\ \Eprint
  {https://arxiv.org/abs/1202.5809} {arXiv:1202.5809 [gr-qc]} \BibitemShut
  {NoStop}%
\bibitem [{\citenamefont {Brito}\ \emph {et~al.}(2016)\citenamefont {Brito},
  \citenamefont {Cardoso}, \citenamefont {Herdeiro},\ and\ \citenamefont
  {Radu}}]{Brito:2016}%
  \BibitemOpen
  \bibfield  {author} {\bibinfo {author} {\bibfnamefont {R.}~\bibnamefont
  {Brito}}, \bibinfo {author} {\bibfnamefont {V.}~\bibnamefont {Cardoso}},
  \bibinfo {author} {\bibfnamefont {C.~A.}\ \bibnamefont {Herdeiro}},\ and\
  \bibinfo {author} {\bibfnamefont {E.}~\bibnamefont {Radu}},\ }\href
  {https://doi.org/https://doi.org/10.1016/j.physletb.2015.11.051} {\bibfield
  {journal} {\bibinfo  {journal} {Physics Letters B}\ }\textbf {\bibinfo
  {volume} {752}},\ \bibinfo {pages} {291} (\bibinfo {year}
  {2016})}\BibitemShut {NoStop}%
\bibitem [{\citenamefont {Finster}\ \emph {et~al.}(1999)\citenamefont
  {Finster}, \citenamefont {Smoller},\ and\ \citenamefont
  {Yau}}]{Finster:1999}%
  \BibitemOpen
  \bibfield  {author} {\bibinfo {author} {\bibfnamefont {F.}~\bibnamefont
  {Finster}}, \bibinfo {author} {\bibfnamefont {J.}~\bibnamefont {Smoller}},\
  and\ \bibinfo {author} {\bibfnamefont {S.~T.}\ \bibnamefont {Yau}},\
  }\href@noop {} {\bibfield  {journal} {\bibinfo  {journal} {Phys. Rev.}\
  }\textbf {\bibinfo {volume} {D59}},\ \bibinfo {pages} {104020} (\bibinfo
  {year} {1999})}\BibitemShut {NoStop}%
\bibitem [{\citenamefont {Herdeiro}\ \emph {et~al.}(2017)\citenamefont
  {Herdeiro}, \citenamefont {Pombo},\ and\ \citenamefont
  {Radu}}]{Herdeiro:2017}%
  \BibitemOpen
  \bibfield  {author} {\bibinfo {author} {\bibfnamefont {C.~A.}\ \bibnamefont
  {Herdeiro}}, \bibinfo {author} {\bibfnamefont {A.~M.}\ \bibnamefont
  {Pombo}},\ and\ \bibinfo {author} {\bibfnamefont {E.}~\bibnamefont {Radu}},\
  }\href@noop {} {\bibfield  {journal} {\bibinfo  {journal} {Physics Letters
  B}\ }\textbf {\bibinfo {volume} {773}},\ \bibinfo {pages} {654–662}
  (\bibinfo {year} {2017})}\BibitemShut {NoStop}%
\bibitem [{\citenamefont {Daka}\ \emph {et~al.}(2019)\citenamefont {Daka},
  \citenamefont {Phan},\ and\ \citenamefont {Kain}}]{Daka:2019}%
  \BibitemOpen
  \bibfield  {author} {\bibinfo {author} {\bibfnamefont {E.}~\bibnamefont
  {Daka}}, \bibinfo {author} {\bibfnamefont {N.~N.}\ \bibnamefont {Phan}},\
  and\ \bibinfo {author} {\bibfnamefont {B.}~\bibnamefont {Kain}},\ }\href@noop
  {} {\bibfield  {journal} {\bibinfo  {journal} {Phys. Rev.}\ }\textbf
  {\bibinfo {volume} {D100}},\ \bibinfo {pages} {084042} (\bibinfo {year}
  {2019})}\BibitemShut {NoStop}%
\bibitem [{\citenamefont {Sun}\ \emph {et~al.}(2024)\citenamefont {Sun},
  \citenamefont {Cui}, \citenamefont {Huang},\ and\ \citenamefont
  {Wang}}]{Sun2024a}%
  \BibitemOpen
  \bibfield  {author} {\bibinfo {author} {\bibfnamefont {S.-X.}\ \bibnamefont
  {Sun}}, \bibinfo {author} {\bibfnamefont {S.-Y.}\ \bibnamefont {Cui}},
  \bibinfo {author} {\bibfnamefont {L.-X.}\ \bibnamefont {Huang}},\ and\
  \bibinfo {author} {\bibfnamefont {Y.-Q.}\ \bibnamefont {Wang}},\ }\href
  {https://doi.org/10.1140/epjc/s10052-024-12942-z} {\bibfield  {journal}
  {\bibinfo  {journal} {Eur. Phys. J. C}\ }\textbf {\bibinfo {volume} {84}},\
  \bibinfo {pages} {699} (\bibinfo {year} {2024})},\ \Eprint
  {https://arxiv.org/abs/2310.10267} {arXiv:2310.10267 [gr-qc]} \BibitemShut
  {NoStop}%
\bibitem [{\citenamefont {Liang}\ \emph {et~al.}(2024)\citenamefont {Liang},
  \citenamefont {Sun}, \citenamefont {Ren},\ and\ \citenamefont
  {Wang}}]{Liang2024a}%
  \BibitemOpen
  \bibfield  {author} {\bibinfo {author} {\bibfnamefont {C.}~\bibnamefont
  {Liang}}, \bibinfo {author} {\bibfnamefont {S.-X.}\ \bibnamefont {Sun}},
  \bibinfo {author} {\bibfnamefont {J.-R.}\ \bibnamefont {Ren}},\ and\ \bibinfo
  {author} {\bibfnamefont {Y.-Q.}\ \bibnamefont {Wang}},\ }\href
  {https://doi.org/10.1140/epjc/s10052-023-12345-6} {\bibfield  {journal}
  {\bibinfo  {journal} {Eur. Phys. J. C}\ }\textbf {\bibinfo {volume} {84}},\
  \bibinfo {pages} {14} (\bibinfo {year} {2024})},\ \Eprint
  {https://arxiv.org/abs/2306.11437} {arXiv:2306.11437 [gr-qc]} \BibitemShut
  {NoStop}%
\bibitem [{\citenamefont {Chapman}\ and\ \citenamefont
  {Leiter}(1976)}]{Chapman:1976}%
  \BibitemOpen
  \bibfield  {author} {\bibinfo {author} {\bibfnamefont {T.~C.}\ \bibnamefont
  {Chapman}}\ and\ \bibinfo {author} {\bibfnamefont {D.~J.}\ \bibnamefont
  {Leiter}},\ }\href@noop {} {\bibfield  {journal} {\bibinfo  {journal}
  {American Journal of Physics}\ }\textbf {\bibinfo {volume} {44}},\ \bibinfo
  {pages} {858} (\bibinfo {year} {1976})}\BibitemShut {NoStop}%
\bibitem [{\citenamefont {Pollock}(2010)}]{Pollock:2010}%
  \BibitemOpen
  \bibfield  {author} {\bibinfo {author} {\bibfnamefont {M.~D.}\ \bibnamefont
  {Pollock}},\ }\href@noop {} {\bibfield  {journal} {\bibinfo  {journal} {Acta
  Physica Polonica B}\ }\textbf {\bibinfo {volume} {41}},\ \bibinfo {pages}
  {1827} (\bibinfo {year} {2010})}\BibitemShut {NoStop}%
\bibitem [{\citenamefont {Collas}\ and\ \citenamefont
  {Klein}(2018)}]{Collas:2018}%
  \BibitemOpen
  \bibfield  {author} {\bibinfo {author} {\bibfnamefont {P.}~\bibnamefont
  {Collas}}\ and\ \bibinfo {author} {\bibfnamefont {D.}~\bibnamefont {Klein}},\
  }\href@noop {} {\emph {\bibinfo {title} {The {D}irac equation in general
  relativity: a guide for calculations}}}\ (\bibinfo  {publisher} {Springer},\
  \bibinfo {address} {Switzerland},\ \bibinfo {year} {2018})\BibitemShut
  {NoStop}%
\bibitem [{\citenamefont {Shapiro}(2022)}]{Shapiro:2022}%
  \BibitemOpen
  \bibfield  {author} {\bibinfo {author} {\bibfnamefont {I.~L.}\ \bibnamefont
  {Shapiro}},\ }\href@noop {} {\bibfield  {journal} {\bibinfo  {journal}
  {Universe}\ }\textbf {\bibinfo {volume} {8}},\ \bibinfo {pages} {586}
  (\bibinfo {year} {2022})}\BibitemShut {NoStop}%
\bibitem [{\citenamefont {Itzykson}\ and\ \citenamefont
  {Zuber}(1980)}]{Itzykson1980}%
  \BibitemOpen
  \bibfield  {author} {\bibinfo {author} {\bibfnamefont {C.}~\bibnamefont
  {Itzykson}}\ and\ \bibinfo {author} {\bibfnamefont {J.-B.}\ \bibnamefont
  {Zuber}},\ }\href@noop {} {\emph {\bibinfo {title} {Quantum Field Theory}}}\
  (\bibinfo  {publisher} {Dover},\ \bibinfo {address} {N.Y.},\ \bibinfo {year}
  {1980})\BibitemShut {NoStop}%
\bibitem [{\citenamefont {Kaku}(1993)}]{Kaku1993}%
  \BibitemOpen
  \bibfield  {author} {\bibinfo {author} {\bibfnamefont {M.}~\bibnamefont
  {Kaku}},\ }\href@noop {} {\emph {\bibinfo {title} {Quantum Field Theory: A
  Modern Introduction}}}\ (\bibinfo  {publisher} {Oxford University Press},\
  \bibinfo {address} {New York, oxford},\ \bibinfo {year} {1993})\BibitemShut
  {NoStop}%
\bibitem [{\citenamefont {Weinberg}(1995)}]{Weinberg1995}%
  \BibitemOpen
  \bibfield  {author} {\bibinfo {author} {\bibfnamefont {S.}~\bibnamefont
  {Weinberg}},\ }\href@noop {} {\emph {\bibinfo {title} {The Quantum Theory of
  Fields}}}\ (\bibinfo  {publisher} {Cambridge University Press},\ \bibinfo
  {year} {1995})\BibitemShut {NoStop}%
\bibitem [{\citenamefont {Jhangiani}(1975)}]{Jhangiani:1975}%
  \BibitemOpen
  \bibfield  {author} {\bibinfo {author} {\bibfnamefont {V.}~\bibnamefont
  {Jhangiani}},\ }\href@noop {} {\bibfield  {journal} {\bibinfo  {journal}
  {Foundations of physics}\ }\textbf {\bibinfo {volume} {7}},\ \bibinfo {pages}
  {111} (\bibinfo {year} {1975})}\BibitemShut {NoStop}%
\bibitem [{\citenamefont {Fleury}\ \emph {et~al.}(2023)\citenamefont {Fleury},
  \citenamefont {Hammad},\ and\ \citenamefont {Sadeghi}}]{Fleury:2023}%
  \BibitemOpen
  \bibfield  {author} {\bibinfo {author} {\bibfnamefont {N.}~\bibnamefont
  {Fleury}}, \bibinfo {author} {\bibfnamefont {F.}~\bibnamefont {Hammad}},\
  and\ \bibinfo {author} {\bibfnamefont {P.}~\bibnamefont {Sadeghi}},\
  }\href@noop {} {\bibfield  {journal} {\bibinfo  {journal} {Symmetry}\
  }\textbf {\bibinfo {volume} {15}},\ \bibinfo {pages} {432} (\bibinfo {year}
  {2023})}\BibitemShut {NoStop}%
\bibitem [{\citenamefont {Burgess}(2022)}]{Burgess:2022}%
  \BibitemOpen
  \bibfield  {author} {\bibinfo {author} {\bibfnamefont {M.}~\bibnamefont
  {Burgess}},\ }\href@noop {} {\emph {\bibinfo {title} {Classical covariant
  fields}}}\ (\bibinfo  {publisher} {Cambridge Universtity Press},\ \bibinfo
  {address} {Cambridge, UK},\ \bibinfo {year} {2022})\BibitemShut {NoStop}%
\bibitem [{\citenamefont {Dolan}\ and\ \citenamefont
  {Dempey}(2015)}]{Dolan:2015}%
  \BibitemOpen
  \bibfield  {author} {\bibinfo {author} {\bibfnamefont {S.~R.}\ \bibnamefont
  {Dolan}}\ and\ \bibinfo {author} {\bibfnamefont {D.}~\bibnamefont {Dempey}},\
  }\href {https://doi.org/10.1088/0264-9381/32/18/184001} {\bibfield  {journal}
  {\bibinfo  {journal} {Class. Quantum Grav.}\ }\textbf {\bibinfo {volume}
  {32}},\ \bibinfo {pages} {184001} (\bibinfo {year} {2015})}\BibitemShut
  {NoStop}%
\bibitem [{\citenamefont {Alcubierre}(2008)}]{Alcubierre08a}%
  \BibitemOpen
  \bibfield  {author} {\bibinfo {author} {\bibfnamefont {M.}~\bibnamefont
  {Alcubierre}},\ }\href@noop {} {\emph {\bibinfo {title} {Introduction to
  $3+1$ Numerical Relativity}}}\ (\bibinfo  {publisher} {Oxford Univ. Press},\
  \bibinfo {address} {New York},\ \bibinfo {year} {2008})\BibitemShut {NoStop}%
\bibitem [{\citenamefont {Synge}(1960)}]{Synge60}%
  \BibitemOpen
  \bibfield  {author} {\bibinfo {author} {\bibfnamefont {J.~L.}\ \bibnamefont
  {Synge}},\ }\href@noop {} {\emph {\bibinfo {title} {Relativity: The general
  theory}}}\ (\bibinfo  {publisher} {North Holland},\ \bibinfo {address}
  {Amsterdam},\ \bibinfo {year} {1960})\BibitemShut {NoStop}%
\bibitem [{\citenamefont {Newman}\ and\ \citenamefont
  {Penrose}(1966)}]{Newman-Penrose-1966}%
  \BibitemOpen
  \bibfield  {author} {\bibinfo {author} {\bibfnamefont {E.~T.}\ \bibnamefont
  {Newman}}\ and\ \bibinfo {author} {\bibfnamefont {R.}~\bibnamefont
  {Penrose}},\ }\href {https://doi.org/10.1063/1.1931221} {\bibfield  {journal}
  {\bibinfo  {journal} {J. Math. Phys.}\ }\textbf {\bibinfo {volume} {7}},\
  \bibinfo {pages} {863} (\bibinfo {year} {1966})}\BibitemShut {NoStop}%
\bibitem [{\citenamefont {Forger}\ and\ \citenamefont
  {Römer}(2004)}]{Forger:2004}%
  \BibitemOpen
  \bibfield  {author} {\bibinfo {author} {\bibfnamefont {M.}~\bibnamefont
  {Forger}}\ and\ \bibinfo {author} {\bibfnamefont {H.}~\bibnamefont
  {Römer}},\ }\href@noop {} {\bibfield  {journal} {\bibinfo  {journal} {Annals
  of Physics}\ }\textbf {\bibinfo {volume} {309}},\ \bibinfo {pages} {306}
  (\bibinfo {year} {2004})}\BibitemShut {NoStop}%
\end{thebibliography}%


\end{document}